\documentclass[final,3p,times,authoryear]{elsarticle}

\usepackage{amssymb}

\usepackage[table]{xcolor}
\usepackage{soul}
\usepackage{subcaption}
\usepackage{graphicx}
\usepackage{multirow}
\usepackage{makecell}
\usepackage{pdflscape}
\usepackage{longtable}
\usepackage{bm}
\usepackage{amsmath}
\usepackage{tabularx,booktabs}
\usepackage{longtable}
\usepackage{multicol}
\usepackage{url}

\newcolumntype{C}[1]{%
 >{\vbox to 11.5ex\bgroup\vfill\centering}%
 p{#1}%
 <{\egroup}}

\journal{Transportation Research Part C: Emerging Technologies}

\begin{document}

\begin{frontmatter}



\title{Anomaly detection and classification in traffic flow data from fluctuations in the flow-density relationship}

\author[label1]{Kieran Kalair\corref{cor1}}
\ead{k.kalair@warwick.ac.uk}
\ead[url]{https://warwick.ac.uk/fac/sci/mathsys/people/students/2016intake/kalair/}
\cortext[cor1]{Corresponding author}

\author[label1,label2,label3]{Colm Connaughton}
\ead{C.P.Connaughton@warwick.ac.uk}
\ead[url]{https://www.colmconnaughton.net}
\address[label1]{Centre for Complexity Science, University of Warwick, Gibbet Hill Road, Coventry CV4 7AL, United Kingdom.}
\address[label2]{Mathematics Institute, University of Warwick, Gibbet Hill Road, Coventry CV4 7AL, United Kingdom.}
\address[label3]{London Mathematical Laboratory, 8 Margravine Gardens, London, W6 8RH, United Kingdom}




\begin{abstract}
We describe and validate a novel data-driven approach to the real time detection and classification of traffic anomalies based on the 
identification of atypical fluctuations in the relationship between density and flow.
For aggregated data under stationary conditions, flow and density are related by the fundamental diagram. 
However, high resolution data obtained from modern sensor networks is generally non-stationary and disaggregated.
Such data consequently show significant statistical fluctuations. 
These fluctuations are best described using a bivariate probability distribution in the density-flow plane.
By applying kernel density estimation to high-volume data from the UK National Traffic Information Service (NTIS), we empirically construct these distributions for London's M25 motorway.
Curves in the density-flow plane are then constructed, analogous to quantiles of univariate distributions.
These curves quantitatively separate atypical fluctuations from typical traffic states. 
Although the algorithm identifies anomalies in general rather than specific events, we find that fluctuations outside the 95\% probability curve correlate strongly with the spikes in travel time associated with significant congestion events. 
Moreover, the size of an excursion from the typical region provides a simple, real-time measure of the severity of detected anomalies. 
We validate the algorithm by benchmarking its ability to identify labelled events in historical NTIS data against some commonly used methods from the literature. 
Detection rate, time-to-detect and false alarm rate are used as metrics and found to be generally comparable except in situations when the speed distribution is bi-modal. 
In such situations, the new algorithm achieves a much lower false alarm rate without suffering significant degradation on the other metrics. 
This method has the additional advantages of being self-calibrating and adaptive: the curve marking atypical behaviour is different for each section of road and can evolve in time as the data changes, for example, due to long-term roadworks.
\end{abstract}


\end{frontmatter}

\section{Introduction}\label{sec:Introduction}

An important problem in highway traffic management is event detection: monitoring the data streams from embedded sensors that monitor the infrastructure in real-time and raising flags to alert operators to potential problems as they happen.
A high level understanding of the typical behaviour of traffic on a section of road is provided by the fundamental diagram which assumes a functional relationship between vehicle density and vehicle flow.
See \cite{GreenshieldsFD} for a review. 
This relationship is valid for `near steady-state' conditions \cite{bivariate_relations_in_nearly_stationary_highway_traffic}, where data has been aggregated on an appropriate time-scale and therefore describes the average behaviour of steady-state traffic.
In this paper, we explore how analysis of the fluctuations in the high resolution, non-aggregated and non-steady-state flow-density data can be used to perform real time anomaly detection, by identifying periods of time when the flow pattern on a stretch of road is atypical. 
Our objective is to provide a systematic, data driven definition of `atypical' behaviour and to suggest a principled way of tracking the severity of events as they occur in real time. 

We emphasise from the outset that anomaly detection and event detection are not necessarily the same thing. 
Network operators, road users and other stakeholders are generally interested in {\em events} such as collisions, obstructions or lane closures.
Such events may cause disruption resulting in an atypical flow pattern that can be picked up by an anomaly detection system. 
However, this is not necessarily the case a-priori.
For example, a broken down vehicle during a period of low demand may not impact upon the traffic flow at all. This event would therefore not result in any anomaly.
Conversely,  anomalies like so-called ``phantom traffic jams" are sometimes observed that are not associated with any underlying event or cause.
In this paper, we investigate the correspondence between events and anomalies in detail using labelled event data to establish that the overlap between the two is sufficient to be of use in practice. 

We present below a purely data-driven method for identifying significant deviations from the typical behaviour of the traffic on a road section using time series data from UK motorways provided by the National Traffic Information Service (NTIS).
The key idea is exploit the large volume of data to arrive at a robust understanding of the range of typical fluctuations about the density-flow relationship that is tailored to each road section and to interpret big excursions from this range as proxies for significant events. 
We adopt a ``macroscopic" perspective meaning that the objective is to detect when the collective behaviour of the traffic on a road section is unusual in some sense. 
This macroscopic perspective means that we do not directly detect individual events like collisions, stationary vehicles or lane closures but rather changes in flow patterns that can be a consequence of such events.
To validate our approach, we measure how periods identified as atypical relate to existing labels of incidents on roads provided by NTIS, and benchmark this comparison against some other available methods from the literature.
Although we are initially agnostic about {\em why} a particular configuration is atypical we nevertheless find our criteria typically detect the most extreme congestion events.
Our entirely unsupervised, data-driven methodology offers comparable or better performance than some existing models based on learning patterns from labelled data.

The main advantages of our approach, compared to specifically trained incident detection methods are as follows.
Firstly, calibration is straightforward, requiring no labelled data to train, instead only a representative sample of data taken from a stretch of road is required. 
Since road incidents are rare in absolute terms, this means that collecting a dataset to calibrate the method on is far easier than collecting a representative set of incidents on a specific section of road. 
We show in \ref{subsec:Stability} that 3 weeks of data is sufficient to identify stable periods of typical behaviour, where as it is highly unlikely that the same time-period would provide a sufficient number of events on a single section of road to adequately train a deep learning model for example.
Secondly, it is self adaptive. One can account for long-term changes in the behaviour on certain sites by simply reapplying the methodology on a new representative sample. 
Thirdly, it is fully interpretable, with a clear reason as to why a data-point is marked atypical, and allows for direct comparison between different locations around a network with reference to the typical behaviour at each site.
Finally, it is shown to also capture periods of labelled event data, showing that physical incidents on the network such as accidents and breakdowns correspond to a subset of the atypical periods a section of road experiences.
In terms of practical application, this analysis should be thought of as a filter that could be help human operators of the smart motorways infrastructure to prioritise their attention.
Regardless of cause, it may be useful to inform operators which parts of their infrastructure are experiencing the most atypical situations, and hence where intervention might be able to help the system to return  to normal.

The remainder of this paper is structured as follows.
First we review the literature on automatic incident detection, then give an overview of our dataset and detail how we define typical behaviour on a link.
We then consider time-scales of atypical events in our data, and compare this to existing event flags.
After, we discuss when to raise atypical event flags, and methodologies to determine severity of events, and consider how our events correlate with travel time spikes.
Finally, we consider how our method performs if one were to use it only to detect labelled events in the dataset, comparing it to existing event detection methodologies from the literature.
We then summarize our findings and results.

\section{Literature review}\label{sec:LitReview}

Multiple approaches have been proposed for the task of anomaly detection as applied to traffic incident detection.
They broadly fall into two classes.
The first class uses data from individual vehicles.  Examples include the use of data from automatic vehicle identification systems  as in \cite{automatic_vehicle_identification_technology_based_freeway_incident_detection}, the identification of single vehicles using cameras  in \cite{Image_Texture_Congestion} or the use of GPS and social media data from navigation apps like Waze in \cite{evaluating_the_reliability_coverage_and_added_value_of_crowdsourced_traffic_incident_reports_from_waze}.
The second class uses time series data from embedded loop sensors, usually aggregated measures of vehicle counts/occupancy and/or speed. 
These methods do not identify individual vehicles and are often statistical in nature.
Our method is of the second class and in this section we summarise some of the relevant literature.

Firstly, basic pattern matching is an approach which has proven to be successful in some contexts.
The idea is to determine some features or thresholds in time series data that indicate event and non-event windows.
Early and well known examples of this are the California algorithms and their variants, discussed in \cite{california_algorithm_original} and \cite{incident_detection_a_bayesian_approach}.
These algorithms compare occupancy values at adjacent sensors, looking for when the pair-wise comparison metrics pass some thresholds.
As a specific example, for two sensors $i$ and $i+1$, where $i+1$ is downstream of $i$, the California Algorithm \# 7 computes the occupancy difference at time $t$ between sensor $i$ and $i+1$, as-well as the relative occupancy difference between the two.
Finally, it compares the downstream occupancy to some threshold, and outputs an incident free, tentative incident, incident occurred or incident continuing flag.
Modifications are made to the logic for other variants.
To apply this in practice, one needs labelled time-series data at the single sensor (loop) level, with labels specifying when and where incidents occurred.

A second popular approach is the so-called McMaster algorithm, based on catastrophe theory, discussed in \cite{catastrophe_theory_and_patterns_in_thirty_second_freeway_traffic_data_implications_for_incident_detection}, and further developed in \cite{mcmaster_distinguising_between_incident_congestion_and_recurrent_congestion_a_proposed_logic}, \cite{congestion_identification_aspects_of_the_mcmaster_incident_dete} and \cite{online_testing_of_the_mcmaster_incident_detection_algorithm_under_recurrent_congestion}.
The initial observation for the approach is that in uncongested regimes, occupancy-flow data typically fits a linear relationship with little scatter.
From this uncongested data, one can construct a lower bound of occupancy-flow data through some parametric form, and also define critical occupancies and flows.
Together, these segment the occupancy-flow diagram into sections that are then considered to represent system states, either uncongested, bottleneck flow or congested.
If a sensor is considered to be congested, then the next sensor downstream is examined, determining if this is also congested, or appears uncongested.
If the downstream detector is uncongested, then there is likely an incident between the two examined detectors, otherwise there is likely a problem somewhere further downstream that has impacted further upstream.
Such an approach requires calibration of particular thresholds, which may differ from station to station and also requires fitting some parametric form of the lower occupancy-flow bound.
Subsequent work on the McMaster algorithm has explored how to correctly calibrate these parameters, with \cite{traffic_congestion_pattern_detection_using_an_improved_mcmaster_algorithm} considering a particle-swarm approach. 

However, calibrating any form of incident detection model is generally difficult because of data quality problems \cite{a_wavelet_based_freeway_incident_detection_algorithm_with_adapting_threshold_parameters}. 
Recently, alternative methodologies have emerged for clustering of data into typical and anomalous states to discover non-recurrent congestion as in \cite{spatio_temporal_clustering_for_non_recurrent_traffic_congestion_detection_on_urban_road_networks}, where higher than expected journey times are distinguished from typical ones using expected journey times scaled by a congestion factor.
Further, specific vehicle trajectories were clustered in \cite{Online_Trajectory_Clustering}, creating a probabilistic tree-like clustered structure that was used to identify anomalous events.  
Additionally \cite{HMM_No_Tracking} use a Gaussian Mixture Hidden Markov Model to extract features and classify traffic states from video data into categories of: empty, open flow, mild congestion, heavy congestion and stopped.

A third important approach is the standard normal deviate (SND) methodology, an early use being in \cite{incident_detection_on_urban_freeways}.
These methods attempt to construct a mean value $\mu$ for a traffic variable on a particular time interval, and some measure of variation $\sigma$. 
Then, one can question if new data, for example speed, is below some threshold value of $\mu - c\sigma$, determining the parameter $c$ to best fit available data.
Of course, if one simply uses a mean and standard deviation for this, there is a risk that outliers may influence these thresholds significantly.
To account for this, one can pre-filter a dataset to remove any known incident periods before computation as in \cite{using_probe_measured_travel_times_to_detect_major_freeway_incidents_in_houston_texas}, or use robust summary statistics as in \cite{delays_caused_by_incidents_data_driven_approach}.
Extensions of this idea include incorporation of spatial information, seen in \cite{quantification_of_nonrecurrent_congestion_delay_caused_by_freeway_accidents_and_analysis_of_causal_factors}, where the total delay time caused by traffic accidents was examined by considering the propagation backwards in space and forwards in time of traffic queues.
Further, \cite{data_driven_parallelizable_traffic_incident_detection_using_spatial_temporally_denoise_robust_thresholds} recently considered robust construction of noise thresholds incorporating smoothing, first determining a robust threshold for each particular interval of a day, then smoothing these using a bilateral filter to incorporate spatial-temporal correlations, assuming nearby windows of space and time should have reasonably similar thresholds.
Closely related to the SND family of algorithms are various approached based on change-point detection.
Such ideas were considered in \cite{Evaluating_Recurring_Traffic_Congestion_using_Change_Point_Regression_and_Random_Variation_Markov_Structured_Model}, where the authors use a Markov structured model to predict the transitions between a congested and free flow state.
Additionally, \cite{application_of_time_series_analysis_techniques_to_freeway_incident_detection} based their methodology on forecasting with ARIMA models, classifying an incident period as one where the data lay outside the confidence bounds of their forecasting model.
Approaches such as this ones assume that the confidence limits provided by a forecasting model are indeed representative of the true uncertainty present in the data, and as a result require extensive time-series validation.
The question of what is an optimal model to forecast traffic states, in the wide range of potential scenarios is also a complex one.
These methods have the advantage that they can be applied without any changes to both the individual loop level, or at the aggregated, macroscopic `link' level. 

Finally we mention that machine learning and deep learning methodologies have seen significant developments in their application to event detection in recent years.
Some of these use traditional datasets but new methodologies, for example \cite{hetero_convlstm_a_deep_learning_approach_to_traffic_accident_prediction_on_heterogenous_spatial_temporal_data} which incorporated traffic data along with weather information and spatial structure into a convolutional-LSTM model to predict the time and location of traffic accidents. 
Additional deep learning approaches include \cite{traffic_monitoring_and_anomaly_detection_based_on_simulation_of_luxembourg_road_network}, which used a convolutional neural network for anomaly detection on the Luxembourg road network and \cite{a_deep_learning_approach_for_traffic_incident_detection_in_urban_networks} which again used a convolutional neural network but for incident detection on urban roads in London.
For such methods, computational intensity and data quality issues continue to act as barriers.  

The approach proposed in this paper takes some ideas from this broad range of literature, but has distinct advantages and applicability.
Like the McMaster algorithm, we segment the density-flow diagram into distinct regions that separate the typical from the atypical.
Unlike the McMaster algorithm, we do not pre-specify how the segmentation of the diagram should be done. 
Instead, the segmentation is defined by the data itself in a parameter-free way, similar to the philosophy underpinning the SND algorithm.
As is the case for SND, using data to determine its own bounds requires that statistically robust approaches should be used.
As such, our method should be seen as combining the idea of segmenting the density-flow diagram with the robust approaches in the literature, whilst side-stepping the two difficult problems of obtaining labelled data and calibrating an event detection system.

\section{National Traffic Information Service (NTIS) data resources}
\label{sec:data}

\subsection{Speed and flow data}\label{subsec:data}
 
Before describing our approach, we briefly summarise the data used. 
This data was obtained from Highways England's National Traffic Information Service (NTIS) 
\footnote{Technical details of the NTIS data feeds are available at \url{http://www.trafficengland.com/services-info}}.
NTIS represents the UK Strategic Road Network graphically as a set of nodes and links with a link representing a section of road. 
Within a link there are no slip roads and no changes to the number of lanes. 
Each link has a number of induction loop sensors embedded in the road surface that report average vehicle counts and speeds at 1 minute intervals.  
Data from individual loops is then aggregated to provide the average speed (in km/hr) and average flow (in vehicles/hr) on each link at 1 minute intervals. 
Derived estimates of link level travel times are also provided.
In this work, we focus on these link-level time series.
Specifically, we use data from the M25 London orbital, one of the busiest smart motorways in the UK.
In total, we collect data starting from April 7th 2017 and ending November 1st 2018, covering the M25 clockwise, with links varying in length between 200 and 10,000 meters.
The majority of this data is used to validate the models performance on event flags, with the actual methodology we propose requiring far less to implement.
As a result, we will discuss only a subset of it for most of the work, and then consider our methodologies performance on a much longer set in section \ref{sec:Validation}.

\subsection{NTIS event flags}\label{subsec:events}

The main purpose of this work is to propose a new method for automatically detecting unusual periods in NTIS time series data. 
Since the NTIS system already has methods to flag certain types of events, it is important to discuss these event flags since we will make some comparisons later.
Recorded in the system are various categories of events, including accidents, obstructions, breakdowns and events denoted `Deviation from Profile' (DFP).
DFP events are declared when the travel time on a link exceeds some profile time by a pre-determined threshold, for at-least 5 consecutive minutes, with the algorithm used to generate profiles not being publicly known.
Each of these represent non-recurrent congestion events on the network, and we annotate our time-series with these flags.
The system also provides flags specifying when maintenance work took place and describing weather conditions, however we do not consider these as non-recurrent congestion events during validation. 
All categories of flag are manually entered into the dataset by operators, after being informed of problems via phone-call by local traffic management centres.
As a result, it is known that the dataset contains missing flags where operators did not have time to register an event.
Also, we are informed by system experts that the flags that are entered have a potential variable delay on the start time. 

\section{Data-driven characterisation of the flow-density relationship and identification of atypical configurations}
\label{sec:method}

\subsection{Using Kernel Density Estimation to model the flow-density relationship}\label{subsec:Interp}

\begin{figure}[ht!]
	\captionsetup{width=\textwidth}
	\begin{subfigure}{\textwidth}
		\includegraphics[width=0.32\textwidth]{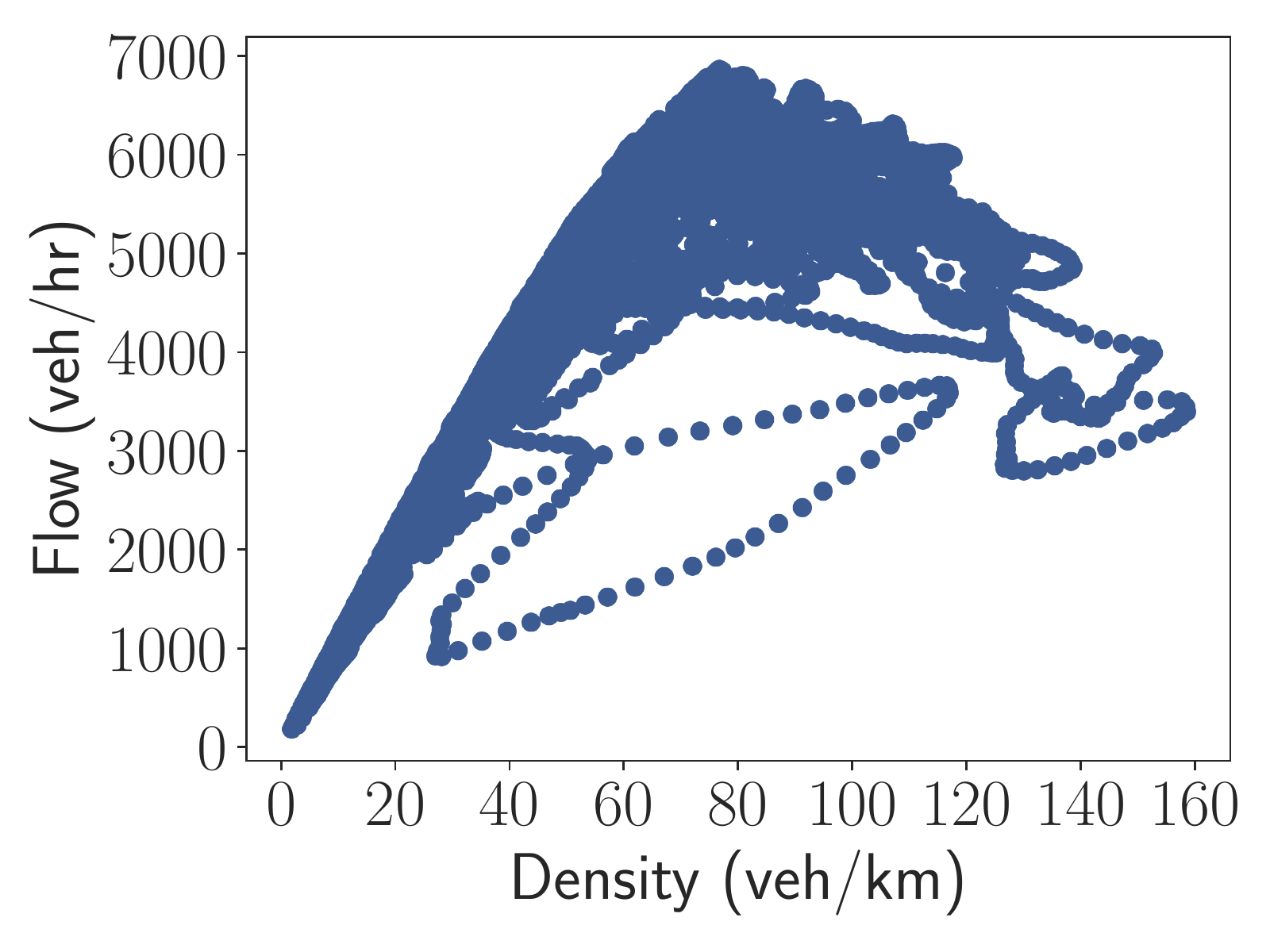}
		\includegraphics[width=0.32\textwidth]{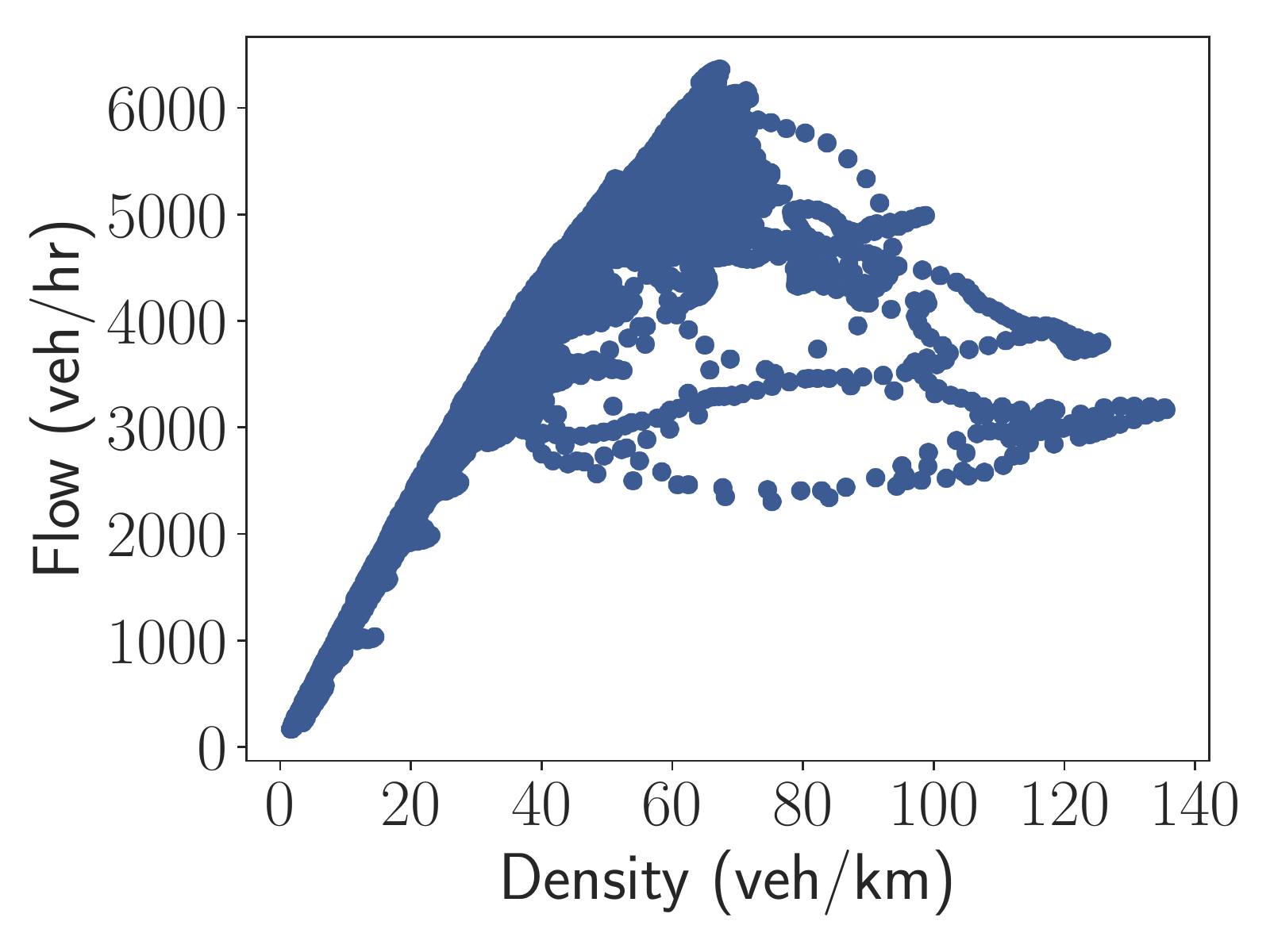}
		\includegraphics[width=0.32\textwidth]{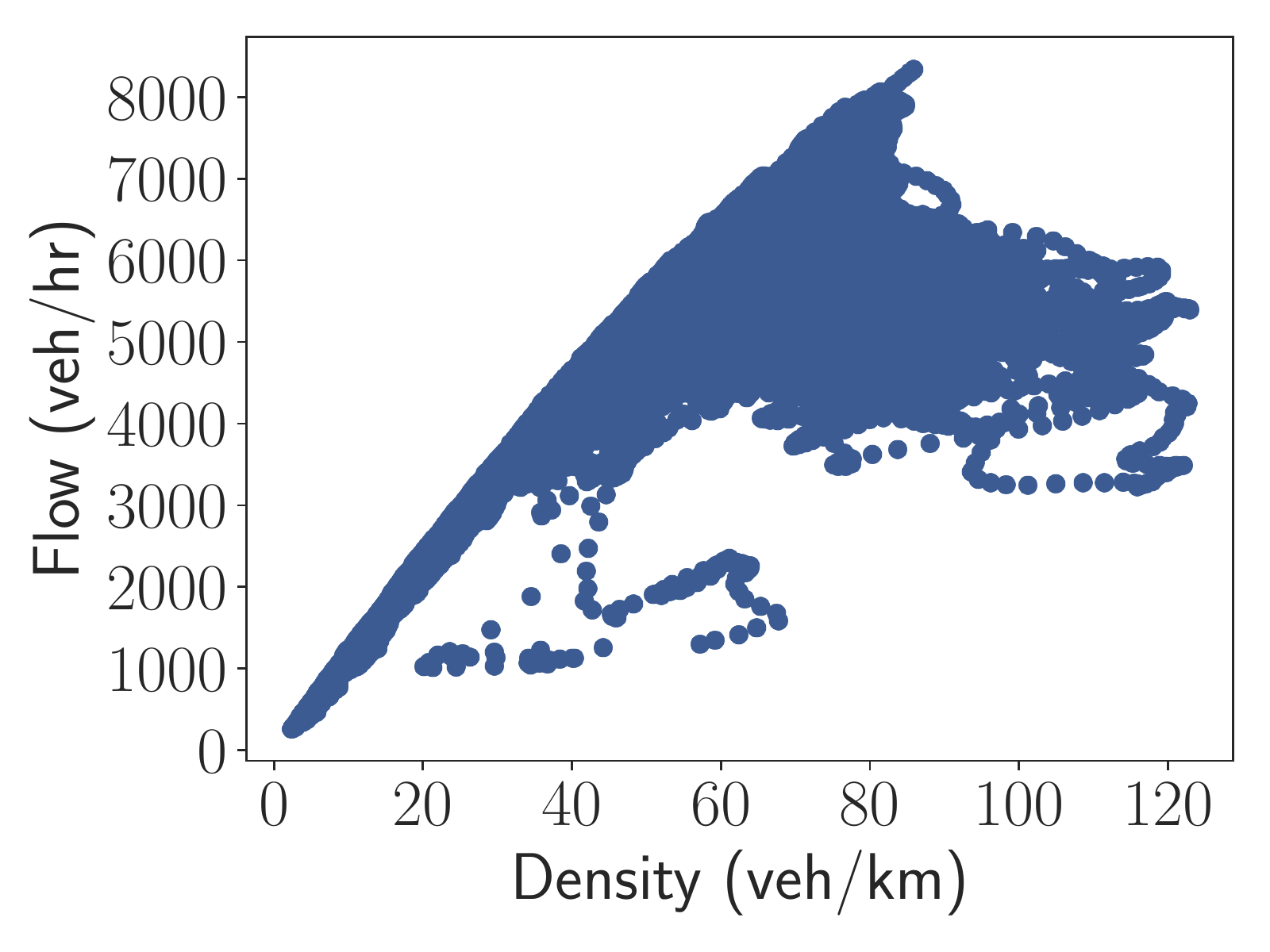}
		\caption{ Example density-flow series using time mean speed data and a density inferred from this. }\label{fig:TMSExampleFundDiag}
	\end{subfigure}
	\begin{subfigure}{\textwidth}
		\includegraphics[width=0.32\textwidth]{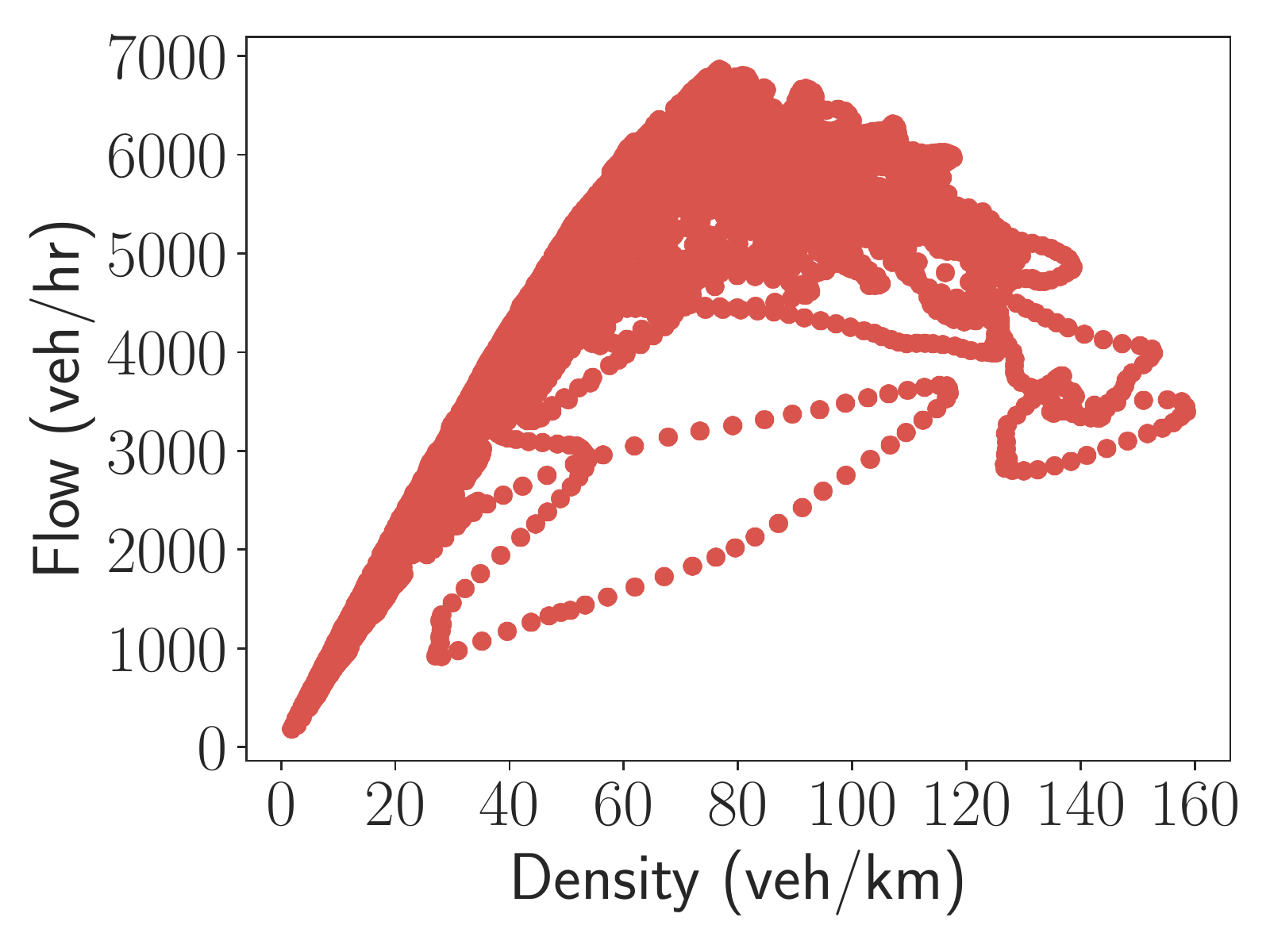}
		\includegraphics[width=0.32\textwidth]{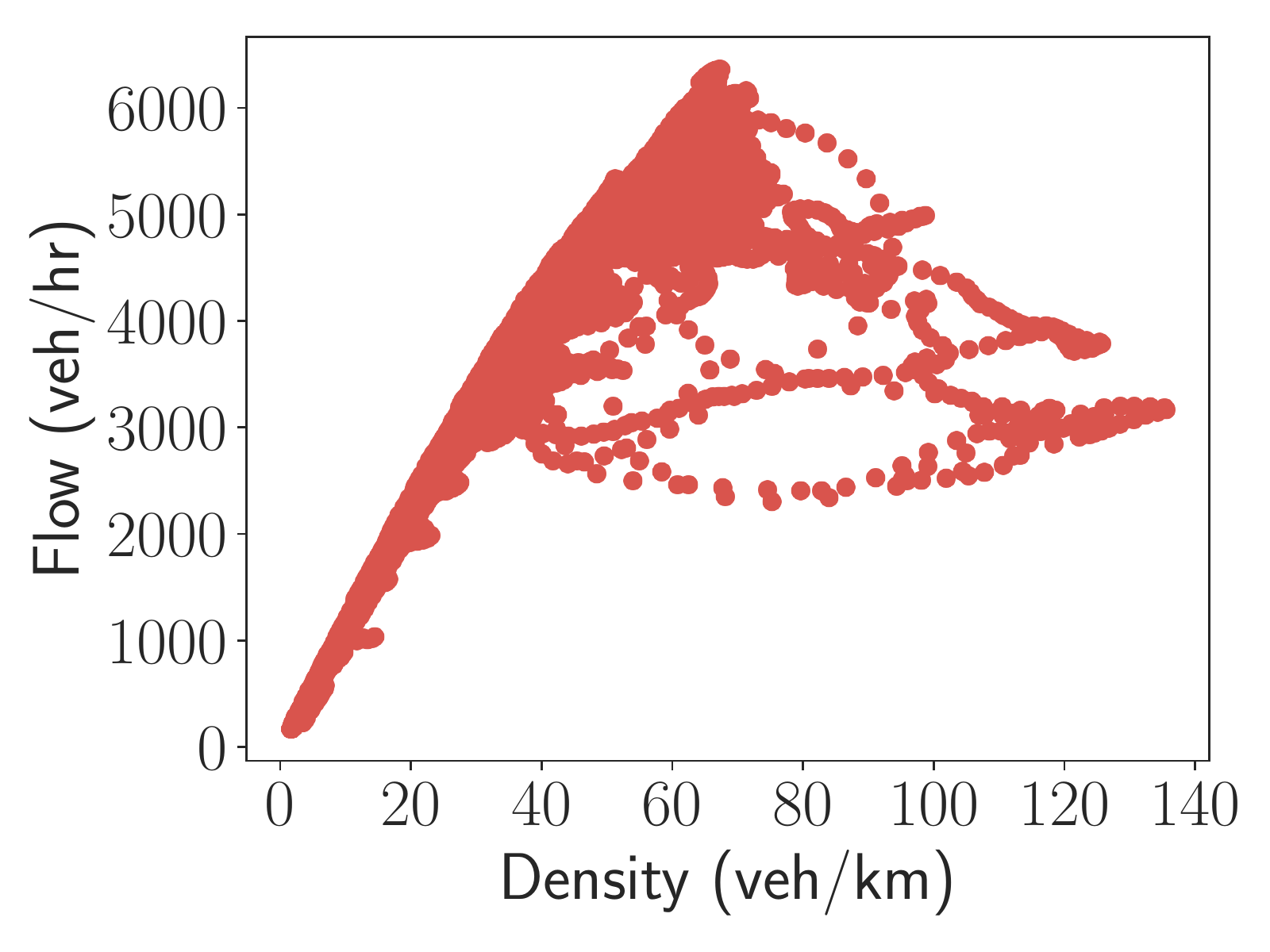}
		\includegraphics[width=0.32\textwidth]{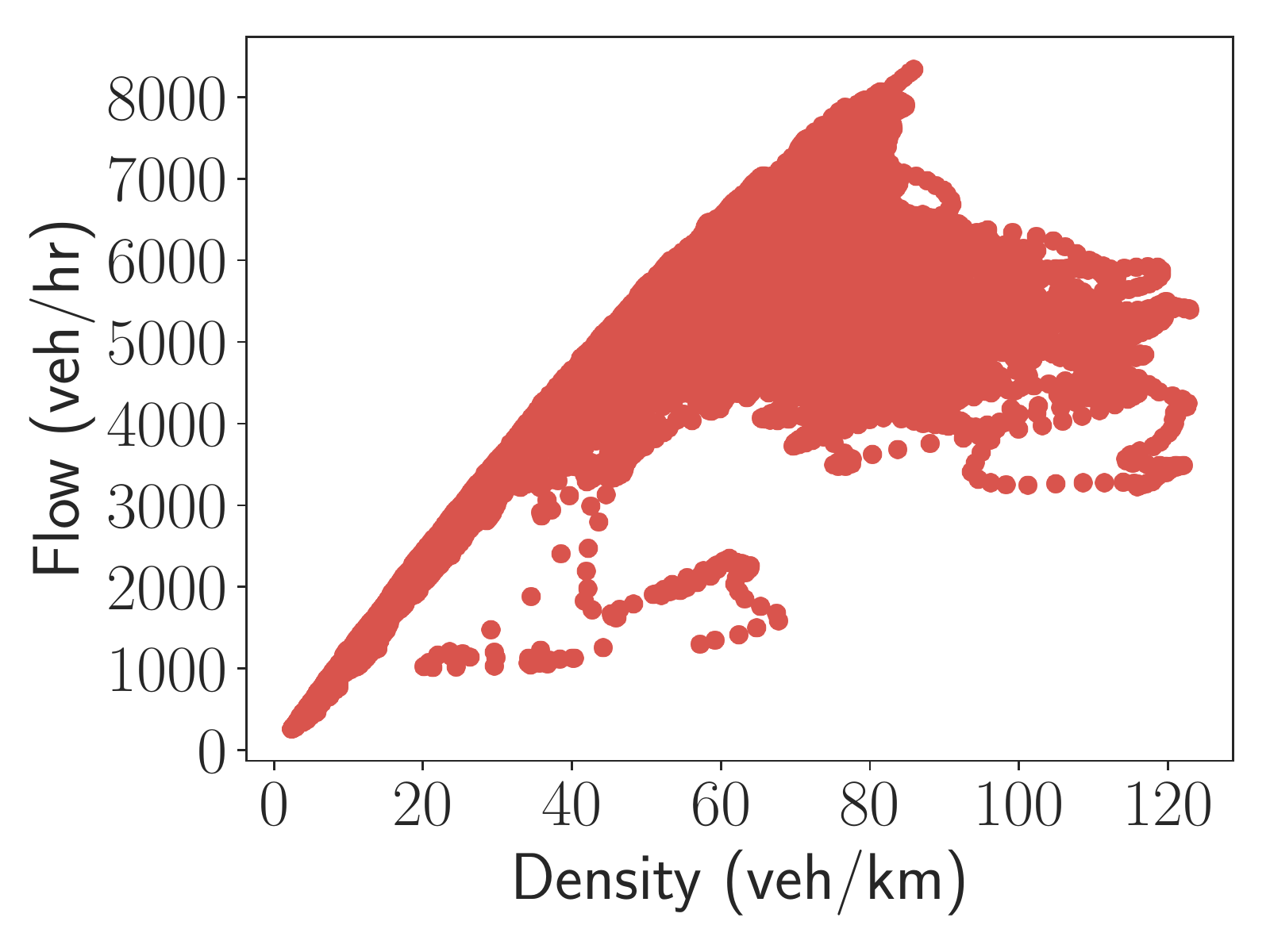}
		\caption{ Example density-flow series using approximated space mean speed data and a density inferred from this. }\label{fig:SMSExampleFundDiag}
	\end{subfigure}
	\caption[SMS vs TMS Diagrams]{Example density-flow series using 10 weeks of M25 data, taken from 3 distinct links. Left column: link between junctions 2 and 3 (3636 metres long). Middle column: link between junctions 3 and 4 (4525 metres long). Right column: link between junctions 8 and 9 (10,009 metres long).  }\label{fig:ExampleFundDiags}
\end{figure}

We note at the outset that NTIS does not provide direct measurements of density.
Throughout this paper, when we refer to `density' we are really referring to a proxy measure obtained by dividing the average flow by the average speed.
NTIS only provides temporally averaged, or `time-mean', speeds. 
It has been noted that this proxy quantity for density can therefore be biased \cite{relation_amoung_average_speed_flow_and_density_and_analogous_relation_between_density_and_occupancy}.
To obtain quantitative estimates of parameters like critical density, for example, one must be careful to account for this bias.
We do not need to measure such quantities however since our objective is to distinguish between typical and atypical fluctuations in the traffic state. 
Direct use of a proxy measure for density is therefore appropriate provided this does not introduce significant artificial scatter into the data.
To test if this is the case, we consider the empirical relation derived in \cite{on_the_estimation_of_space_mean_speed_from_inductive_loop_detector_data}, where NTIS data is taken along with microscopic measurements and a formula to convert from time-mean quantities to space-mean quantities is given.
Applying this to our dataset, we see qualitatively similar properties hold, there is still significant variation, and our methods give generally the same conclusions at each step.
Based in these calculations, we use the raw data provided by NTIS from here onwards.
Example density-flow series using the raw NTIS data are given in Fig. \ref{fig:TMSExampleFundDiag}, and the transformed data using the methods in \cite{on_the_estimation_of_space_mean_speed_from_inductive_loop_detector_data} are given in Fig. \ref{fig:SMSExampleFundDiag}.

Whilst a classical way to consider the properties of a link are to inspect the corresponding fundamental diagram, the raw disaggregated density-flow series shows statistically significant variability.
The reasons for this variability are discussed at length in \cite{bivariate_relations_in_nearly_stationary_highway_traffic}.
Our methods rely on utilizing this variability in data, retaining valuable information in the joint distribution of flow and density that is typically lost when this is replaced by a single curve representing the average relationship.

By modelling the flow-density relationship as a probability distribution, we can incorporate this variability which would typically be removed when aggregating and filtering parts of the raw data.
The price to be paid is additional analytical complexity: reasonable functional forms are required for the joint distribution of flow and density.
We suggest sidestepping this question by using the data to construct the required distribution directly.

A simple and computationally efficient way to do this is using Kernel Density Estimation (KDE). 
Given some samples from an unknown  distribution, $p(\bm{x})$, with $\bm{x} \in  \mathbb{R}^d$, KDE generates a non-parametric estimate of $p(\bm{x})$ by ``smearing out'' the samples using a predefined kernel.
Given $N$ samples,  $\bm{X}_i \in \mathbb{R}^d$, $i \in \{ 1, 2, ..., N \}$, the kernel density estimate is: 
\begin{equation}
\hat{p}_{\bm{\Sigma}}\left( \bm{x} \right) = \frac{1}{N}\sum_{i=i}^N k_{\bm{\Sigma}}\left( \bm{x} - \bm{X}_i \right)
\end{equation}
where $\Sigma$ is a $d \times d$ positive definite matrix called the bandwidth matrix and the kernel, $k_{\bm{\Sigma}}(\bm{x})$, is the multivariate normal distribution:
\begin{equation}
k_{\bm{\Sigma}}(\bm{x}) = \frac{1}{(2\pi)^{\frac{d}{2}}\left| \bm{\Sigma} \right|^{\frac{1}{2}} } \, e^{ -\frac{1}{2}  \bm{x}' \bm{\Sigma}^{-1} \bm{x} }.
\end{equation}
In denoting the estimated density by $\hat{p}_{\bm{\Sigma}}\left( \bm{x} \right)$,  we are suppressing explicit dependence on the samples, $\bm{X}_1,\ldots, \bm{X}_N$ for the sake of notational compactness. 
The estimate $\hat{p}_{\bm{\Sigma}}\left( \bm{x} \right)$ is strongly dependent on the choice of bandwidth matrix,  $\bm{\Sigma}$, since it controls the amount of smoothing applied to the data. 
Choosing $\bm{\Sigma}$ involves a trade-off between under-smoothing and over-smoothing the data. 
See \cite{multivariate_kernel_smoothing_and_its_applications} for an extensive discussion.  
One principled way to select the optimal trade-off is to select the $\bm{\Sigma}$ that minimizes the mean integrated square error (MISE), defined as: 
\begin{equation}\label{equ:MISE}
\text{MISE} ( \hat{p}_{\bm{\Sigma}} ) = \mathbb{E}\left[ \int_{\mathbb{R}^d} \left(\hat{p}_{\bm{\Sigma}}(\bm{x}) - p(\bm{x}) \right)^2 d\bm{x} \right],
\end{equation}
where the expectation value is with respect to the distribution of the samples,  $\bm{X}_i$, $i \in \{ 1, 2, ..., N \}$. 
This optimization cannot be done directly however since $p(\bm{x})$ is unknown. 
There are two approaches to address this problem. 
The first is to adopt a cross-validation approach whereby a subset of the data is used to estimate $\hat{p}_{\bm{\Sigma}}(\bm{x} )$ and remaining data is then used to estimate the MISE in some way. 
The second is to exploit the fact that Eq.~(\ref{equ:MISE}) simplifies significantly in the limit $N \to \infty$ when the number of samples become large. 
In this limit, an analytic formula for the optimal bandwidth can be found although this formula still depends on the (unknown) second derivative of $p(\bm{x})$. 
For example, in the univariate case where the bandwidth is a scalar, $\sigma$, the asymptotic mean integrated square error (AMISE) is:
\begin{equation}\label{equ:AMISE}
AMISE = \frac{R(k)}{N\sigma} + \frac{\sigma^4}{4}\left( m_2(k) \right)^2R(p'')
\end{equation}
where $R(k) = \int k^2(x) \, dx$, $m_2(k) = \int x^2k(x) \, dx$ and $R(p'') = \int p''^2(x) \, dx$ with $p''(x)$ denoting the second derivative of the unknown density, $p(x)$ and $k$ denoting the kernel, in this case dependent on $\sigma$.
AMISE clearly has a minimum as a function of $\sigma$. 
The location of this minimum is the optimal bandwidth and can be calculated analytically:
\begin{equation}\label{equ:AMISESolution}
\sigma_{optimal} = \left( \frac{R(k)}{m^2_2(k)R(p'')} \right)^{\frac{1}{5}}N^{-\frac{1}{5}}.
\end{equation}
Details of these calculations  can be found in \cite[chapter 3]{silverman2018density}.
Methods based on this second approach and its extension to the multivariate case are called `plug-in' methods. 
Different strategies have been suggested for self-consistent estimation of the unknown second derivatives of $p(\bm{x})$ and most modern approaches to KDE are based on the `plug-in' approach, see \cite{jones1996brief}.
In our work we use the method of Chac\'{o}n and Duong, detailed for the multivariate case in \cite{multivariate_kernel_smoothing_and_its_applications,multivariate_plug_in_bandwidth_selection_with_unconstrained_pilot_bandwidth_matrices}.  
The R implementation we use is described in \cite{ks_package}.

\begin{figure*}[ht!]
	\begin{subfigure}[t]{.47\textwidth}
		\raisebox{2.5mm}{\includegraphics[width=\textwidth]{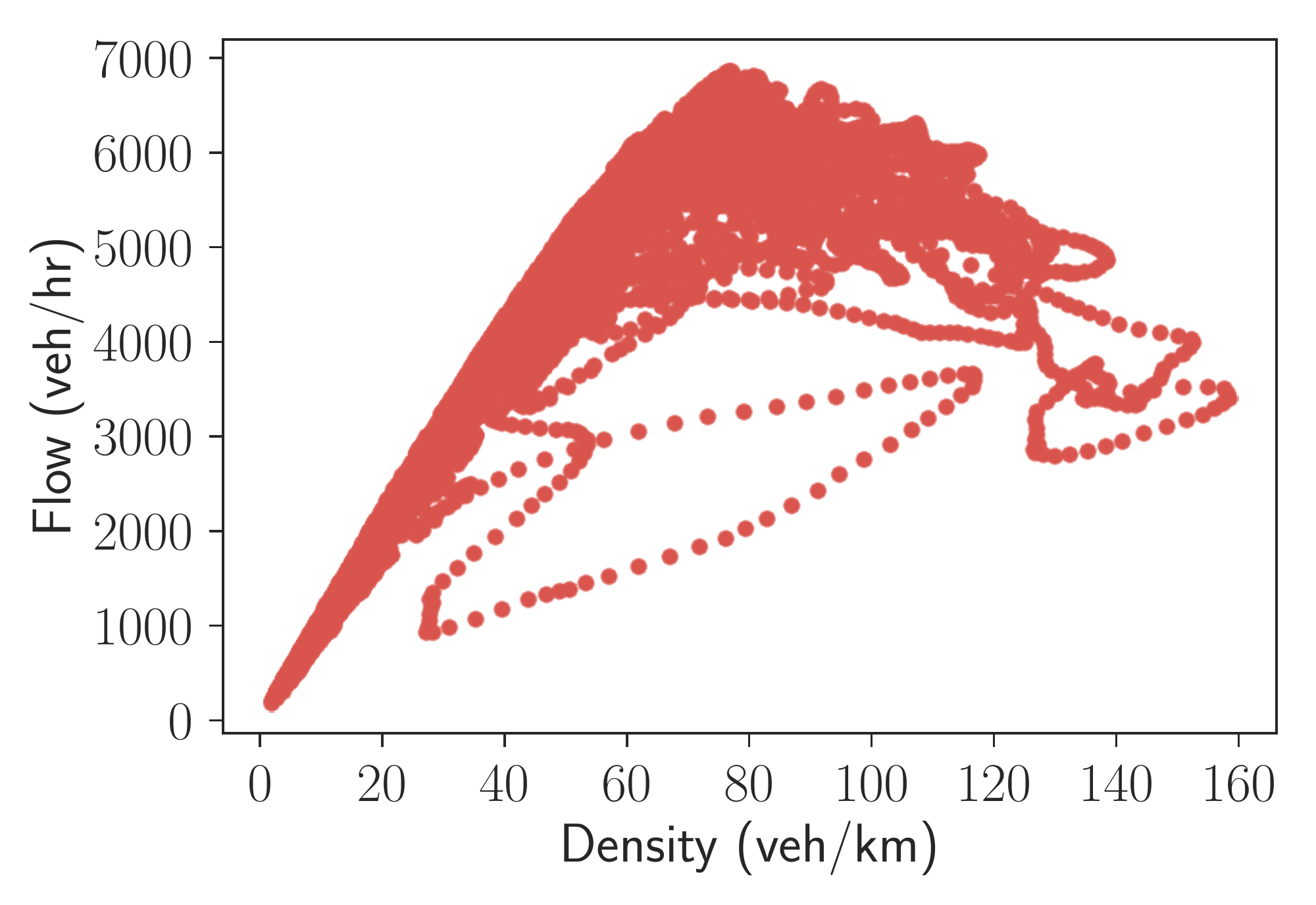}}
		\caption[ Example KDE: data ]{ Scatterplot generated from 10 weeks of data from a representative M25 link between junctions 2 and 3.  \label{fig:ScatterVsKDE:A}}
	\end{subfigure}
	\begin{subfigure}[t]{.47\textwidth}
		\raisebox{0mm}{\includegraphics[width=\textwidth]{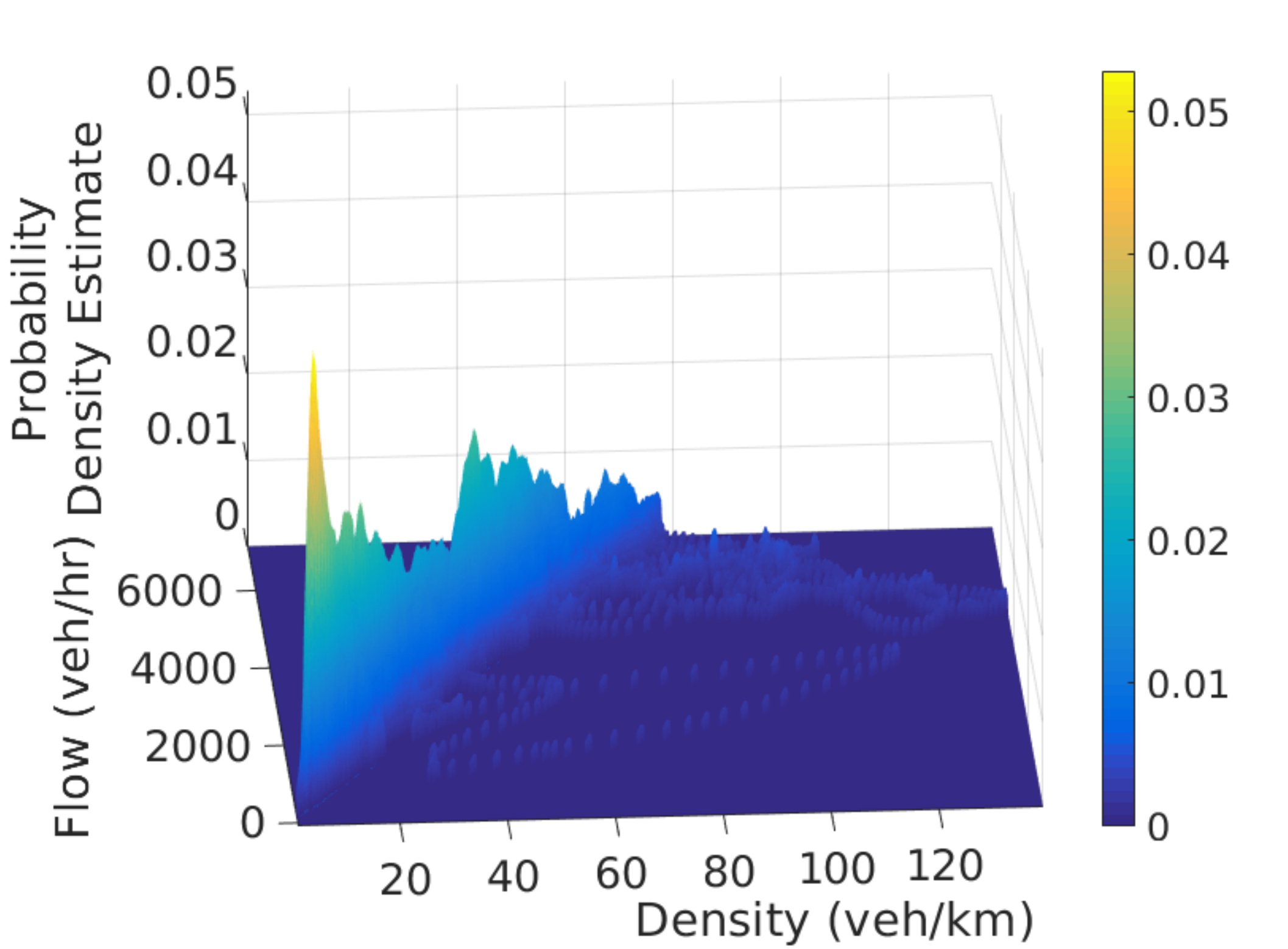}}
		\caption[ Example KDE: pdf ]{ Kernel Density Estimate of the joint distribution of flow and density generated from this data. \label{fig:ScatterVsKDE:B}}
	\end{subfigure}
	\caption[KDE example]{Representative example of the use of KDE to obtain a representation of the joint distribution of flow and density for a single link of the M25  using NTIS data.}\label{fig:ScatterVsKDE}
\end{figure*}

Fig. \ref{fig:ScatterVsKDE} shows an illustrative example of this process. Fig. \ref{fig:ScatterVsKDE:A} shows a scatterplot of the data from a representative M25 link between junctions 2 and 3. Fig. \ref{fig:ScatterVsKDE:B} shows a 3D rendering of the corresponding joint distribution obtained using KDE. The latter makes clear that  most of the mass is centred around two modes, a feature that is much less evident from the scatterplot. We refer to these as the low density and high density modes.
Note that this multi-modal structure is not a sampling bias since the data is uniformly sampled at 1 minute intervals across the shown 10 week data collection window. 
Rather it reflects the fact that this link spends a lot more time in the low density regime compared to the high density regime.

\subsection{Separating typical and atypical configurations}\label{subsec:Contours}

We now turn to the central point of the paper: using the KDE to systematically draw a distinction between typical and atypical behaviour of the traffic on a link. 
The idea is to calculate a level curve of  the KDE representation, $\hat{p}_{\bm{\Sigma}}(\bm{x})$, of the joint distribution of flow and density that encloses a predefined portion, $1-\alpha$, of the total probability. 
Level curve here means a curve satisfying $\hat{p}_{\bm{\Sigma}}(\bm{x}) = \text{constant}$. 
Points inside this level curve are considered typical whereas points outside are considered atypical. 
By adjusting the parameter $\alpha$ we can adjust the relative frequency between typical and atypical. 
This is essentially a two-dimensional bivariate analogue of a one-dimensional confidence interval for a univariate distribution. 
This construction is a topic of active research in the statistics literature because inference of level curves is analytically and computationally highly non-trivial. 
See \cite{chen2017density} and the references therein. 
In our application however, the large volume of data available from NTIS means practically useful level curves can be obtained without addressing the much harder problem of quantifying the uncertainty in these curves.

\begin{figure}[ht!]
	\begin{subfigure}{.47\textwidth}
		\includegraphics[width=\textwidth]{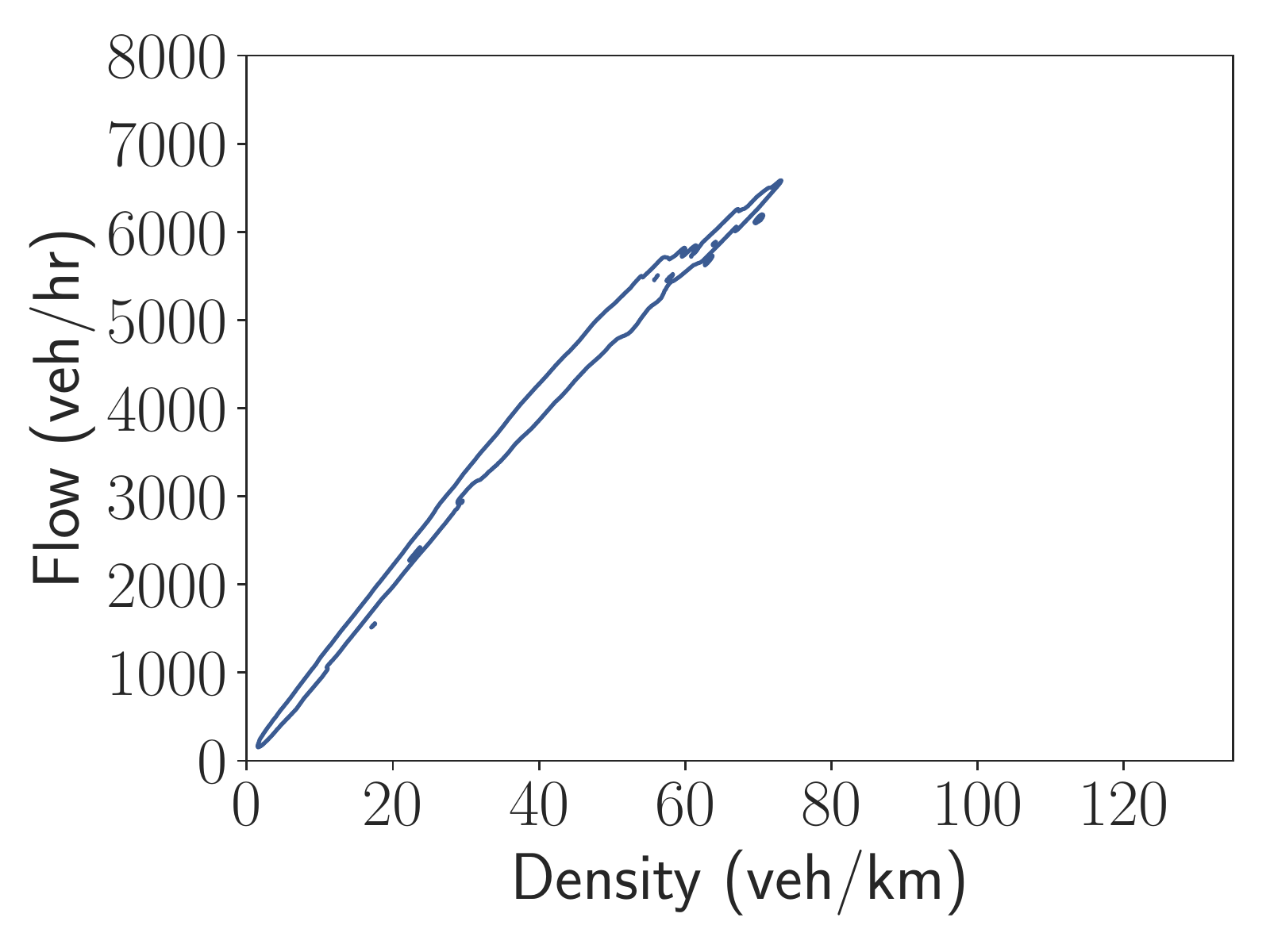}
		\caption{Data taken between junctions 2 and 3 on the M25.}\label{fig:Example_Contour_95_Link_1}
	\end{subfigure}
	\begin{subfigure}{.47\textwidth}
		\includegraphics[width=\textwidth]{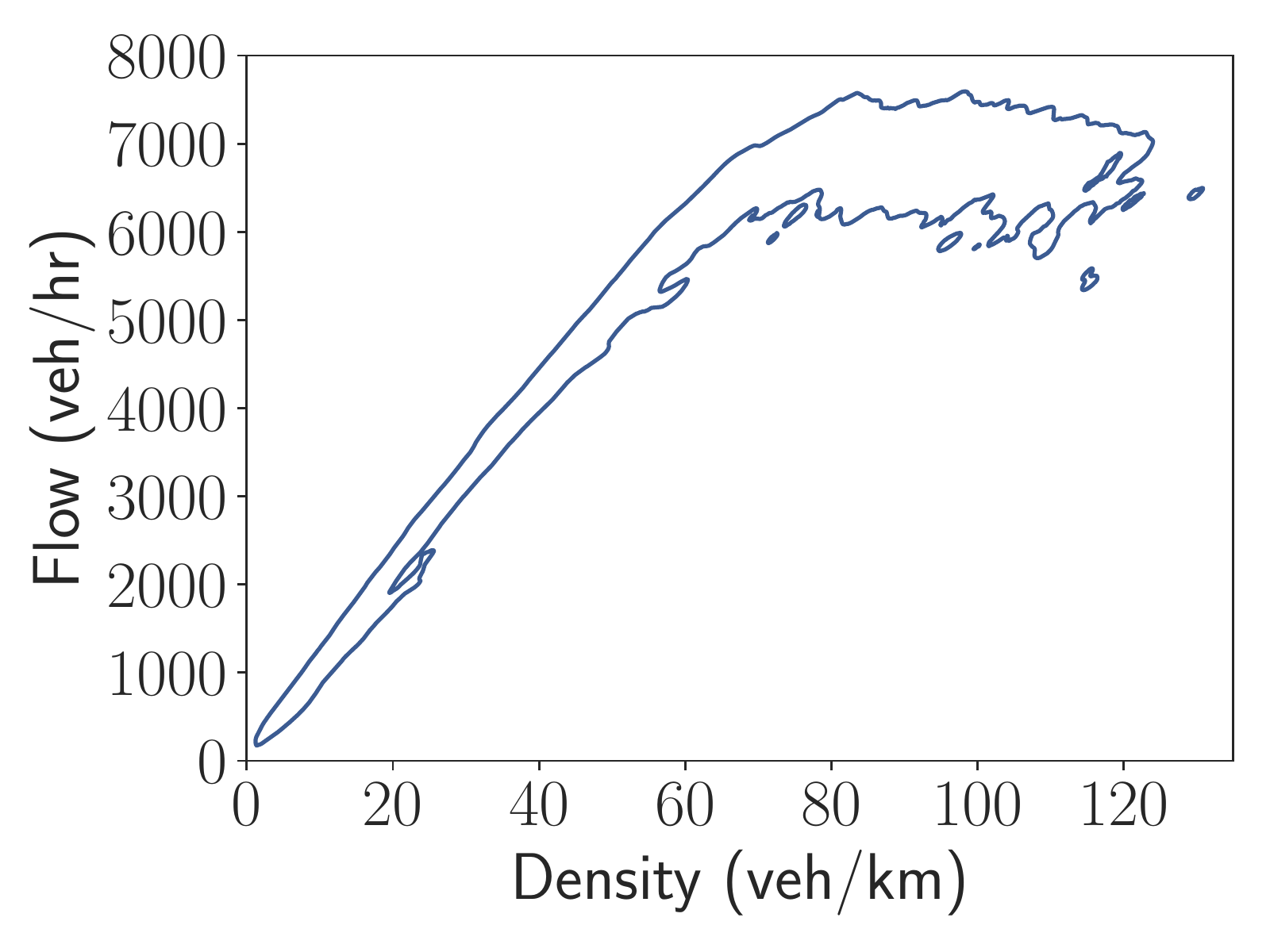}
		\caption{Data taken between junctions 8 and 9 on the M25.}\label{fig:Example_Contour_95_Link_22}
	\end{subfigure}
	\caption[Example Density-Flow Contours]{Level curves corresponding to a threshold $\alpha = 0.05$ created using data from two different links. Each contains 95\% of the probability mass of the respective KDE representations. }\label{fig:ExampleContours}
\end{figure}

Once we have obtained the KDE representation, $\hat{p}_{\bm{\Sigma}}(\rho,f)$, of the joint distribution of density, $\rho$, and flow, $f$,  the computational task is to find $z^*$ such that the integral of all mass above this height equals some threshold value $1 - \alpha$.
We determine $z^*$ by finding the (unique) root of the function:
\begin{equation}\label{equ:RootForHeight}
H(z) = 1 - \alpha - \int_{\rho_{min}}^{\rho_{max}} \int_{f_{min}}^{f_{max}}  \hat{p}_{\bm{\Sigma}}(\rho,f)\, \Theta \left[ \hat{p}_{\bm{\Sigma}}(\rho,f)\ - z \right] \, df\, d\rho.
\end{equation}
Here $\Theta(x)$ is the Heaviside function,
\begin{displaymath}
\Theta(x) = \left\{\begin{array}{ll} 1 & \text{if $x\geq 0$}\\ 0 &  \text{if $x < 0$}\end{array}\right. ,
\end{displaymath}
which serves to threshold the integrand by setting it to zero when $\hat{p}_{\bm{\Sigma}}(\rho,f)$ is lower than $z$.
In equation \ref{equ:RootForHeight}, the root of $H(z)$ will simply be the value at which the integral of our thresholded distribution is equal to the desired threshold.
Note that as our data has bounds on density and flow, we only need to integrate between these.

Having found the desired height $z^*$, we slice the surface $\hat{p}_{\bm{\Sigma}}(\rho,f)$ at this value, thereby defining a level curve $c(\rho,f)$ that contains the desired amount of mass, with points lying outside the curve being outliers at the $\alpha$ threshold.
For example, if we choose $\alpha = 0.05$, we would have 95\% of the data inside $c(\rho, f)$, with the remaining 5\% considered atypically large deviations from the standard behaviour. 
The level curves for taking $\alpha = 0.05$ are shown in Fig. \ref{fig:ExampleContours} for some representative links. 
These two examples are illustrative of how we capture known features of traffic flow, in this case recurrent bottlenecks.
Flow breakdown is generally absent on the link shown in Fig. \ref{fig:Example_Contour_95_Link_1}, however the link shown in Fig.~\ref{fig:Example_Contour_95_Link_22} clearly sustains a high flow as we move from densities around 60 veh/km to 120 veh/km, suggesting it is impacted by a recurrent bottleneck.
As the traffic state evolves in time on any given link, it traces out a trajectory in the $\rho-f$ plane.
NTIS can track this trajectory almost in real time.
When the trajectory makes an excursion outside of the curve delimiting the region of typical behaviour, we call it an `atypical traffic event'.

In order for the concept of an atypical traffic event to be useful, the level curves delimiting the region of typical behaviour should be stable in time. 
Our analysis showed this to be the case.
Some details can be found in \ref{subsec:Stability}. In particular Fig. \ref{fig:StabilityPlots} shows that the curves obtained for a particular link using disjoint data windows as input to the KDE are very similar.

\subsection{Timescales of atypical traffic events and comparison to NTIS events}
\label{subsec:EventTimescales}

\begin{figure}[ht!]
	\begin{subfigure}{.48\textwidth}
		\includegraphics[width=\textwidth]{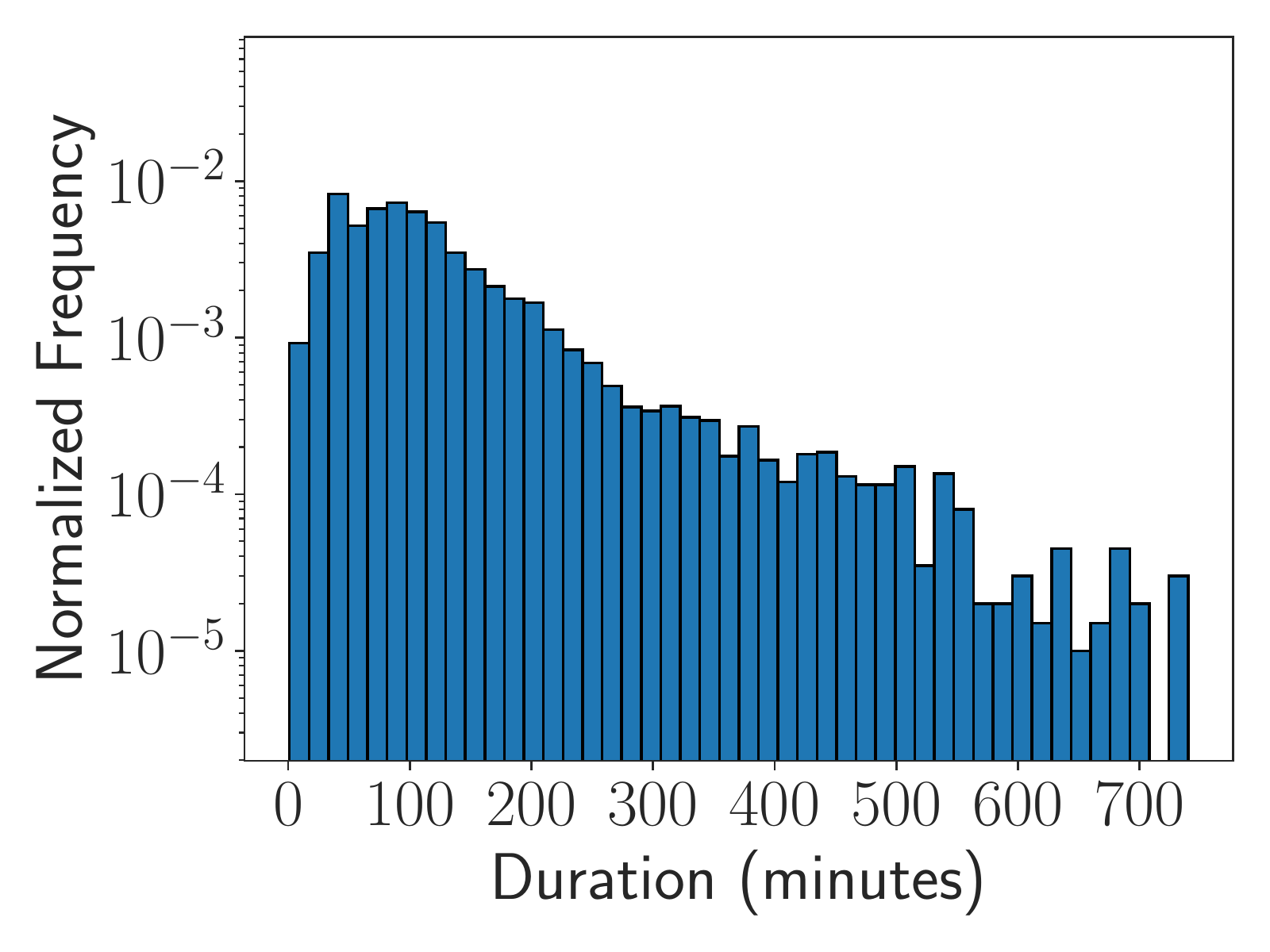}
		\caption{Accident and obstruction events flagged by NTIS.}\label{fig:Acc_Obs_Time_Scales}
	\end{subfigure}
	\begin{subfigure}{.48\textwidth}
		\includegraphics[width=\textwidth]{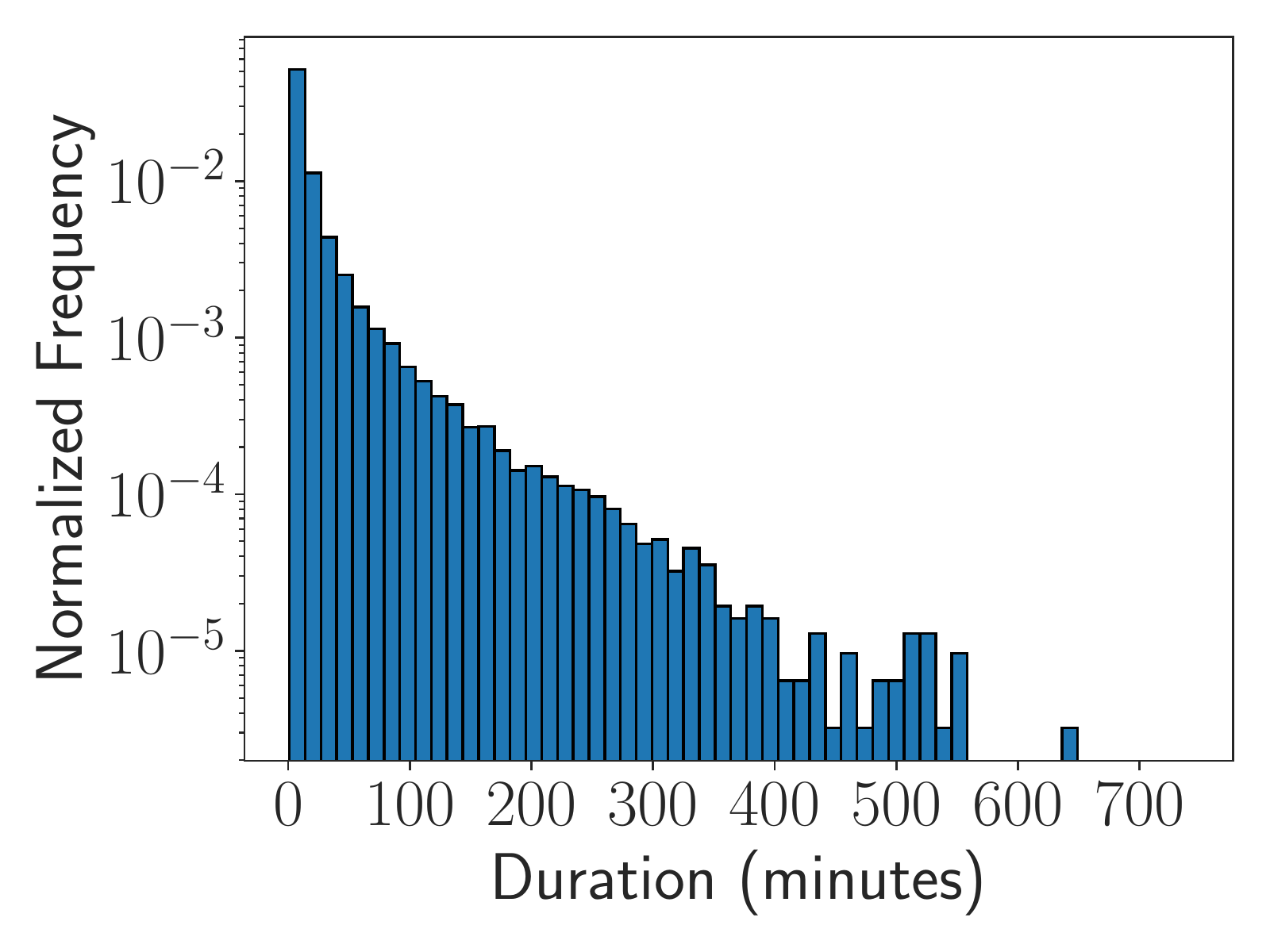}
		\caption{Atypical traffic events generated by our algorithm.}\label{fig:Abnormal_Traffic_Time_Scales}
	\end{subfigure}
	\caption[Event Time-Scale Plots]{ Comparison of the empirical distributions of durations of NTIS accidents and obstructions events and durations of atypical traffic events defined through our methodology. The two distributions are clearly distinct, with atypical traffic event durations being mainly concentrated at small values less than 100 minutes, where as accidents and obstructions peak around 100 minutes, then decay. }\label{fig:EventTimeScales}
\end{figure}

We might expect that many atypical traffic events correspond to small excursions that rapidly return to the typical region.
Very brief excursions are unlikely to be of much practical interest.
It is therefore useful to compare the distribution of durations of atypical traffic events generated by the procedure described above with the distribution of durations of events flagged by NTIS.
We scan all the links in our data and determine the start and end points of each NTIS accident and obstruction event and each atypical traffic event. 
This gives us a set of durations for each category.
The resulting histograms of durations are shown in Fig. \ref{fig:EventTimeScales}. 
From Fig. \ref{fig:Acc_Obs_Time_Scales}, we see the typical duration of an NTIS accident or obstruction event is near 100 minutes, significantly higher than that of an abnormal traffic event Fig. \ref{fig:Abnormal_Traffic_Time_Scales} which simply decays as duration is increased, rather than rising to a peak and decaying after as accident and obstruction durations do.
This clearly highlights two things.
The first is that, a large number of the abnormal traffic events may be so short lived that the actual impact on traffic on a link is minor. 
An infrastructure operator, for example, might not be interested in seeing these minor deviations.
It can therefore make sense to apply secondary thresholding to single out the most significant  ones.
This is done in Sec. \ref{sec:DFTB}.
Secondly,  when an event occurs, it may not perturb the traffic state significantly for a number of minutes, causing only minor obstructions to begin with.  
However these can grow in time so that one eventually sees strongly atypical density-flow behaviour.
There is therefore a natural notion of the severity of an event which can evolve in time. 
An approach for quantifying severity is discussed in Sec. \ref{sec:severity}.

\section{Identification of significant  `Deviation from Typical Behaviour' (DFTB) events}
\label{sec:DFTB}

In this section we describe a protocol for identification of significant abnormal traffic events by filtering out those of short duration. 
We call these 'Deviation from Typical Behaviour' (DFTB)  events to distinguish them from the NTIS 'Deviation from Profile'  (DFP) events previously described in Sec. \ref{subsec:events}.
We expect significant commonality between the two types of event and some comparisons are made below.

\subsection{Determining When to Raise Flags}\label{subsec:WhenToFlag}

\begin{figure}[ht!]
	\centering
	\begin{subfigure}[t]{.48\textwidth}
	\includegraphics[width=\linewidth]{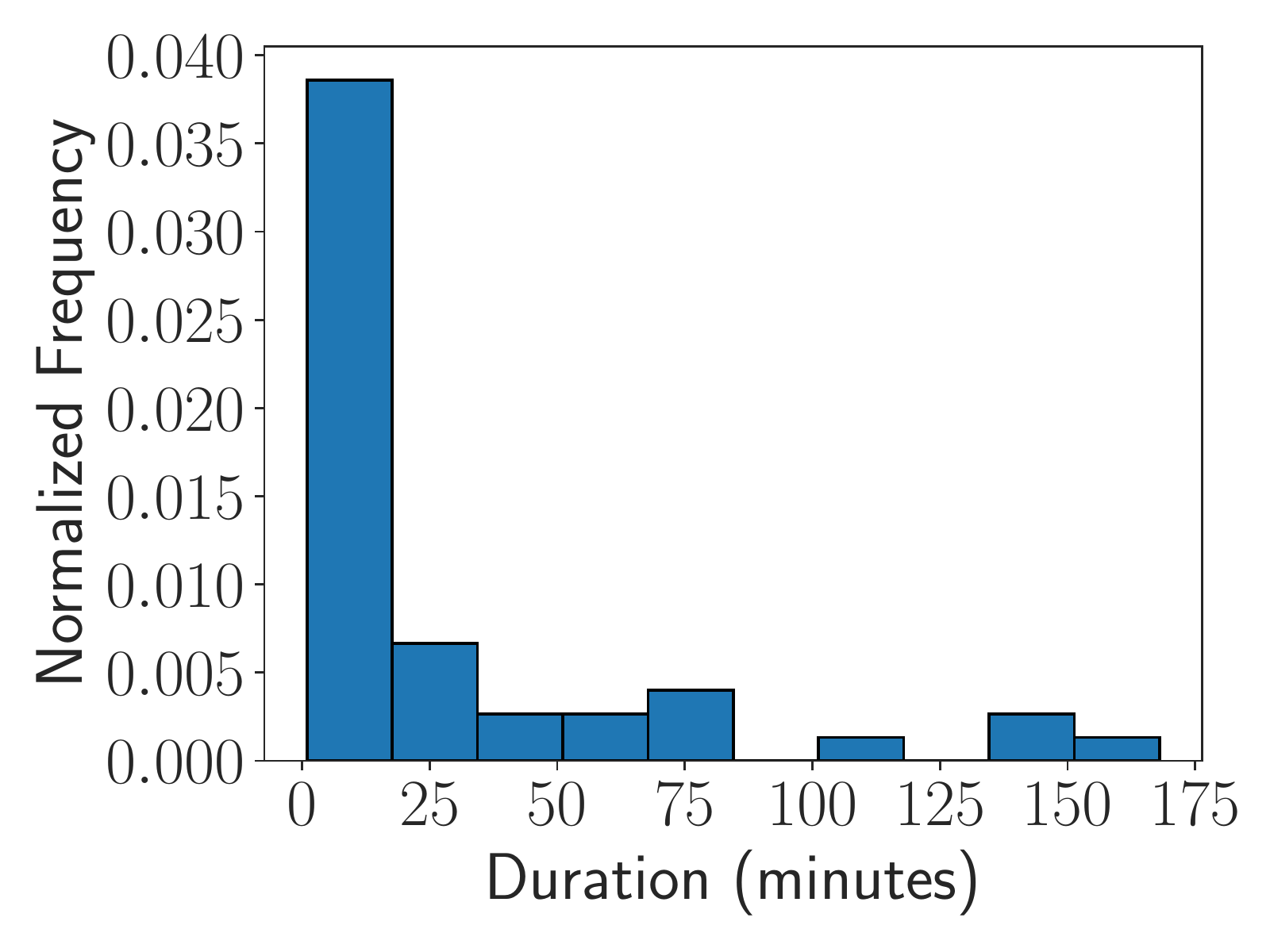}
	\caption[Deviation Durations]{ The duration of atypical traffic events observed in our 3-week training window.  The majority are short term as emphasized by the initial large peak. }\label{fig:DeviationDurations}
	\end{subfigure}
	~	
	\begin{subfigure}[t]{.48\textwidth}
	\includegraphics[width=\linewidth]{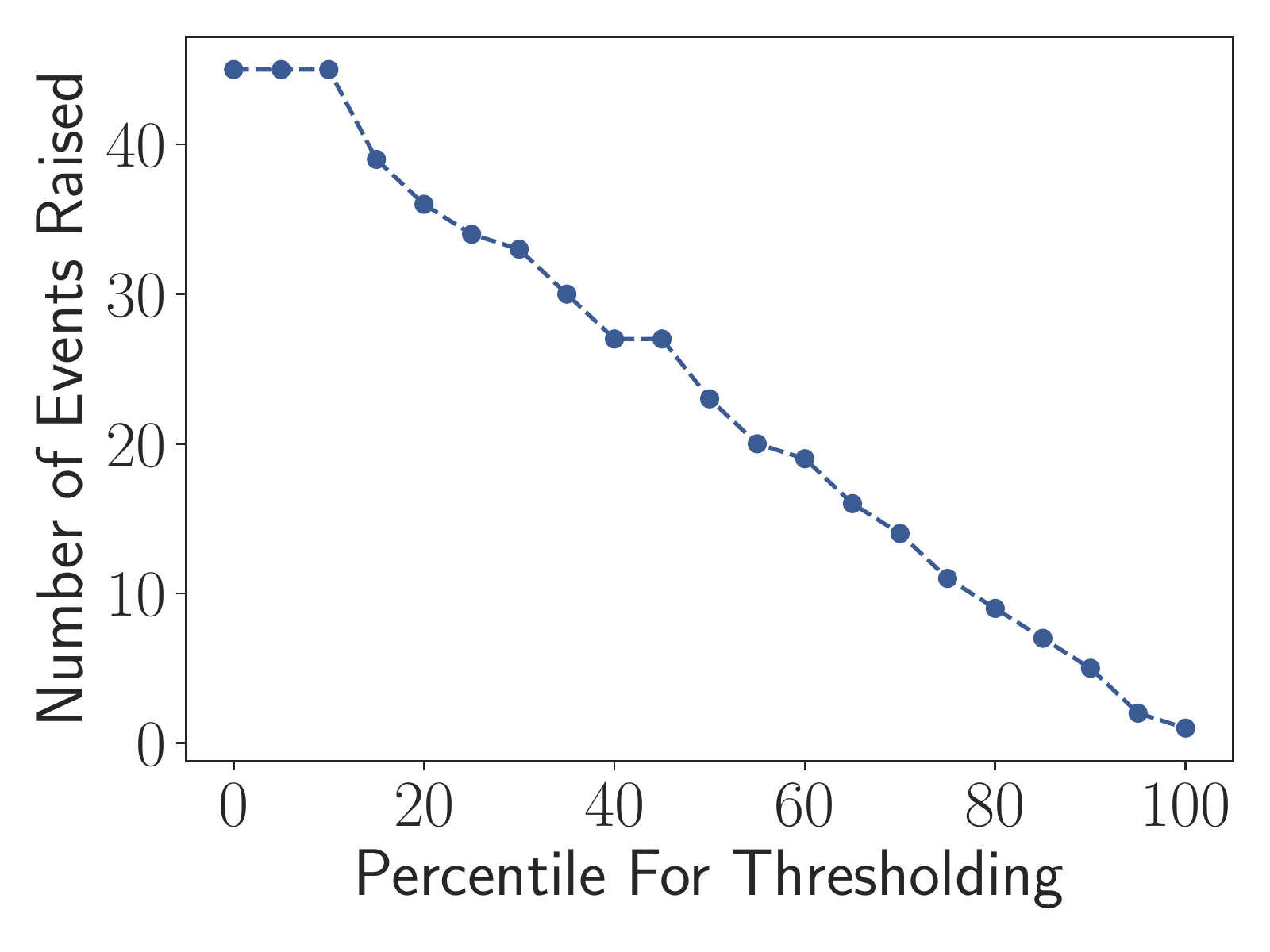}
	\caption[Threshold vs Number of Events]{ The number of DFTB events raised in the training window as a function of the percentile chosen as the threshold for duration. At the extremes, we have every single deviation registering an event (threshold = 0th percentile) or no events raised (threshold = 100th percentile). }\label{fig:ThresholdVsNumEvents}
	\end{subfigure}
	\caption{Analysis of duration of atypical traffic events and number of events raised. Plots are shown for a representative link, between junctions 2 and 3 on the M25.}\label{fig:MergedDurationsAndCounts}
\end{figure}

We propose to make decisions on raising DFTB event flags by defining a time threshold for which trajectories must be outside of the typical region in order be considered significant. 
This filters out the aforementioned minor fluctuations. Throughout this section, we use 3-weeks of data to perform the KDE and to identify the curve enclosing the typical region. 
A further 6 weeks of data is then used as a test set within which we seek to identify events so that decisions are based solely on past data.
To begin, we plot the histogram of durations of atypical traffic events observed on a representative link during the training window. 
A example is shown in Fig. \ref{fig:DeviationDurations}. 
We see a large number of short-duration excursions from the typical region, and far fewer large ones. 
These extreme deviations represent the most severe events on the link, with the less extreme ones being located in the short-term deviations.
By setting the threshold at a given percentile of the duration distribution we obtain a systematic and purely data driven notion of significance that does not rely on any details of individual links. 
Obviously selecting a higher threshold percentile gives fewer event flags.
The relationship between the threshold and the number of events is shown in  in Fig. \ref{fig:ThresholdVsNumEvents}.

\begin{figure*}[ht!]
	\begin{subfigure}{.48\textwidth}
	\includegraphics[width=\textwidth]{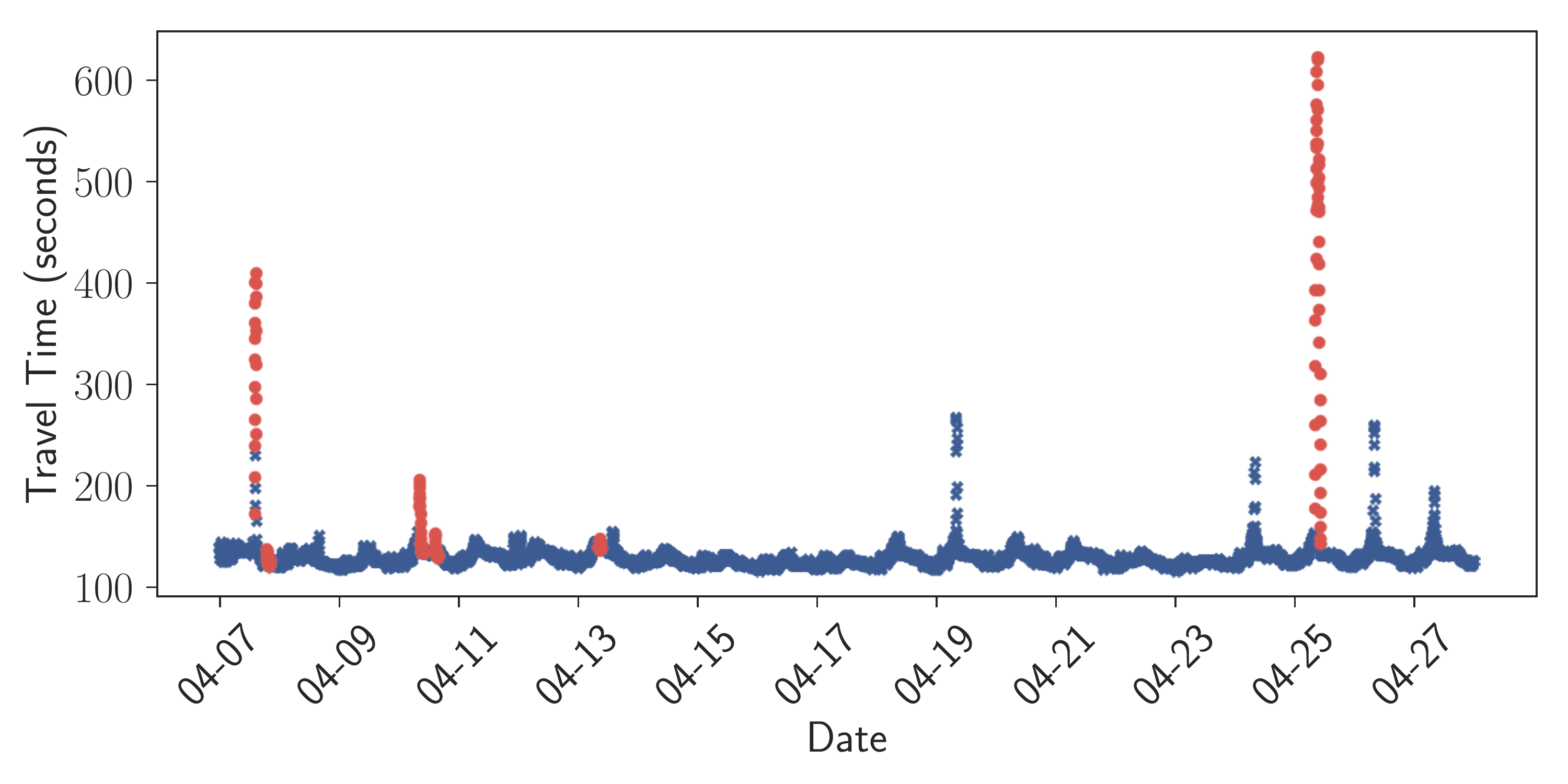}
	\caption{NTIS event flags in the training data.}
	\end{subfigure}
	\hfill
	\begin{subfigure}{.48\textwidth}
	\includegraphics[width=\textwidth]{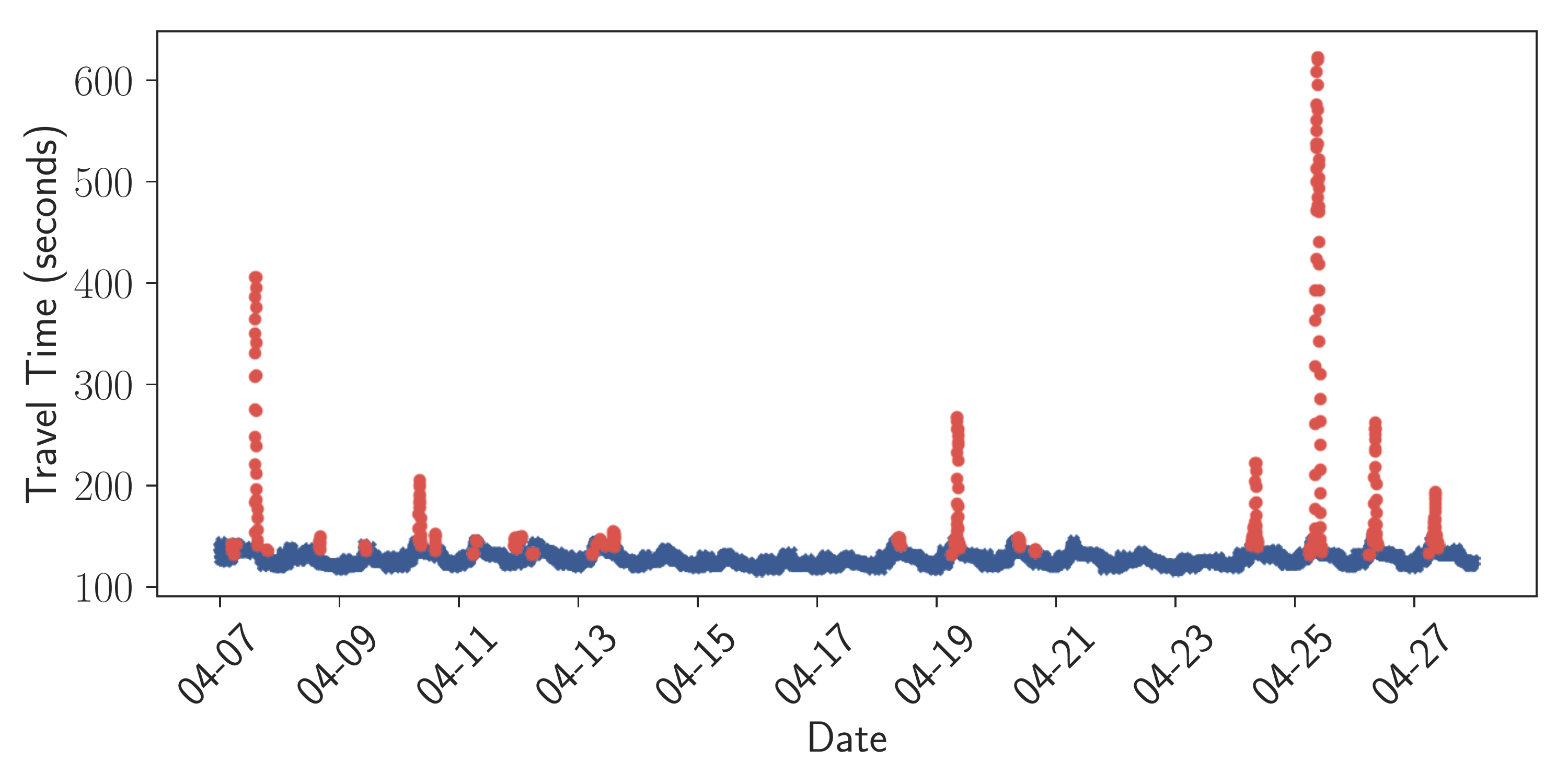}
	\caption{DFTB flags in the training data without thresholding.}
	\end{subfigure}
	\hfill
	
	\begin{subfigure}{.48\textwidth}
	\includegraphics[width=\textwidth]{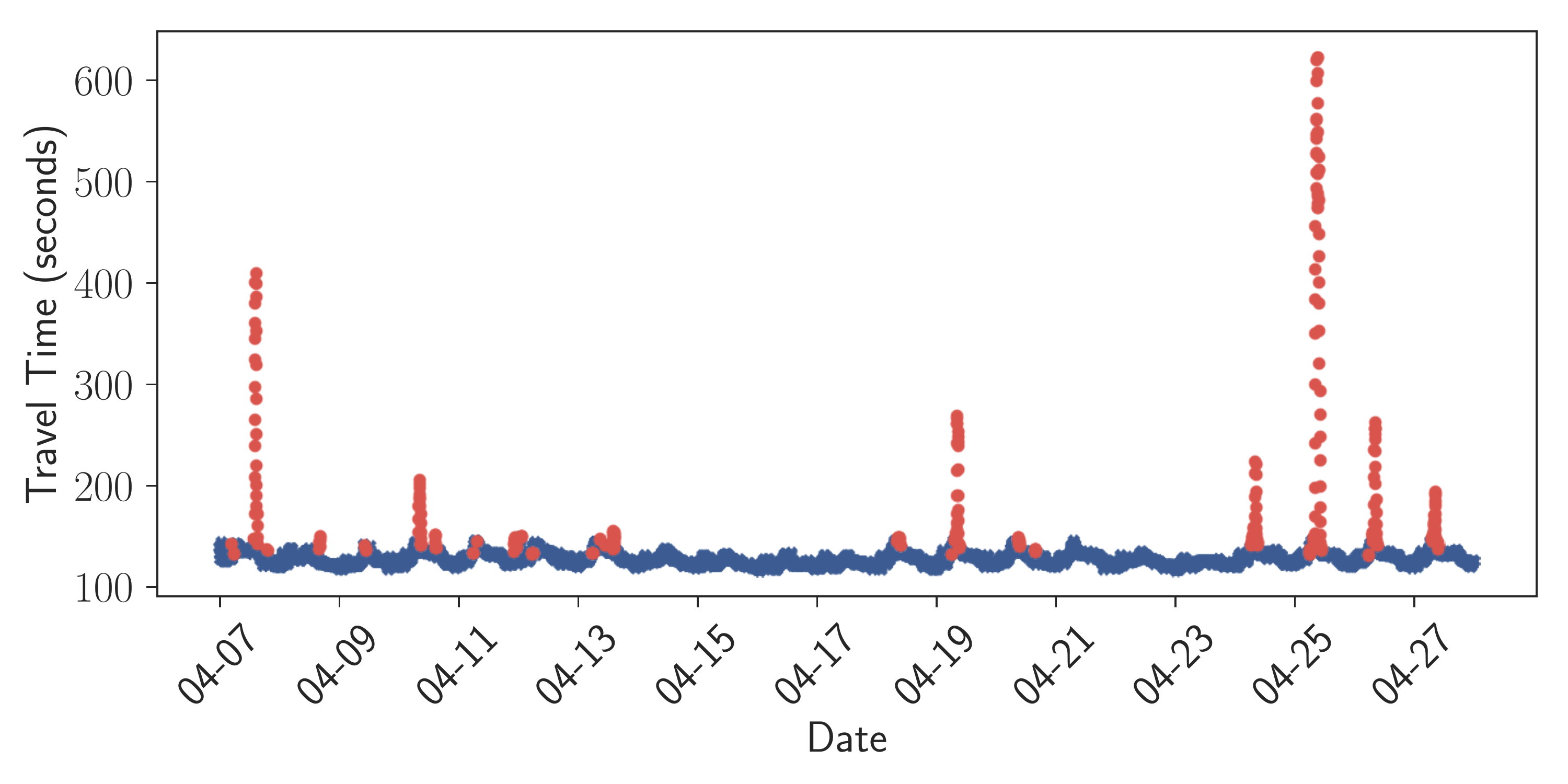}
	\caption{DFTB flags raised in training data with threshold at  40th percentile.}
	\end{subfigure}
	\hfill
	\begin{subfigure}{.48\textwidth}
	\includegraphics[width=\textwidth]{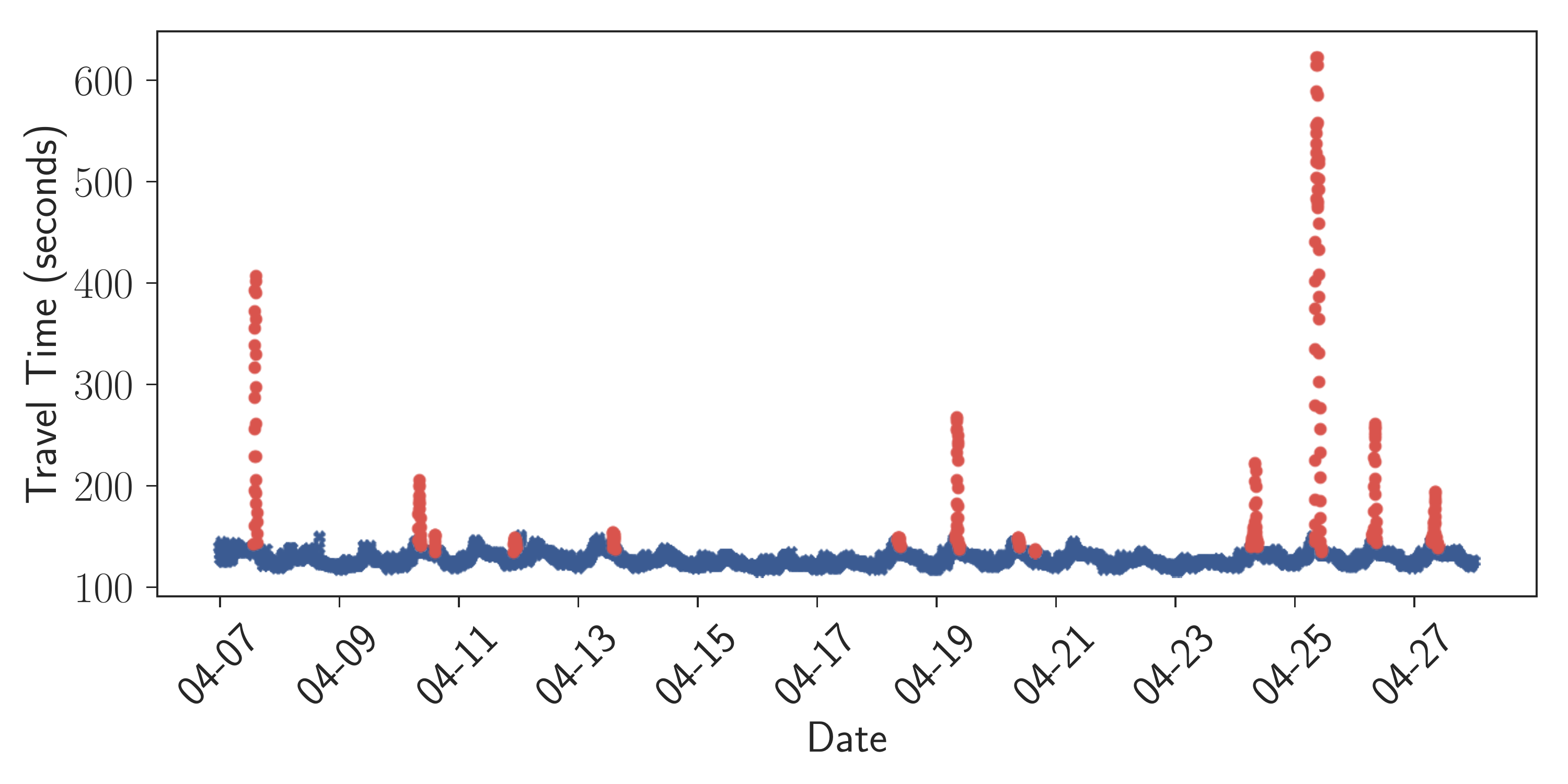}
	\caption{DFTB flags raised in training data with threshold at 80th percentile.}
	\end{subfigure}
	\hfill
	\caption[Travel Times with Labels]{ Travel times on the link studied, separated into cases with and without event flags. On each plot, blue $(\times)$ represents a data-point with no event flag, where as red $(\bullet)$ represents a raised flag. Plot (a) is of travel times coloured and symboled to represent NTIS event flags, where as the rest are done to represent the presence of a DFTB flag at various thresholds. For reference, the threshold values for the 40th and 80th percentiles are 8 and 23 minutes respectively. }\label{fig:TT_TrainData}
\end{figure*}

It is natural to look where DFTB events appear in travel time series and to compare them to events flagged by NTIS.
In Fig. \ref{fig:TT_TrainData}, we plot the travel times for the link in question, separated by colour and symbol into cases with and without DFTB flags raised. 
For reference, we also show the same plot using NTIS flags, using deviation from profile, accident and obstruction flags, and finally show how our flags change for a range of different thresholds.  

From Fig. \ref{fig:TT_TrainData}, it is clear that atypical fluctuations in the flow-density relationship identify spikes in the travel times on a link. 
This identification is indirect since we never modelled travel times.
For low thresholds, we see many DFTB flags are raised during low-travel times. However as we increase the threshold we identify only the large spikes in the series.
Fig. \ref{fig:TT_TrainData} (a) also shows that some DFTB events are associated with significant spikes in travel times that are not associated with any NTIS event.
It would be interesting to investigate these further.

\subsection{Results for DFTB flags in the test data}

\begin{figure}[ht!]
	\begin{subfigure}{.48\textwidth}
	\includegraphics[width=\textwidth]{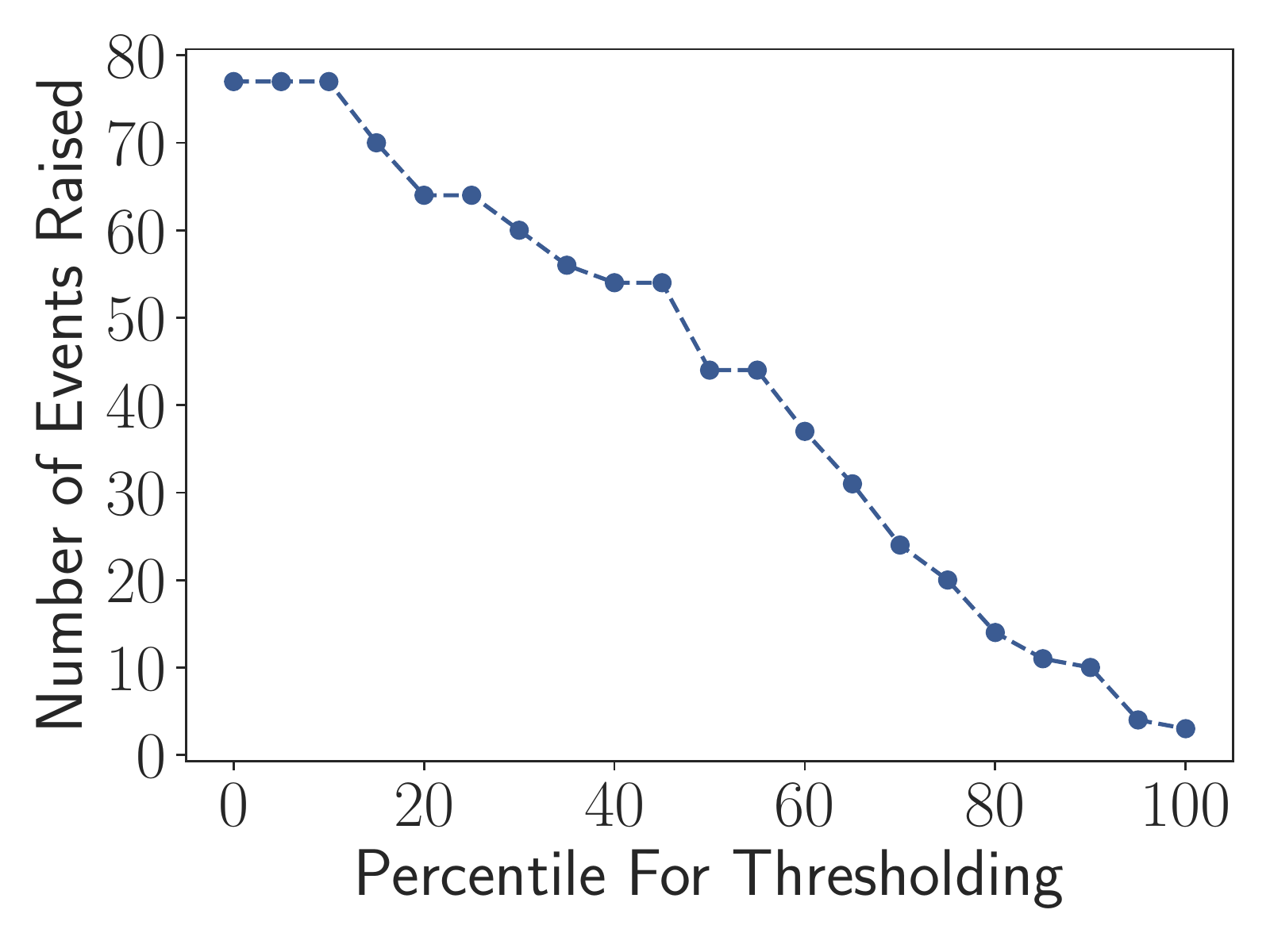}
	\caption{Test set 1: 3 weeks of data starting April 28th 2017}
	\end{subfigure}
	\hfill
	\begin{subfigure}{.48\textwidth}
	\includegraphics[width=\textwidth]{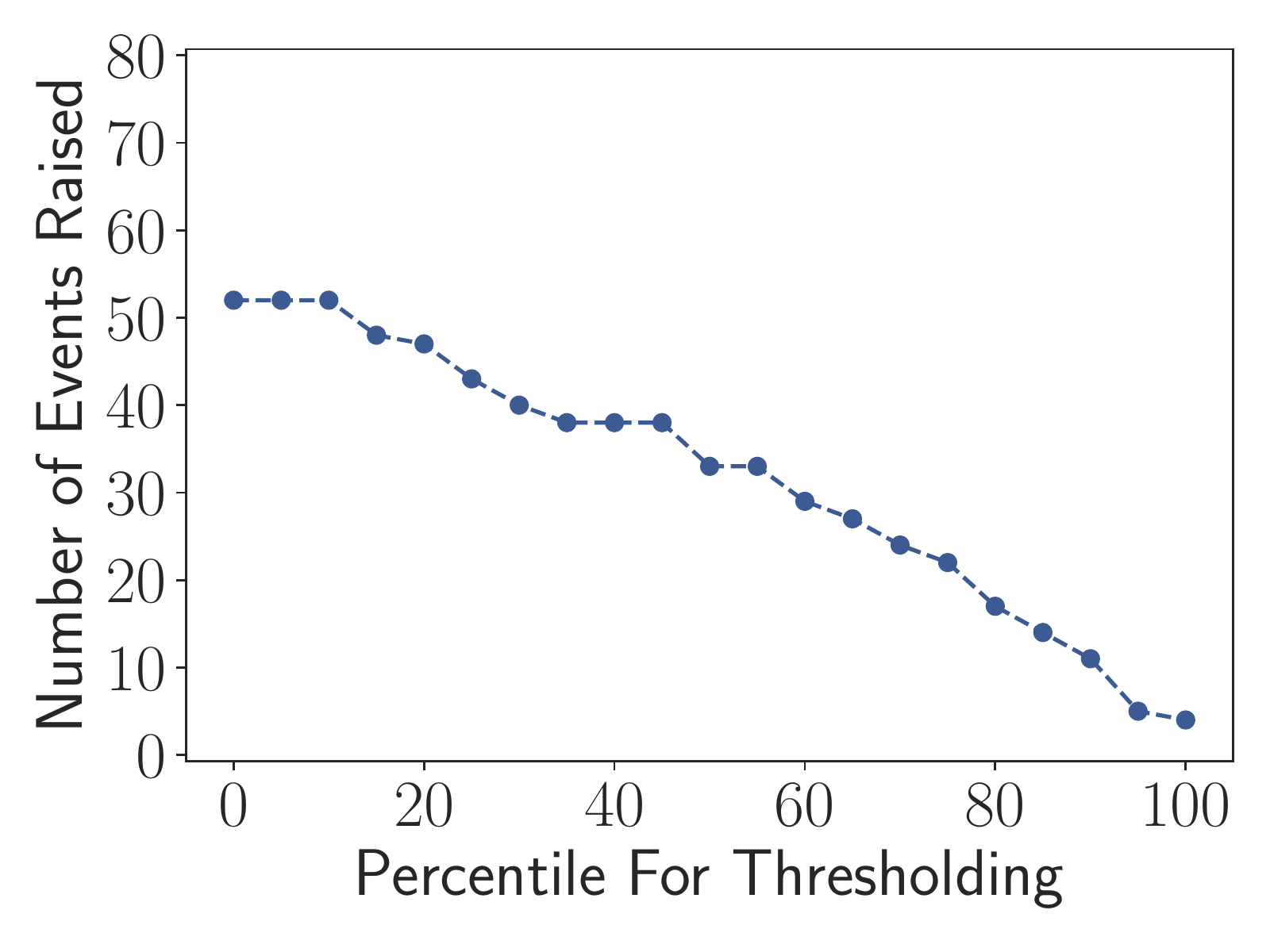}
	\caption{Test set 2: 3 weeks of data starting May 19th 2017}
	\end{subfigure}
	\caption[Test Data Results - Number of Events]{ Plots of the number of DFTB events raised as a function of threshold percentile for each of our test datasets. We see similar relationships to those observed in Fig. \ref{fig:ThresholdVsNumEvents}, with slightly more events raised in these windows at low thresholds. }\label{fig:TestSetsNumEvents}
\end{figure}

The plots in section \ref{subsec:WhenToFlag} are for the training data that was used to determine the contour defining the typical region. 
It is therefore not surprising that the DFTB events selected are in close correspondence with atypical events observed in the travel time series.
We now check that the process generalizes to unseen data, using two subsequent 3 week periods of test data that has not been used to construct the curve enclosing the typical region. 
The first 3 week test period starts April 28th 2017 (test set 1) and the second starts May 19th 2017 (test set 2). 
In Fig. \ref{fig:TestSetsNumEvents}, we show the number of events raised as a function of threshold again, but for each of the two test periods using unseen data. 

\begin{figure*}[ht!]
	\begin{subfigure}{.48\textwidth}
	\includegraphics[width=\textwidth]{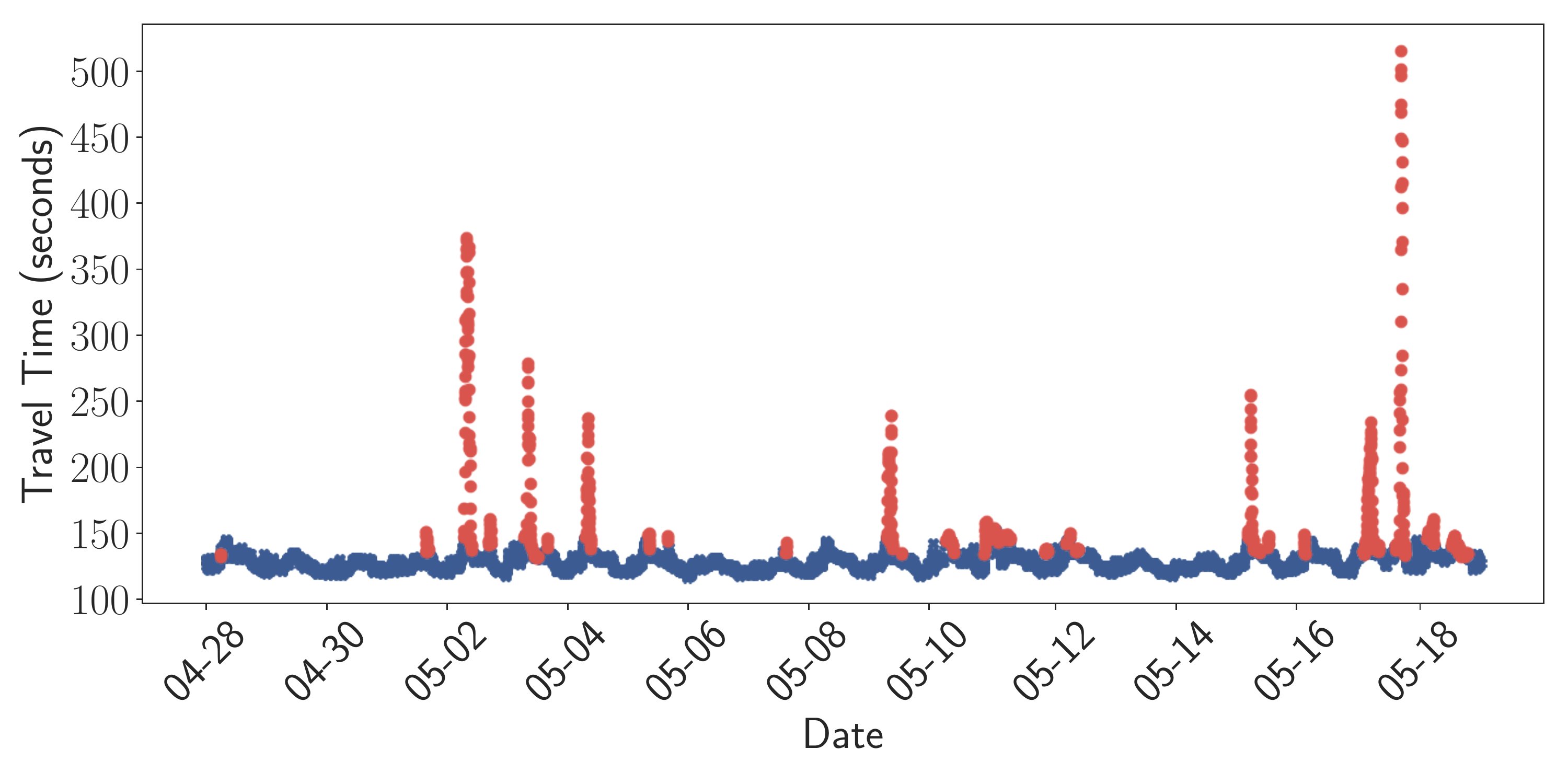}
	\caption{DFTB flags in test set 1 with threshold at the 40th percentile.}
	\end{subfigure}
	\hfill
	\begin{subfigure}{.48\textwidth}
	\includegraphics[width=\textwidth]{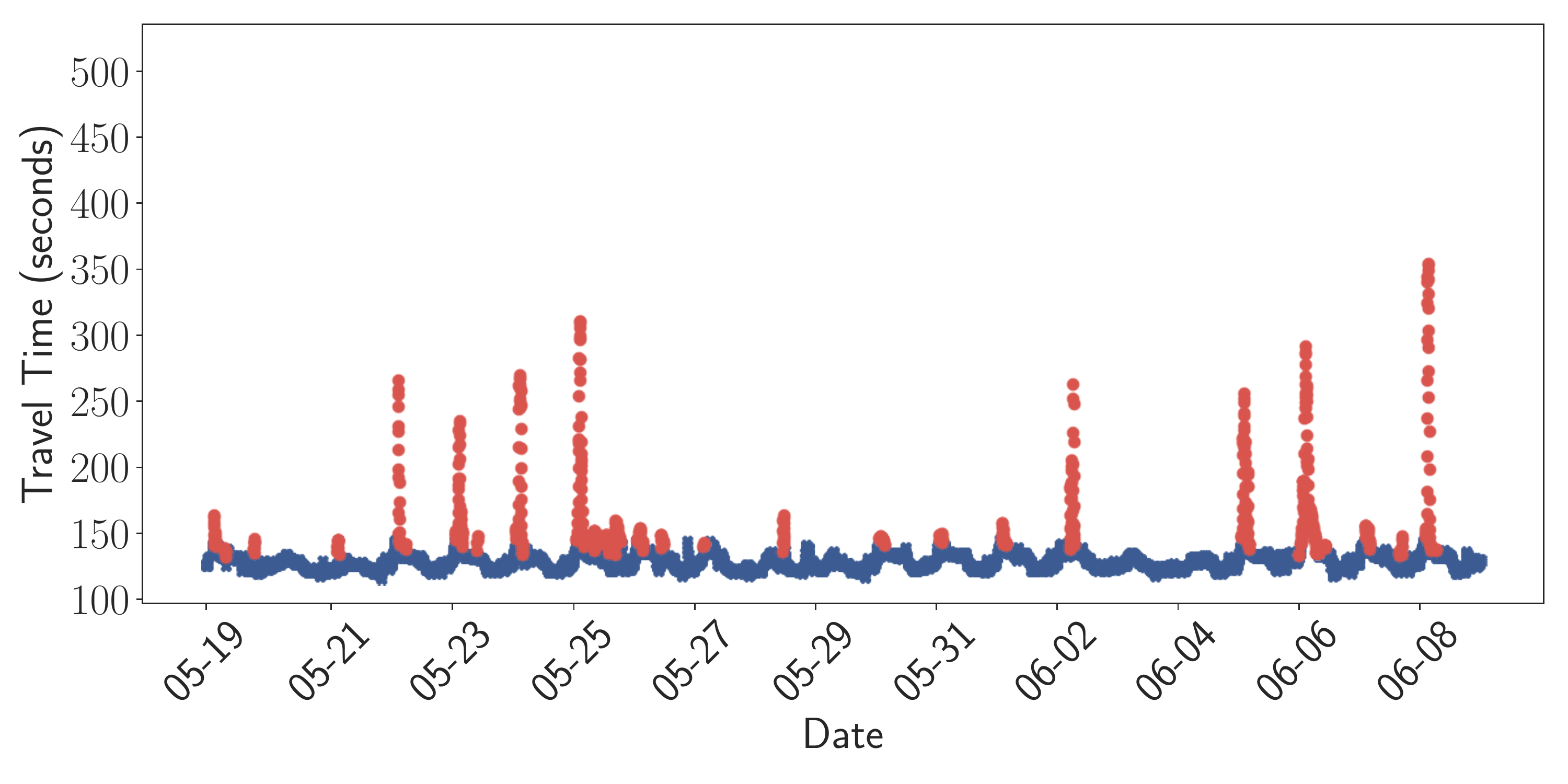}
	\caption{DFTB flags in test set 2 with threshold at the 40th percentile.}
	\end{subfigure}
	
	\begin{subfigure}{.48\textwidth}
	\includegraphics[width=\textwidth]{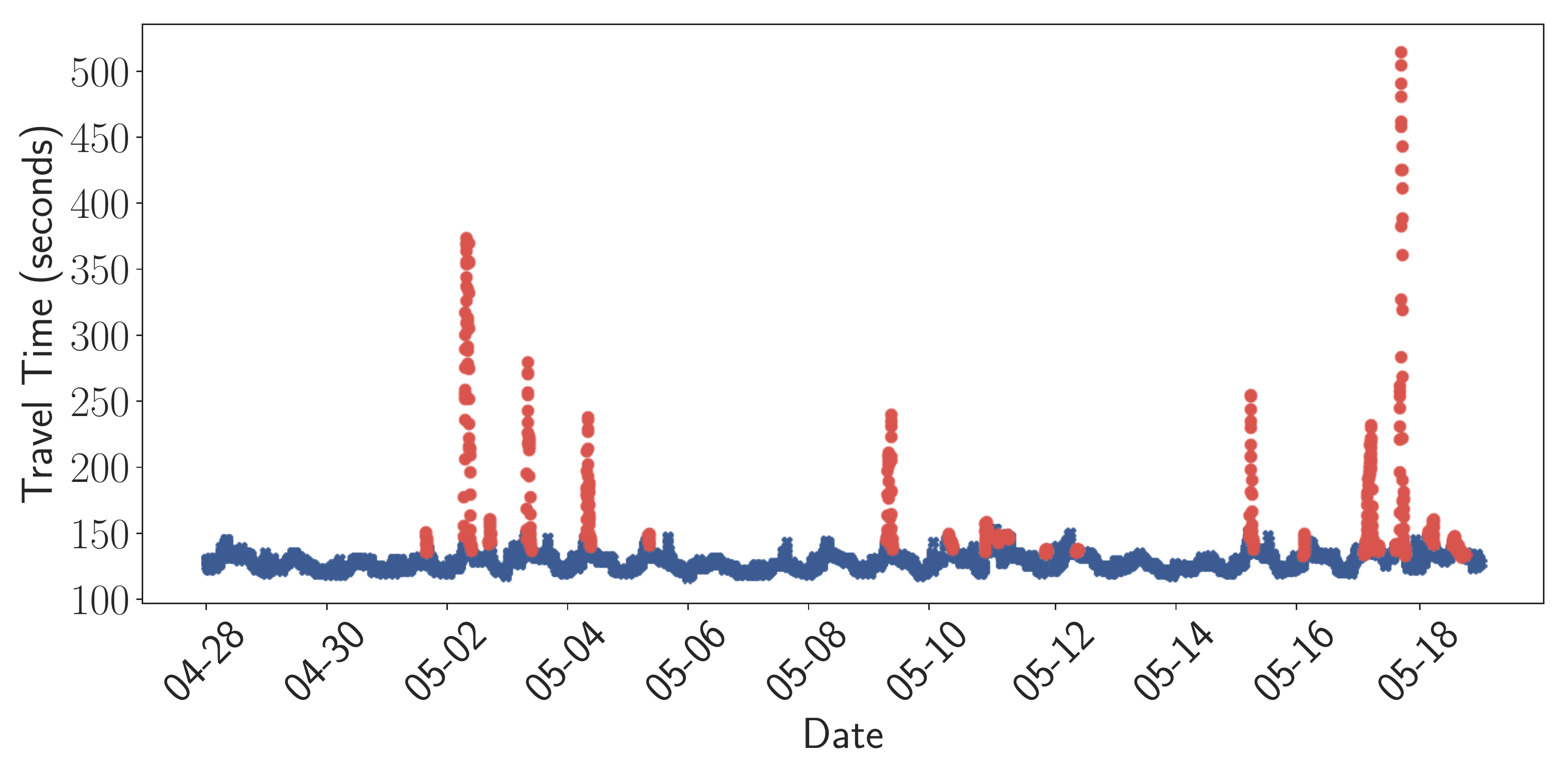}
	\caption{DFTB flags in test set 1 with threshold at the 80th percentile.}
	\end{subfigure}
	\hfill
	\begin{subfigure}{.48\textwidth}
	\includegraphics[width=\textwidth]{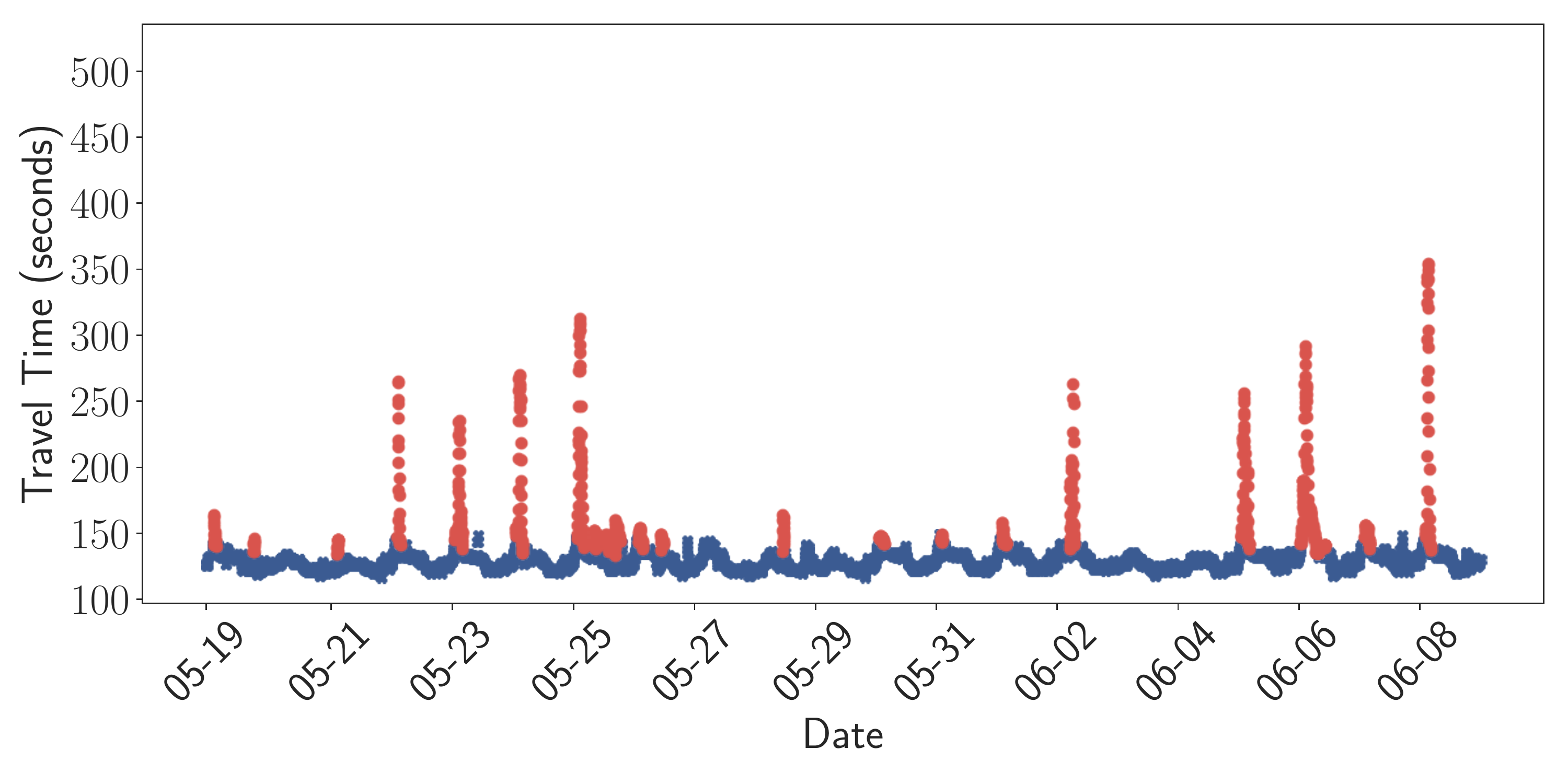}
	\caption{DFTB flags in test set 2 with threshold at the 80th percentile.}
	\end{subfigure}

	\caption[Test Data Travel Times]{ Travel time series for test set 1 (left column) and test set 2 (right column). On the top row, we show results when thresholding at the 40th percentile, and on the bottom row results for thresholding at the 80th percentile. As in the training case, we see the significant spikes in the series are captured by our methods, however thresholding at the 40th percentile also identifies many seemingly small perturbations in the travel times. As we move to the thresholding at the 80th percentile, we our DFTB flags match only the significant spikes and a small number of low travel time periods.
	For reference, the threshold values for the 40th and 80th percentiles are 4 and 23 minutes respectively. }\label{fig:TT_TestData}
\end{figure*}

The series of travel times with colours and symbols used to represent the presence of event flags in are shown in Fig. \ref{fig:TT_TestData}. 
Inspecting our results in Fig. \ref{fig:TestSetsNumEvents} and \ref{fig:TT_TestData}, we see that our results are both quantitatively and qualitatively similar in the training and testing scenarios considered.

\section{Severity ranking of DFTB events in real time}
\label{sec:severity}

A clear disadvantage of duration thresholding for raising DFTB event flags is that the process is retrospective: the duration of an event is only known once traffic conditions return to normal. 
Since NTIS updates every minute, this problem can be avoided by assigning a dynamical severity level to events based on the evolution of the trajectory in the flow density plane.
Minor fluctuations can then be removed almost in real time by filtering on this severity level.

\subsection{Dynamical severity measures for DFTB events}

There are many ways one could ascribe a severity level to an excursion from the typical region. 
Relevant features include how far the trajectory has deviated from the typical region and how long the trajectory has spent outside the typical region. 
Since these could be combined in different ways, perhaps with other features, there is no single `right' quantitative measure of severity.
We argue that any proposed measure should satisfy two practical requirements:
\begin{itemize}
\item[1)] severity should be 0 when trajectory is inside the typical region,
\item[2)] the numerical value of severity should be quickly and intuitively understandable.
\end{itemize}
The first criterion simply reflects that we are only interested in ranking abnormal traffic events. 
The second is a question of normalisation. We choose to normalize our measure so that a severity score of unity implies an event as extreme as the worst observed in the training data. 
Doing so permits an intuitive understanding: a score between 0 and 1 suggests traffic is in a state no more severe than seen in the training window, where as a score above 1 means the situation has developed more so than previously seen.
We tried multiple different definitions of severity. 
We ultimately decided that calculating the current distance of a trajectory to the nearest point on the boundary of the typical region alone provides a useful and computationally feasible  severity measure.
We did this by approximating the boundary of the typical region as a polygon and calculating the distance to the nearest point - see \ref{sec:computational}.
Excursions from the typical region almost always start small, reach some maximal distance and then return to the typical region after some time. 
A severity score based on distance to the typical region means that severity evolves in such a way that maximal severity is reached when the largest perturbation from the contour is observed, and we slowly increase to and decrease from this value in time. 
The duration aspect of severity is captured indirectly since excursions that reach a larger distance from the typical region are generally also of longer duration.

\subsection{Results for severity ranking of DFTB events in test data}

\begin{figure}[ht!]
	\centering
	\begin{subfigure}{0.48\linewidth}
	\includegraphics[width=\textwidth]{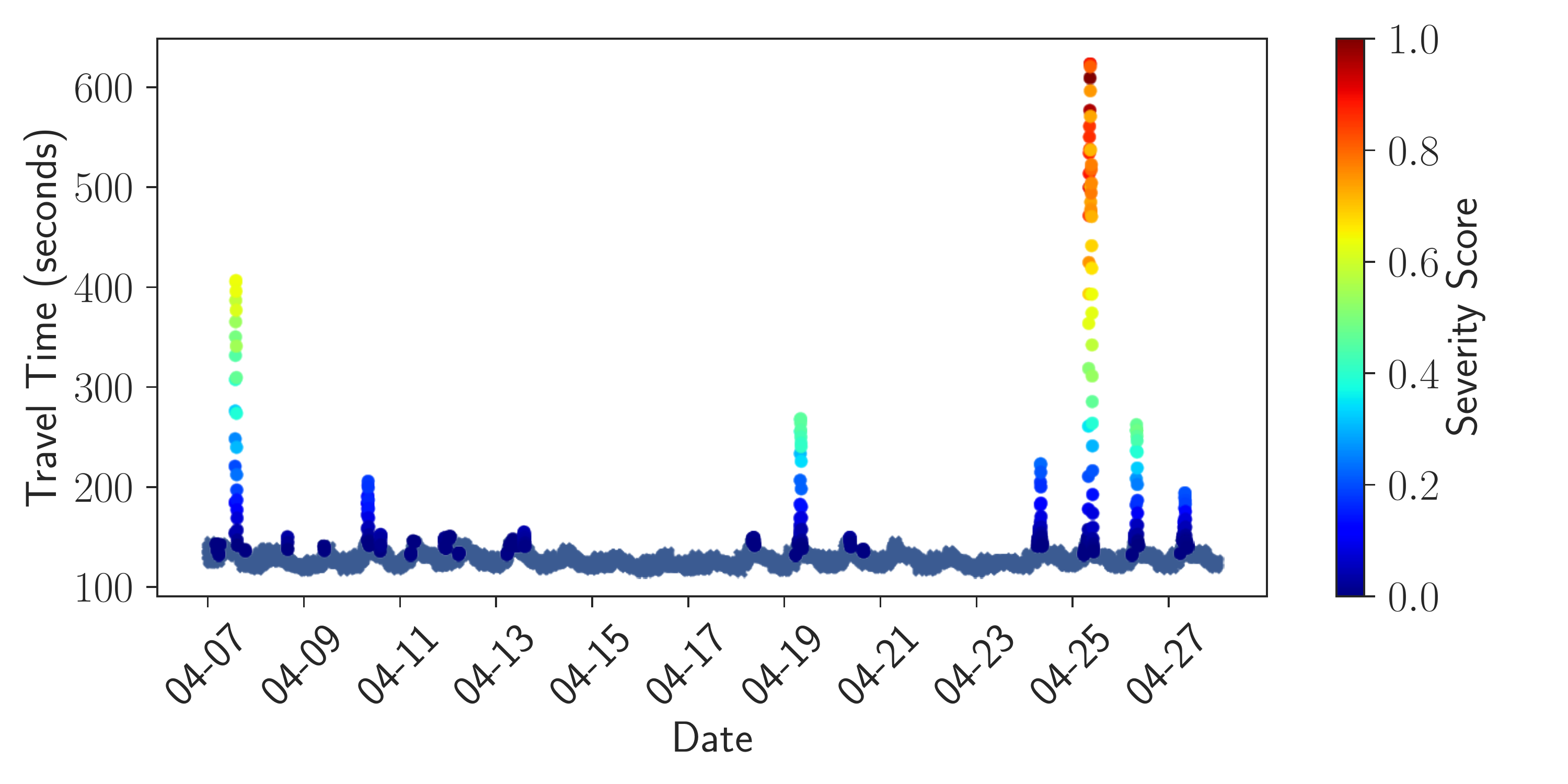}
	\caption{Training data.}
	\end{subfigure}
	\begin{subfigure}{0.48\linewidth}
	\includegraphics[width=\textwidth]{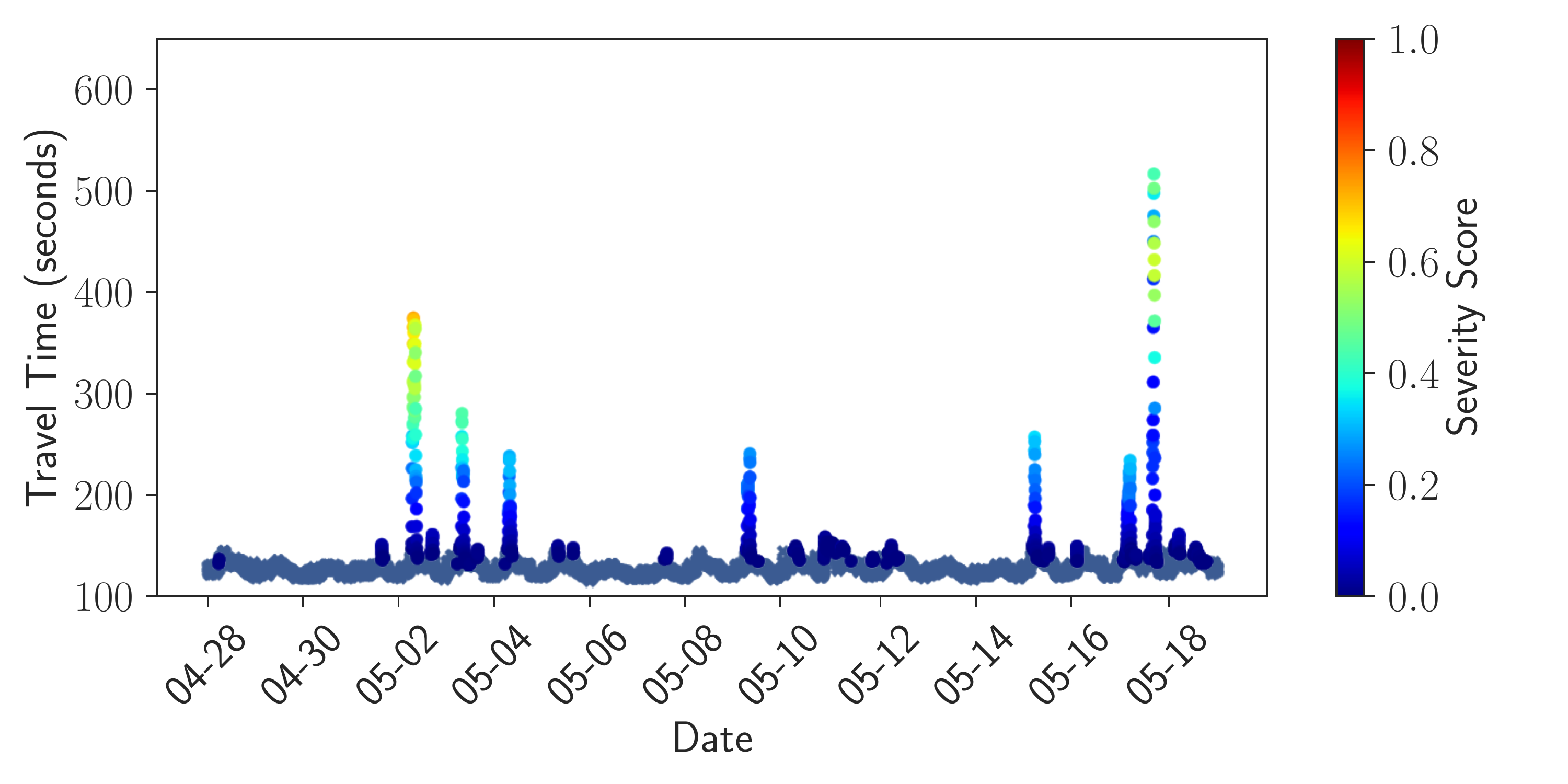}
	\caption{Test set 1}
	\end{subfigure}
	\begin{subfigure}{0.48\linewidth}
	\includegraphics[width=\textwidth]{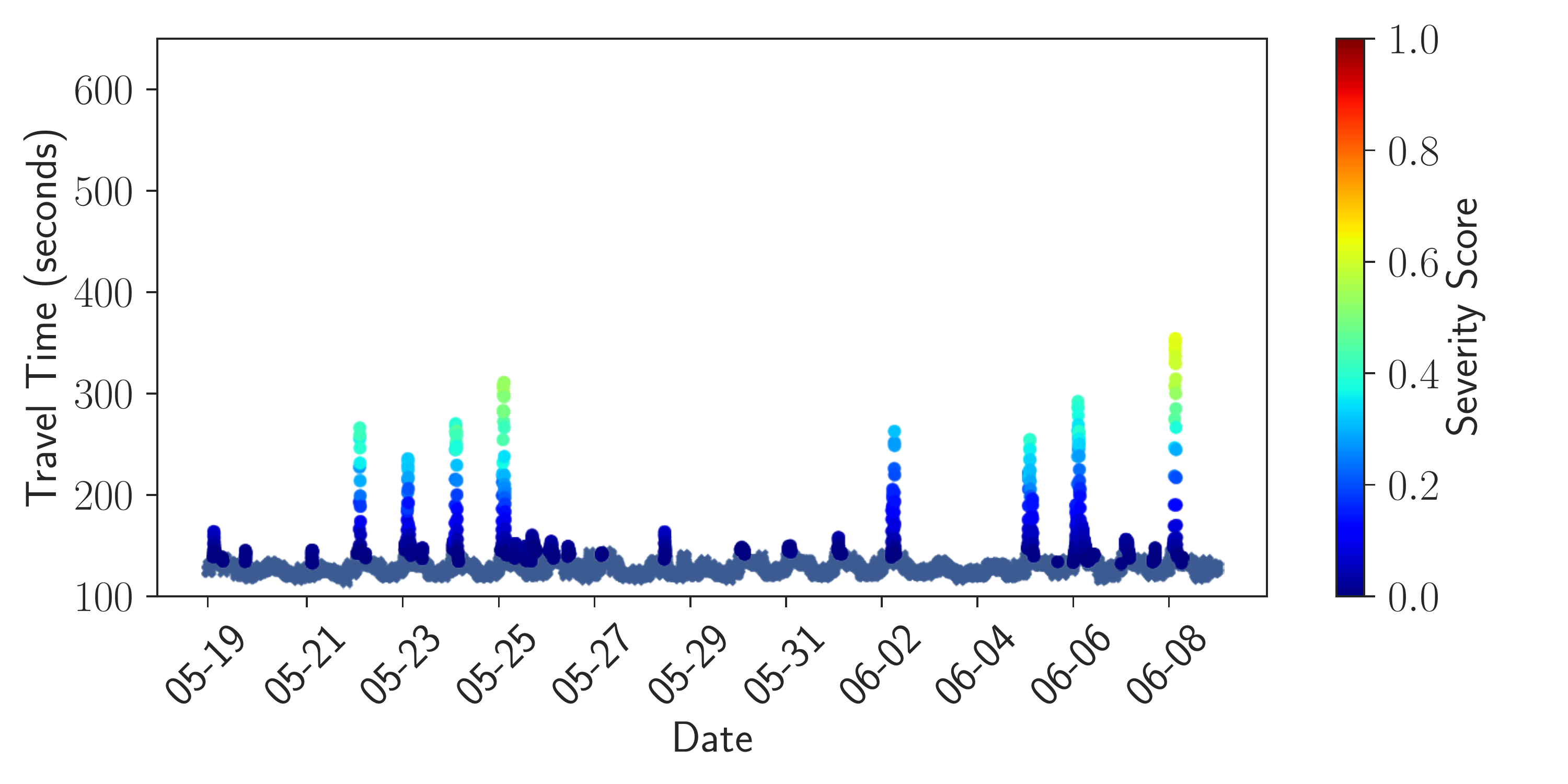}
	\caption{Test set 2}
	\end{subfigure}
	\caption[Comparison of Severity Score Methods]{ Comparison of severity scores in relation to travel times using the training and test data set. We colour the points by their severity score, normalized by the maximum value in the training window. In all cases, we see low severity scores when travel times are only marginally above the typical values, and higher scores when large deviations occur. }\label{fig:TestDataWithSeverity}
\end{figure}

Fig. \ref{fig:TestDataWithSeverity} shows a representative example of how the dynamical severity score of events  raised using the distance based severity method relate to travel times.
Considering Fig. \ref{fig:TestDataWithSeverity}, we clearly see the most severe scores are attained with the largest peaks in the travel time series.
Importantly, in Fig. \ref{fig:TestDataWithSeverity}, DFTB flags are raised, and severity computed, whenever a data-point leaves the typical region, eliminating the need to wait to decide whether or not to raise a flag, but allowing instead thresholding of events on the severity value.
We note that there is a single spike (around 05-18 in Test Set 1) that has the second largest observed travel time, but does not have the second highest severity. This can occur because we do not model travel times directly in our methodology.
This spike simply has a less extreme deviation in-terms of distance from the typical contour than other cases, which may have smaller travel time spikes but larger excursions from the typical region.
Nevertheless, the overall conclusion is that the severity of deviations from the typical region strongly correlate to travel time spikes. 

\section{Validation \& Comparison to Existing NTIS Events}\label{sec:Validation}

It is clear from section \ref{sec:severity} that our methodology identifies travel time spikes though atypical fluctuations with varying severities.
This is one form of validation showing that these atypical situations we would display to operators correspond to practical situations where road users are experiencing delays.
However, another form of validation is to question how many atypical fluctuations in the density-flow relationship correspond to actual event flags recorded in the data.
As discussed in section \ref{subsec:events}, NTIS has multiple event categories, each of which are practically important to traffic operators.
Here, we compare how alerts one might raise at different severity thresholds using our method compare to non-recurrent congestion event flags in the data, and how alternative methodologies perform at the same task.

\subsection{Comparison Models}

A number of methods for incident detection have been investigated throughout the literature, and were discussed in section \ref{sec:LitReview}.
We also note that NTIS only provides event locations at the link-level, and hence we never know which two loop sensors an event actually occurred between.
As such, the classic California algorithms are not appropriate for comparison as they rely on comparing adjacent sensors and flagging when discrepancies emerge between them.
However, in practice our method segments the density-flow diagram like the McMaster method, and defines a robust threshold for normality like the robust variants of the SND method.
We therefore choose to compare to these two methods as closely as possible using the available link level data. 

To implement the SND methodology, one isolates non-overlapping windows of speed time-series, and then computes summary statistics measuring the average and spread of the data.
Initially these were the mean and variance, however more robust statistics, being the median, inter-quartile range (IQR) and median absolute deviation (MAD) have since been applied.
For a given window $w$, if we denote the mean and median speeds as $\mu^{w}_{mean}$ and $\mu^{w}_{median}$ respectively, and the standard deviation, IQR and MAD as $\zeta_{sd}$, $\zeta_{IQR}$ and $\zeta_{MAD}$ respectively, we then compute threshold speeds for that window as: 
\begin{equation}\label{equ:SND}
\begin{split}
\tilde{s}_{1} &= \min\left( s_1^{max}, \mu^{w}_{mean} - c_{1}\zeta_{sd} \right), \\
\tilde{s}_{2} &= \min\left( s_2^{max}, \mu^{w}_{median} - c_{2}\zeta_{IQR} \right), \\
\tilde{s}_{3} &= \min\left( s_3^{max}, \mu^{w}_{median} - c_{3}\zeta_{MAD} \right). \\
\end{split}
\end{equation}
This approach is taken from \cite{data_driven_parallelizable_traffic_incident_detection_using_spatial_temporally_denoise_robust_thresholds}, where one avoids `swamping' of the robust statistics.
As in the cited work, we set the threshold values $s_1^{max}, s_2^{max}$ and $s_3^{max}$ to be 45 mph, informed by the work in \cite{traffic_congestion_and_reliability_trends_and_advanced_strategies}.
To apply this methodology, one then needs to determine optimal values for either $c_1$, $c_2$ or $c_3$ depending on methodology choice. 
We choose to use windows of length 15 minutes, computing a different set of statistics for windows across a week as in \cite{data_driven_parallelizable_traffic_incident_detection_using_spatial_temporally_denoise_robust_thresholds}.
We see consistently better performance on our dataset using robust methods, and generally a slight performance improvement when using the median and IQR so only consider the threshold defined by $\min\left( 45, \mu^{w}_{median} - c_{2}\zeta_{IQR} \right)$, which we refer to as SND (Robust) from now on.
After fitting, we raise alarms when the measured speed on a link falls below this threshold for at-least 3 consecutive minutes.

Another insightful method for comparison is a variant of the McMaster algorithm, in the sense that we take the occupancy-flow diagram at the link level and then segment it into distinct regions.
Occupancy is used here to exactly follow the work in \cite{mcmaster_distinguising_between_incident_congestion_and_recurrent_congestion_a_proposed_logic}.
This has more parameters than the SND methodology, as one has to specify a `lower bound of uncongested data' (LUD), a critical occupancy and a critical flow.
We consider a quadratic form for LUD as in \cite{online_testing_of_the_mcmaster_incident_detection_algorithm_under_recurrent_congestion}, giving us 5 total parameters to fit for the model.
It should be noted that the full McMaster logic, as detailed in \cite{mcmaster_distinguising_between_incident_congestion_and_recurrent_congestion_a_proposed_logic}, contains an initial step of determining a congested, bottle-neck or uncongested state, and then has a second step where one distinguishes between recurrent and incident congestion.
As we are dealing with link-level data, it is not sensible to conduct this secondary check, however we can still compare to a method that in spirit segments the occupancy-flow diagram and determines when the link state is outside reasonable bounds. 
If we consider only links not subject to recurrent congestion, it is sensible to make comparisons between this model and our own.
As such, one should not entirely consider this the McMaster algorithm, however it is also an adept comparison as the proposed methodology similarly attempts to segment the density-flow diagram, however by discovering the bounds on typical behaviour in an entirely data-driven way.

\subsection{Measuring Performance}

To tune the comparison models and measure performance, we first isolate all non-recurrent congestion events on a link.
We then split the data into a training set, which we calibrate our models with, and a test set which we measure performance on.
By calibration, we specifically mean 3 things. 
For the DFTB model, we are aiming to choose the optimal severity threshold that will distinguish between fluctuations representing labelled events and those not.
For the SND model, we compute the median and IQR of the data, and choose the optimal $c_2$ value that distinguishes between typical speed variation and decreases relating to incident flags.
Finally, for the McMaster algorithm, we fit all parameters to segment the diagram in such a way that when flags are raised by the method, they correspond optimally to real incident flags. 
As such it is clear that for the DFTB and SND methods, we are fitting a single parameter, and for the McMaster we are fitting 5.

Since the amount of data used for training could influence model performance,  for a fair comparison we used the same 3-week subset of data used to fit the typical behaviour contour. 
To verify that training window length does not strongly influence the results, we also fit the models using 12 weeks of data and found no significant changes.
To ensure that we evaluate our models on a truly representative set of data, we take our test period to be 45 weeks.

Given this, we then compute the standard performance metrics used throughout incident detection work.
The first of these is detection rate (DR), calculated as the number of labelled events detected divided to the total number of labelled events.
Another is false alarm rate (FAR), calculated as the number of times an event flag was raised when there was no flag in the data, divided by the number of applications of the algorithm.
The final one is mean time to detect (MTTD), being the average time between the start of an event in the data and the identification of the event by the method.
We express DR and FAR as percentages from now on.
As with all multi-objective optimisation problems, there is no mathematical answer to the question of how to trade these performance measures off against each other. This depends on the application.
We choose to combine DR, MTTD and FAR into the single optimization criterion, denoted performance index (PI), that was used in \cite{data_driven_parallelizable_traffic_incident_detection_using_spatial_temporally_denoise_robust_thresholds}, \cite{an_unsupervised_feature_learning_approach_to_improve_automatic_incident_detection} and \cite{support_vector_machine_models_for_reeway_incident_detection}:
\begin{equation}\label{equ:PerformanceIndex}
PI = \left( \epsilon_{DR} - \frac{DR}{100} \right) \cdot \left( \frac{FAR}{100} + \epsilon_{FAR} \right) \cdot \left( MTTD \right)
\end{equation}
where $\epsilon_{DR}$ and $\epsilon_{FAR}$ are constants that avoid trivial minima of the criteria. 
While we recognise that there is arbitrariness in this choice, it does facilitate easy comparison to existing literature.
We set $\epsilon_{DR} = 1.01$ and $\epsilon_{FAR} = 0.001$ as in \cite{data_driven_parallelizable_traffic_incident_detection_using_spatial_temporally_denoise_robust_thresholds}.
We choose the parameters of each model that minimize equation \ref{equ:PerformanceIndex}, and then compare their performance to our anomaly detection methodology. 
In general, calibration of automated incident detection methods is considered to be a difficult task \cite{traffic_management_centre_use_of_incident_detection_algorithms_findings_of_a_nationwide_survey}, yet to fit a set of models over many links, one must comprise and use some reasonable criteria as an optimization objective.

It is important to note that with the discussed methods, one can easily envision how an operator could interpret and change the single parameters of the DFTB and SND methods.
Since both have essentially one variable parameter, the severity threshold in the DFTB case and the $c$ parameter in the SND case, it is entirely interpretable to an operator experienced in traffic management but not in the technical aspects of the model.
An increase in either of these thresholds will allow more variation before flagging, but it is reasonably clear how each of the DR, FAR and MTTD would naturally increase or decrease when doing this.
This is not true however for the McMaster algorithm, as altering a 5 parameter model and understanding how your changes will impact the performance criteria is not as simple. 

\subsection{Validation Results}

The initial comparison we make is to take the  same 3-week dataset we trained the contour of typical behaviour on, and train the comparison models using the NTIS labels.
After choosing the parameter set that has optimal PI in this training set, we then apply the tuned model to the unseen test data.
Results for this comparing the DFTB performance in DR, FAR and MTTD to the robust SND model are give in table \ref{table:Results3Weeks}, considering a variety of link lengths and locations along the M25.
We have selected this subset of links as those that appear to have minimal data quality problems, both in missing time-series data and event flags.
After presenting aggregated statistics in the final rows of table \ref{table:Results3Weeks}, we then consider the typical differences between the models in table \ref{table:Results3WeeksSummary}, where we also test if there is statistically significant evidence that distinguishes between the two models for each performance metric.
\begin{table}[ht!]
		\centering
	    \begin{tabular}{|c|c|c|c|c|c|c|c|} 
	    \hline
	    \multicolumn{2}{|c|}{Link} & \multicolumn{2}{c|}{DR}           & \multicolumn{2}{c|}{FAR}    & \multicolumn{2}{c|}{MTTD}    \\
	    \hline
	    Location    & Length (km) & \makecell{SND \\ (Robust)} & DFTB & \makecell{SND \\ (Robust)} & DFTB & \makecell{SND \\ (Robust)} & DFTB \\        
	    \hline
	    East        & 0.7 & 80.392 & 86.275 & 0.937 & 	0.968 & 5.707 & 4.659 \\ 
	    \hline
	    East        & 1 & 89.130 & 84.783 & 1.990 & 2.250 & 4.341 & 2.589 \\ 
	    \hline
	    South East  & 0.5 & 82.353 & 79.412 & 2.140 & 1.779 & 10.143 & 12.222 \\  
	    \hline
	    South East  & 0.4 & 83.871 & 80.645 & 2.150 & 1.781 & 7.500 & 9.320 \\
	    \hline
	    South East  & 5.1 & 81.967 & 78.689 & 0.741 & 0.545 & 14.260 & 14.771 \\  
	    \hline 
	    South       & 0.7 & 96.552 & 96.552 & 2.664 & 0.712 & 4.214 & 4.250 \\ 
		\hline 
	    South       & 1.2 & 69.091 & 49.091 & 0.706 & 0.041 & 10.789 & 14.741 \\ 
	    \hline
	    South West  & 0.9 & 94.271 & 81.250 & 10.058 & 1.348 & 3.541 & 7.025 \\ 
	    \hline
	    South West  & 6.9 & 44.106 & 46.388 & 2.402 & 0.435 & 14.957 & 12.877 \\  
	    \hline
	    West        & 2.5 & 64.286 & 78.571 & 0.490 & 0.457 & 9.593 & 9.756 \\ 
	    \hline
	    West        & 3.9 & 35.057 & 89.080 & 1.578 & 2.566 & 10.672 & 7.168 \\ 
	    \hline
	    West        & 1.2 & 91.163 & 81.860 & 9.402 & 0.651 & 12.096 & 15.341 \\ 
	    \hline
	    North West  & 1.9 & 98.131 & 90.654 & 5.317 & 1.594 & 7.410 & 7.103 \\ 
	    \hline
	    North West  & 1.3 & 47.945 & 42.466 & 1.489 & 0.334 & 17.371 & 13.903 \\   
	    \hline
	    North West  & 0.9 & 54.286 & 48.571 & 0.522 & 0.408 & 17.842 & 13.412 \\ 
	    \hline
	    North       & 4.4 & 69.143 & 61.714 & 1.032  & 0.736 & 15.289 & 14.120 \\ 
	    \hline
	    North East  & 1.9 & 89.091 & 87.273 & 1.026  & 0.845  & 11.245 & 11.604 \\ 
	    \hline
	    \multicolumn{8}{c}{} \\
		\hline
	    \multicolumn{2}{|c|}{Mean}    & 74.755 & 74.310 & 2.626 & 1.026 & 10.410 & 10.286 \\
	    \hline
	    \multicolumn{2}{|c|}{Median}  & 81.967 & 80.645 & 1.578 & 0.736 & 10.672 & 11.604 \\
	    \hline
	    \multicolumn{2}{|c|}{Std Dev} & 19.581 & 17.429 & 2.914 & 0.730 & 4.538  & 4.141 \\
	    \hline
	    \multicolumn{2}{|c|}{IQR}     & 24.844 & 24.561 & 1.465 & 1.137 & 6.850  & 6.800 \\
	    \hline
	    \end{tabular}
	    \caption{Comparison of model performance across a set of representative links of varying lengths and locations around the M25. All models are shown 3 weeks of training data. }\label{table:Results3Weeks}
	    \vspace{3mm}
	    \begin{tabular}{|c|c|c|c|c|}
	    \hline 
	           & \makecell{DFTB DR - \\ SND (Robust) DR} & \makecell{DFTB FAR - \\ SND (Robust) FAR} & \makecell{DFTB MTTD - \\SND (Robust) MTTD} \\
	    \hline 
	    Mean   & -0.445           & -1.600             & -0.124 \\
	    \hline
	    Median & -3.278           & -0.361             & 0.036 \\
	    \hline
	    \makecell{Wilcoxon signed-rank \\ test p-value} & 0.170 & 0.004 & 0.890 \\
	    \hline
	    \makecell{Sign test p-value} & 0.077	& 0.0127 & $>0.999$ \\
	    \hline
	    \end{tabular}
	    \caption{ Aggregated model performance summary. We consider the mean and median differences between the 3 evaluation metrics, and question if there is statistically significant evidence to suggest the two models have differing performance in each metrics. This is done through both a non-parametric Wilcoxon signed-rank test and a sign test, both considering the measurements to be paired. }\label{table:Results3WeeksSummary}
\end{table}
Inspecting table \ref{table:Results3Weeks}, we see that the mean and median DR for the DFTB method is lower detection rate than the robust SND methodology across the considered links.
The mean and median FAR across the links is lower for the DFTB compared to the robust SND method, and we see that the DFTB results in a marginally lower mean value of MTTD, but a marginally higher median value of the same metric.

Before any further discussion of these performance scores, let us first determine which of the observed differences are statistically significant.
This is to avoid basing conclusions on small differences that could have arisen by chance.
Statistical significance of differences in model performance is a topic that has been  discussed extensively in the machine learning literature, with \cite{statistical_comparsions_of_classifiers_over_multiple_data_sets} providing an overview of the hypothesis tests that are most appropriate for comparing performance across multiple datasets.
This is further discussed in \cite{non_parametric_statistical_analysis_for_multiple_comparison_of_machine_learning_regression_algorithms}.
They conclude that two statistical tests are generally appropriate for model comparison: the Wilcoxon signed-rank test and the sign test.
We therefore apply both of these, and show results in table \ref{table:Results3Weeks}.
To perform a Wilcoxon signed-rank test, we first compute the paired differences in the data, and record the sign of this (+1 or -1).
We then compute the absolute values of each paired difference, and rank them from smallest to largest.
Our test statistic is then the sum of these ranks, multiplied by +1 or -1 depending on the sign of the paired difference.
This statistic is then compared to known distributions under the null hypothesis: assuming the difference in medians between the paired samples is 0.
Hence, in our context, rejecting the null hypothesis under the Wilcoxon signed rank test suggests there is statistically significant evidence that the median difference between the two models is not 0. 

A sign test on the other hand does not account for ranks, and acts as more of a `tournament' between two models.
A `match' in this tournament represents training and then testing each model on a single link.
In each match the two models get the same training data and calibrate to this, then have to predict the same test data.
We then only record the `winner' of each match, asking which model had superior DR, FAR, or MTTD.
If one model consistently wins when we compare the results for each match, then we suspect that the methods are not equally skilled at the particular task.
Formally, the null hypothesis in such a test is that given a pair of measurements, it is equally likely that either the first or second is larger or smaller than the other.
This suggests that rejecting the null hypothesis in our context means one model is consistently producing a larger DR, or lower FAR and MTTD.
Further discussion of both of these tests is given in \ref{appendix:Hyptests} for those interested, but one can also consult \cite{statistical_comparsions_of_classifiers_over_multiple_data_sets} or papers that have since used the discussed methods, such as \cite{a_divide_and_conquer_based_early_classification_approach_for_multivariate_time_series_with_different_sampling_rate_components} and \cite{belief_based_chaotic_algorithm_for_support_vector_data_description} for other examples of modelling works using this approach.

The results from these tests in table \ref{table:Results3WeeksSummary} suggest there is sufficient evidence to reject the null hypothesis of the median difference between the DFTB and SND FAR being 0 (Wilcoxon signed-rank test).
Further, there is sufficient evidence to reject the null hypothesis that the probability the FAR for the DFTB method is larger than that of the robust SND method equals 0.5 (sign test).
None of the other tests in table \ref{table:Results3WeeksSummary} suggest any statistically significant results.    
This analysis is based on a 0.05 significance level, with a p-value falling below 0.05 indicating statistically significant results, providing evidence to reject the null hypothesis.

If we compare to the McMaster algorithm with the discussed link-level limitation, we see that it consistently has a higher MTTD compared to the DFTB model on our dataset, taking a mean of 4.5 and median of 4 minutes longer to detect events than the DFTB does. 
It also has a lower DR, 6.7\% lower in mean and 2.2\% lower in median.
Finally, it achieves a better false alarm rate, with a mean difference of 0.19\% and a median difference of 0.28\%, but this is at the expense of the other two metrics, and being far harder to fit.

We also note that the FAR reported above might seem higher than in the literature. 
We are aware some events can be missed in the system, an indeed informed of this by system experts.
Additionally, we have not pre-filtered the data to identify only events that blocked lanes as in \cite{data_driven_parallelizable_traffic_incident_detection_using_spatial_temporally_denoise_robust_thresholds} and other works.
We find FAR to be consistent across models showing this is a systematic feature of the data and not the models.
Our typical MTTD values are slightly larger than those discussed in \cite{data_driven_parallelizable_traffic_incident_detection_using_spatial_temporally_denoise_robust_thresholds}, however again this may relate to the data quality problems, and could be further impacted by the definition of deviation from profile events that could be minor for some time then develop into significant events on the link.

Finally, one has to also consider from a practical perspective, what might be important to a traffic operator who would be in an environment using any of these models?
A survey of such operators was conducted and analysed in \cite{traffic_management_centre_use_of_incident_detection_algorithms_findings_of_a_nationwide_survey}, and one of the principle reasons for limited integration and operational use of automated incident detection systems was difficulty of algorithm calibration.
Difficulty of calibration can be thought of in two ways.
The first is deciding on an exact optimization criteria that captures what peak performance means exactly to an operator.
Whilst we have used PI for this, it is an open problem to determine what might be more appropriate.
However, our model, as with the SND methodology, is fully interpretable and since both have a single parameter, calibration is much more reasonable than say, a deep learning model that might have billions of parameters that an operator could not fesable adjust without expert knowledge. 

\subsection{Analysis in Bi-Modal Speed Cases}

\begin{figure}[ht!]
	\begin{subfigure}{.32\textwidth}
	\includegraphics[width=\textwidth]{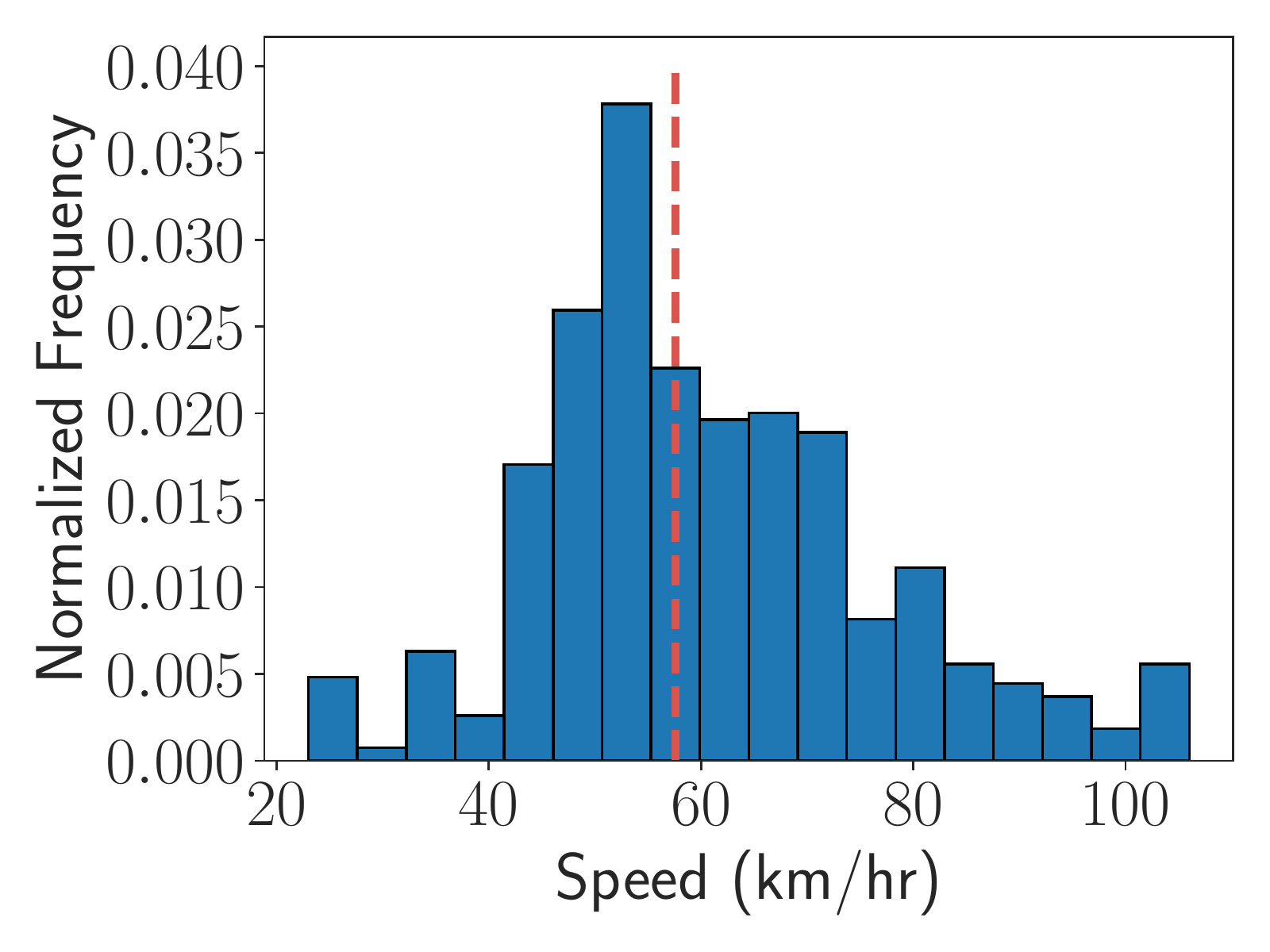}
	\caption{An example uni-modal speed distribution.}\label{fig:UniModalExample}
	\end{subfigure}
	~
	\begin{subfigure}{.32\textwidth}
	\includegraphics[width=\textwidth]{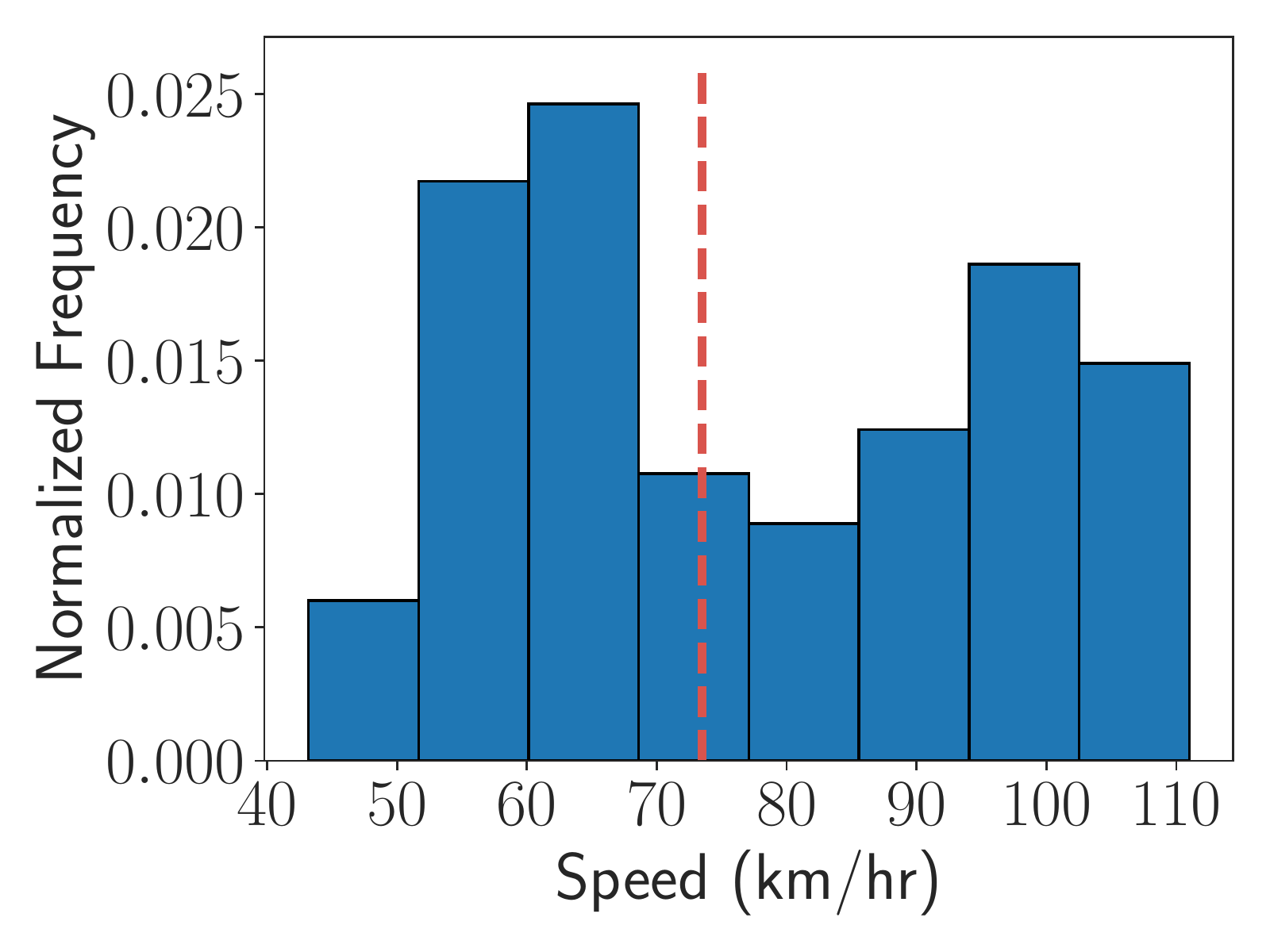}
	\caption{An example bi-modal speed distribution (1).}\label{fig:BiModalExample1}
	\end{subfigure}
	~
	\begin{subfigure}{.32\textwidth}
	\includegraphics[width=\textwidth]{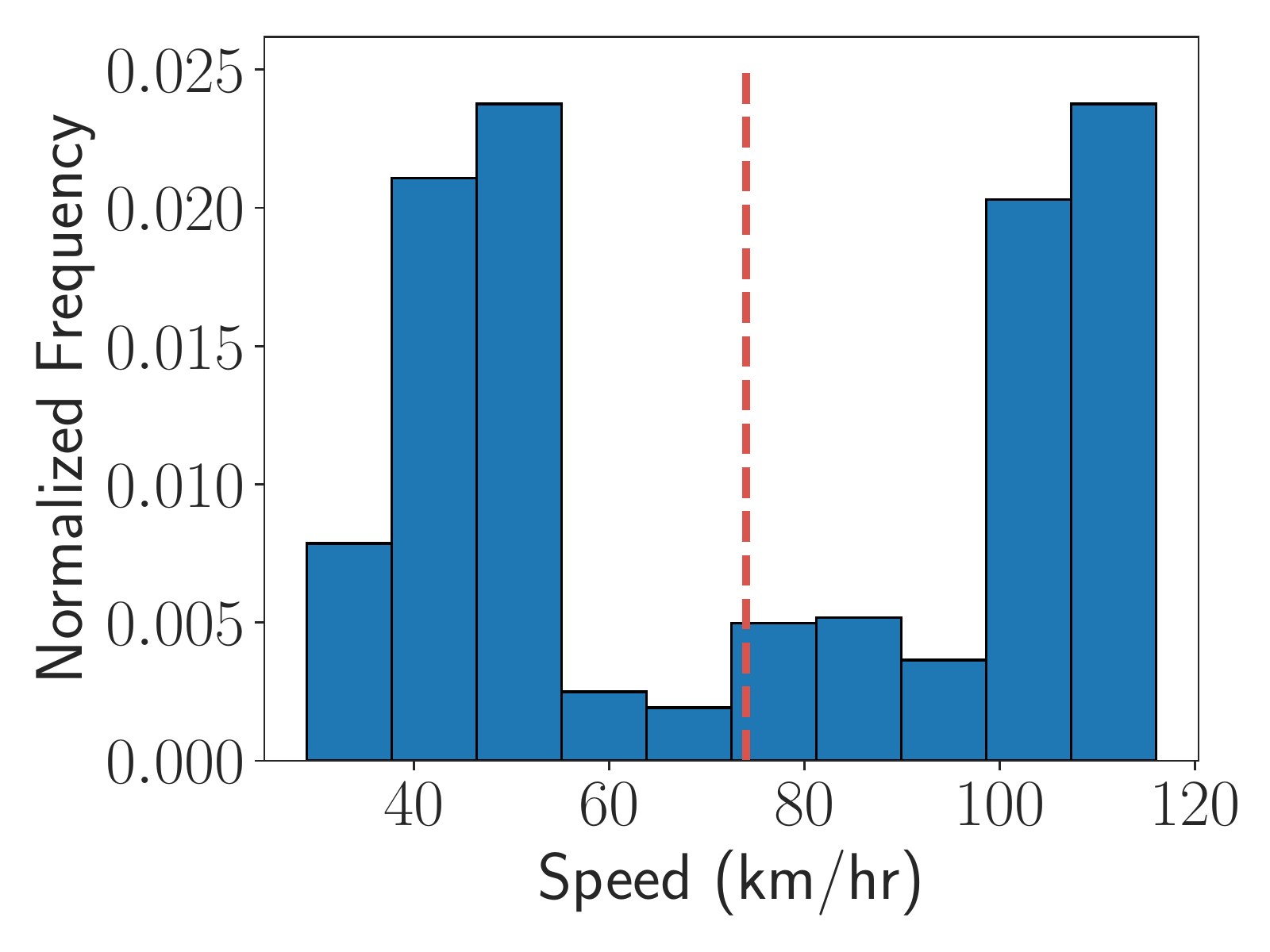}
	\caption{An example bi-modal speed distribution (2).}\label{fig:BiModalExample2}
	\end{subfigure}
	\caption[Speed Distributions in Example Time Windows]{ Distribution of speed in 3 example time-windows, each of length 15 minutes. We take data for 45 weeks of data and plot the resulting histogram for each bin, marking the median of the data with a dashed red line. }\label{fig:BiModalBinExample}
\end{figure}

The above performance tests indicate that our algorithm is generally comparable to well calibrated robust SND in terms of performance on NTIS events data although there are a few instances where there are very large differences, particularly in the FAR. 
We now investigate why this is so.
In the process, we identify situations in which our bivariate approach provides some fundamental advantages over SND.
Again referring to table \ref{table:Results3Weeks}, we see that robust SND method attains a $10.058\%$ FAR on the first link in the south-west investigated, which is enormous even considering the data-quality problems we might have. 
First, one might say this is just a `poor' training window choice, not representative of the long-term link behaviour, however this does not appear to be the case.
If we extend the training period, even providing 12 and 24 weeks of data, we still see similar false alarm rates from the SND method.
Additionally, the DFTB achieves a FAR of 1.3\%, at the expense of a higher MTTD and lower DR.
This suggests that, on this link, some property of the speed leads to it not being well modelled by a single location and scale parameter pair.
We investigate this by taking all false alarms raised by the robust SND method, and considering what time-bin each lies in through the week.
We then consider the distribution of the speeds in each bin.
If the SND assumptions were valid, the speeds in each bin would be uni-modal, and a sensible measure of outliers is $\mu - c\sigma$.
This bin could still contain extreme values or a heavy tail, which are accounted for through the use of robust statistics.
However, if the speed distribution in a bin is bi-modal, then it is not appropriate to judge outliers using a single location and scale parameter.
To illustrate this we refer to Fig. \ref{fig:BiModalBinExample}, where we plot the distribution of speeds in 3 time-windows. 
In Fig. \ref{fig:UniModalExample}, we show an example where the use of the SND threshold is reasonable, where as in Fig. \ref{fig:BiModalExample1} and \ref{fig:BiModalExample2} we show cases where there appears to be bi-modality in our data.

Inspecting Fig. \ref{fig:UniModalExample}, we see an example time-window on the link where it is completely reasonable to judge outliers as values being some threshold away from the median of the distribution.
However, in Fig. \ref{fig:BiModalExample1}, we see a bi-modal distribution, one mode around 100 km/hr and another at 60 km/hr.
Similarly, in Fig. \ref{fig:BiModalExample2}, we see one mode at 110 km/hr and another at 50 km/hr.
This behaviour does not appear to be the result of roadworks, as we have removed points marked with these flags before plotting the distributions.
With this bi-modal behaviour, optimizing the SND parameters essentially faces a dilemma.
Since the inter-quartile range, median absolute deviation and standard deviation will all be very large, the method can choose a very small $c$ value to still flag in these bi-modal cases, but then be more prone to false alarms elsewhere, or choose a larger value and almost completely ignore flagging in these bins.

\begin{figure}[ht!]
	\begin{subfigure}{.48\textwidth}
	\includegraphics[width=\textwidth]{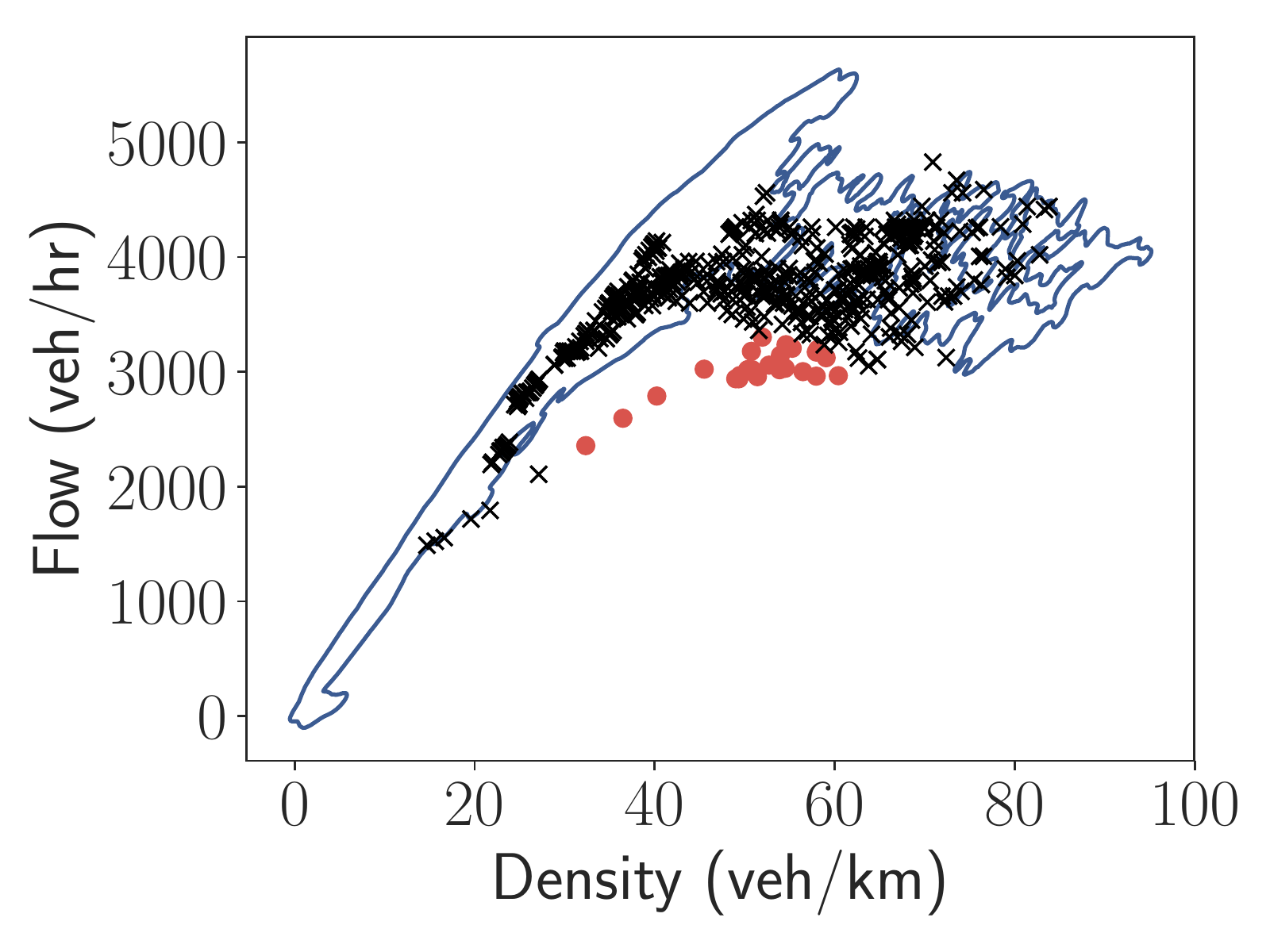}
	\caption{ Example flags for the window (1) shown in Fig. \ref{fig:BiModalExample1}, plotted on-top of the contour of typical behaviour used in the DFTB method. }\label{fig:BiModalExample1DFTB}
	\end{subfigure}
	~
	\begin{subfigure}{.48\textwidth}
	\includegraphics[width=\textwidth]{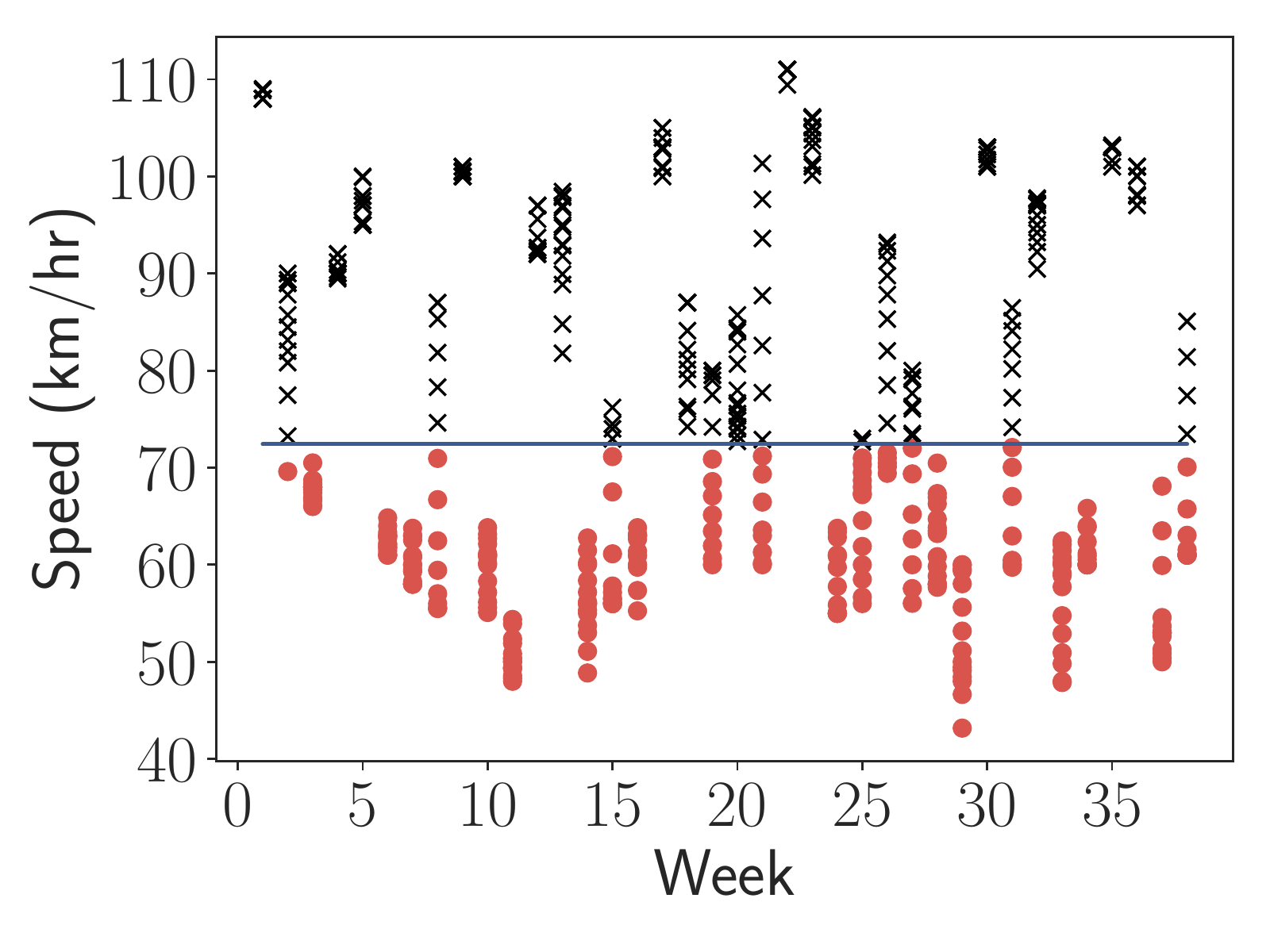}
	\caption{ Example flags for the window (1) shown in Fig. \ref{fig:BiModalExample1}, plotted with the speed threshold used in the robust SND method. }\label{fig:BiModalExample1SND}
	\end{subfigure}

	\begin{subfigure}{.48\textwidth}
	\includegraphics[width=\textwidth]{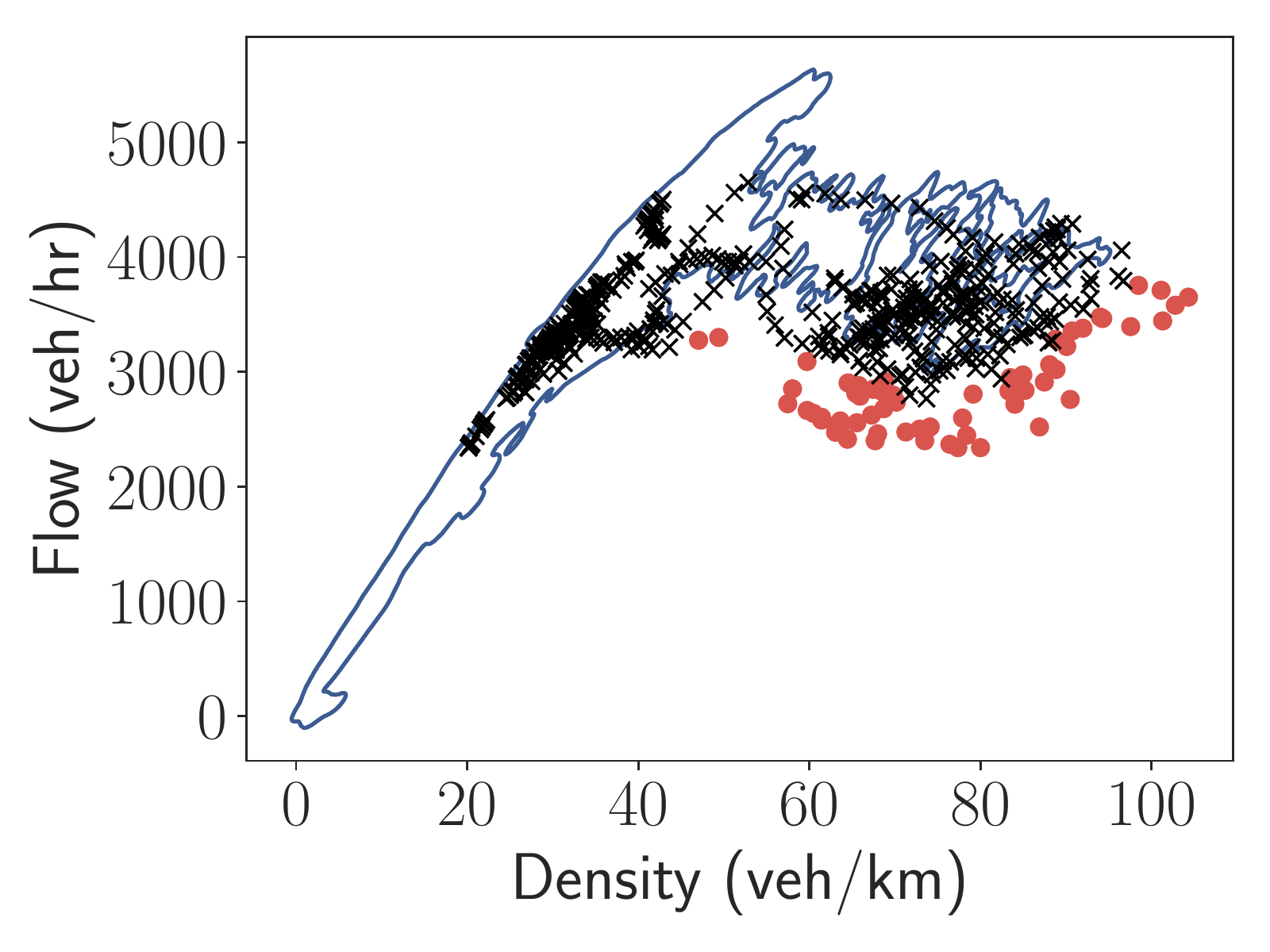}
	\caption{ Example flags for the window (2) shown in Fig. \ref{fig:BiModalExample2}, plotted on-top of the contour of typical behaviour used in the DFTB method. }\label{fig:BiModalExample2DFTB}
	\end{subfigure}
	~
	\begin{subfigure}{.48\textwidth}
	\includegraphics[width=\textwidth]{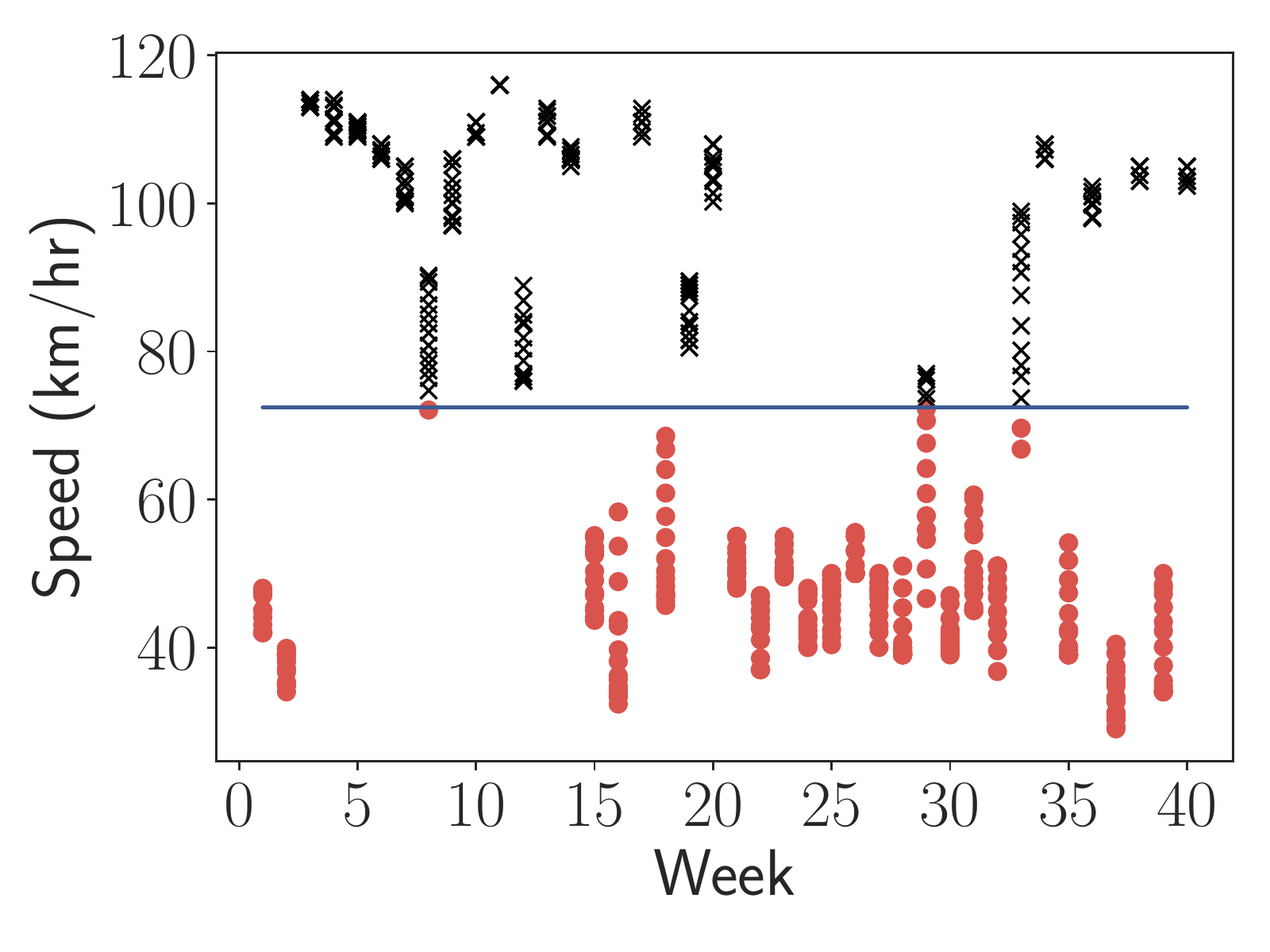}
	\caption{ Example flags for the window (2) shown in Fig. \ref{fig:BiModalExample2}, plotted with the speed threshold used in the robust SND method. }\label{fig:BiModalExample2SND}
	\end{subfigure}
	\caption[Example flags in a specific time-window for each method]{ Example flags in a specific time-window for each method. We show the threshold for each method with a solid blue line, points flagged as `typical' with black ($\times$), and points flagged as atypical in red ($\bullet$). Notice that points can be outside of the contour of typical behaviour, but still be typical as they are not far enough away to have a severity past some threshold. Note that these SND speed thresholds have been set based on optimal performance in the training window, and the very large spread of data in some bins appears to have set the $c$ parameter to a very small value, which in turn has lead to quite high thresholds for these two bins, but over the entire dataset gave better performance. }\label{fig:CompareBimodalData}
\end{figure}

We can further show this point by plotting the actual data inside each bin, along with the differing thresholds for them.
To do this, we take bins (1) and (2) as shown in Fig. \ref{fig:BiModalExample1} and \ref{fig:BiModalExample2}, and collect all the reported data during them for the specific link.
We then plot this data, along with the thresholds from the DFTB and robust SND methodology, with results shown in Fig. \ref{fig:CompareBimodalData}.

The main interpretation one should make from Fig. \ref{fig:CompareBimodalData} is that data-points in the shown time-windows clearly fall in two regimes of the density-flow diagram. 
Some have densities around 20-40 veh/km and flows between 2000 and 4500 veh/hr.
Additionally, there is a further cluster of data-points at high density high flow points, around densities of 60-80 veh/km and flows of around 3000-4000 veh/hr.
A weakness of the SND method is highlighted as it cannot capture the two regime component in the bin as the DFTB method does.

Exactly what causes this bi-modality is unclear without a significant amount more data on the physical properties of the road and environment throughout the year.
One could see if there is some sort of seasonal bottleneck impacting the link, for example once every 2 weeks or month where the flow becomes restricted, speed drops but not due to an incident.
If this were true, one could adapt the SND methodology to use a different number of bins and so forth, but we see that our model does not need this adaptation.
By using a different definition of a typical threshold, we are able to distinguish better in these particular bi-modal states than a simple uni-variate approach as with the SND algorithm, however all of the complexity in defining our threshold is automated, with the kernel smoothing, bandwidth selection and so forth, so the end user still receives a single parameter model that is fully interpretable. 

\section{Summary \& conclusion}

In this work, we have explored how empirical analysis of fluctuations in the flow-density relationship can be combined with streaming NTIS data to identify significant deviations from typical behaviour of traffic flow on a road section almost in real time. 
Starting by considering a kernel density estimate of the raw density-flow data, we showed how to define a contour of typical behaviour in which standard traffic behaviour lies, and investigated the deviations from this contour.
We showed how the time-scales of these deviations can be used to infer the presence of significantly atypical behaviour on a link, and hence identify problematic links.
To avoid creating a delay in raising events, we looked at methods to automatically and in real time assess the severity of a deviation, finding the distance of a density-flow data-point from the typical behaviour contour offered a severity measure that directly correlated with the spikes in travel time series.
Our methods indirectly identify spikes in the travel time series and event flags raised by the NTIS system.   

This approach may present some practical advantages over existing methods of event detection. 
One advantage is that our approach is entirely model free in the sense that the boundary of the region of typical behaviour for each link is autonomously learned entirely from data.
There is no need for complicated time series analysis of travel times or calibration of parameters or travel time profiles at the level of individual links in order to identify events.
Furthermore, by relearning the boundary of the typical region at periodic intervals, the system could be made self-adaptive to automatically capture long term changes in the behaviour of traffic on a link.
A second advantage is that identification of events and assignment of a severity score can be done almost in real time since the process operates directly on the speed and flow data. 
From a practical perspective, we receive data from the NTIS system on a minute-scale, and the computation of a severity takes significantly less than this.
An obvious future extension would be to try this approach directly on loop level data rather than on aggregated link-level data as we have done here. 
This may provide a more granular view of what is happening on the network and may further increase the speed with which information about significant deviations can be extracted.

\section*{Acknowledgements}

The authors are grateful to Dr. Steve Hilditch and Thales UK for sharing expertise on the operation of NTIS and UK transportation systems more generally.  We also thank Ayman Boustati and Alvaro Cabrejas Egea  for help with data acquisition and pre-processing. This work was supported in part by the EPSRC and MRC under grant number EP/L015374/1.

\appendix

\section{Computational aspects of our implementation}
\label{sec:computational}

Two points should be noted from an implementation standpoint.
Firstly, it is possible, although rare, to find deviations from the typical behaviour contour that are low-density and high-flow. 
This is essentially atypically good periods for a link, and should not be considered in an event detection system.
From the formulation of our system, such deviations always occur on the top left of the contour, with deviations associated with atypically poor behaviour being associated with trajectories leaving the contour to the right, to higher density values.
As a result, we can simply check, when a deviation occurs, if we have exited the contour to the left or right, and if we have exited to the left we do not raise any event flags.
In practice, such an adjustment is rare but ensures we never label ideal situations as events to act on.

Secondly, the contours are typically made up of a single large component, and occasionally a number of smaller components scattered. 
If these components are negligibly small relative to the total area of all contour components, we can neglect them for computational speed and simplicity.
To implement this, we discard any contour components that make up less than $5\%$ of the total area of the entire contours. 

Numerically, we determine the contours of typical behaviour using the ks package in R \cite{ks_package}, and detect data-points as inside or outside of them using the Matlab function `inpolygon' although there are many ray casting implementations that could be equally used.
As the contour has no explicit functional form, we determine the distance by first defining it on a high-resolution set of (x,y) points, then finding the closest point in this set to the data in question.

\section{Stability of flow-density relationship over time}\label{subsec:Stability}

\begin{figure*}[ht!]
	\begin{subfigure}{.32\textwidth}
		\includegraphics[width=\textwidth]{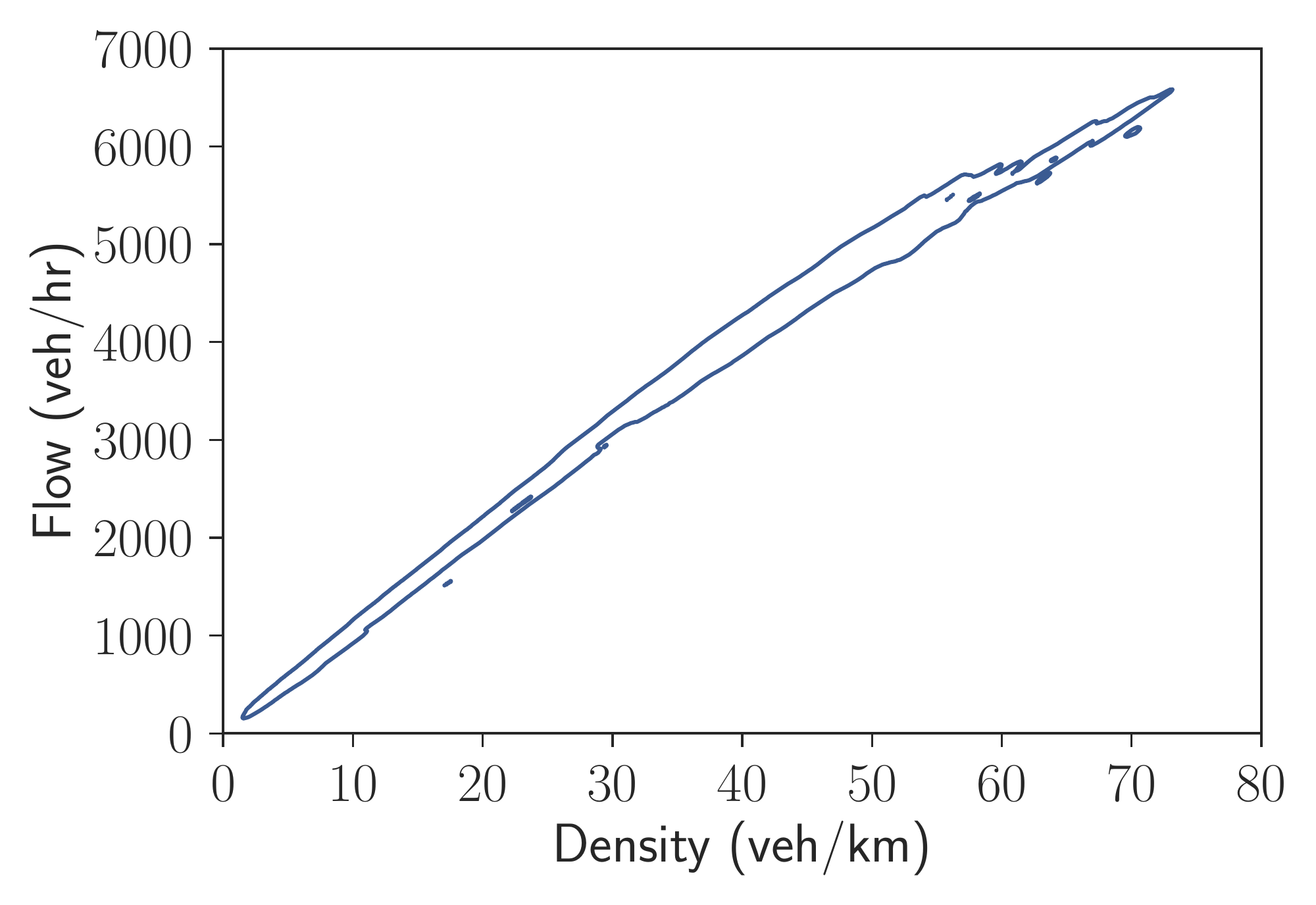}
	\end{subfigure}
	\begin{subfigure}{.32\textwidth}
		\includegraphics[width=\textwidth]{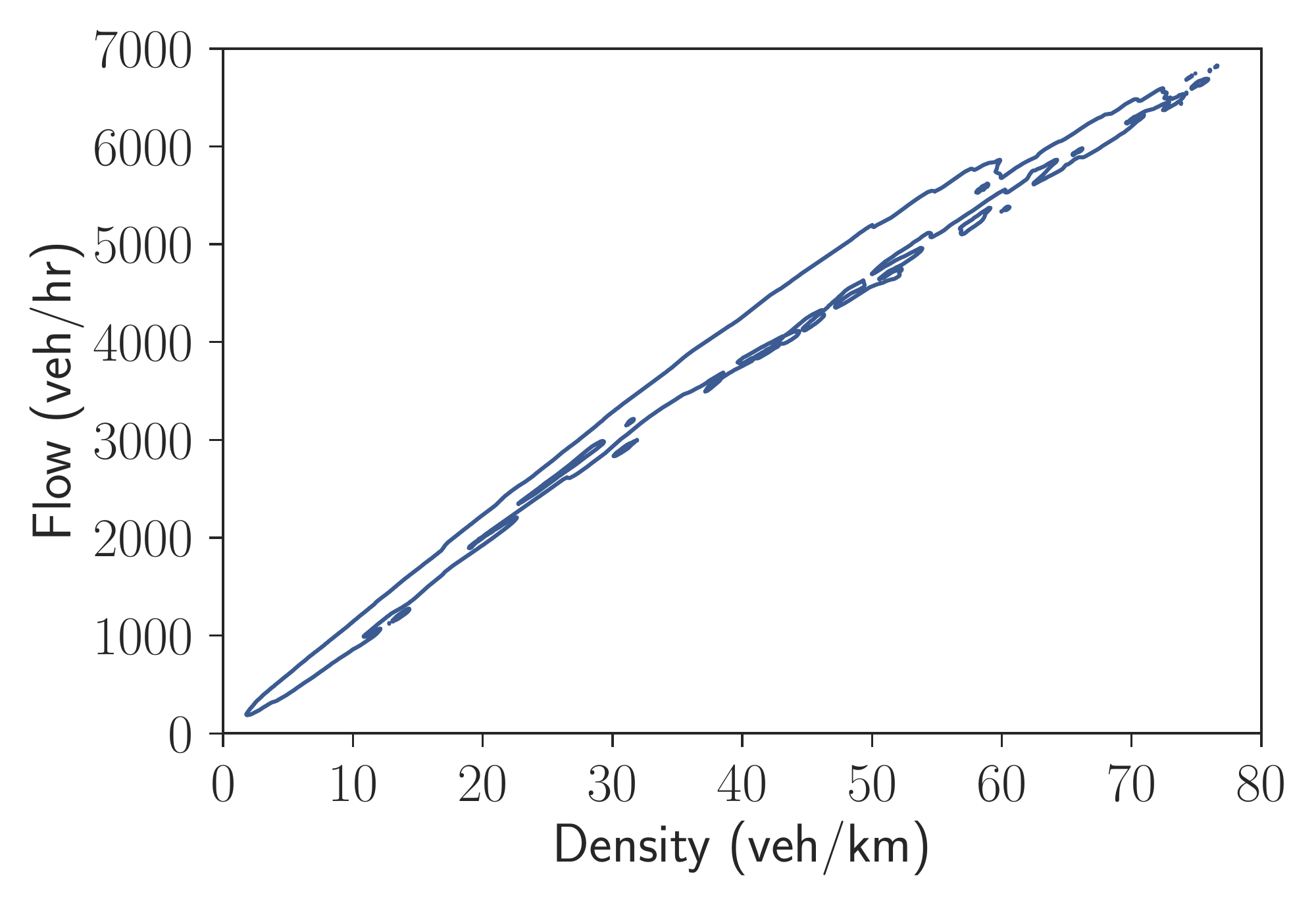}
	\end{subfigure}
	\begin{subfigure}{.32\textwidth}
		\includegraphics[width=\textwidth]{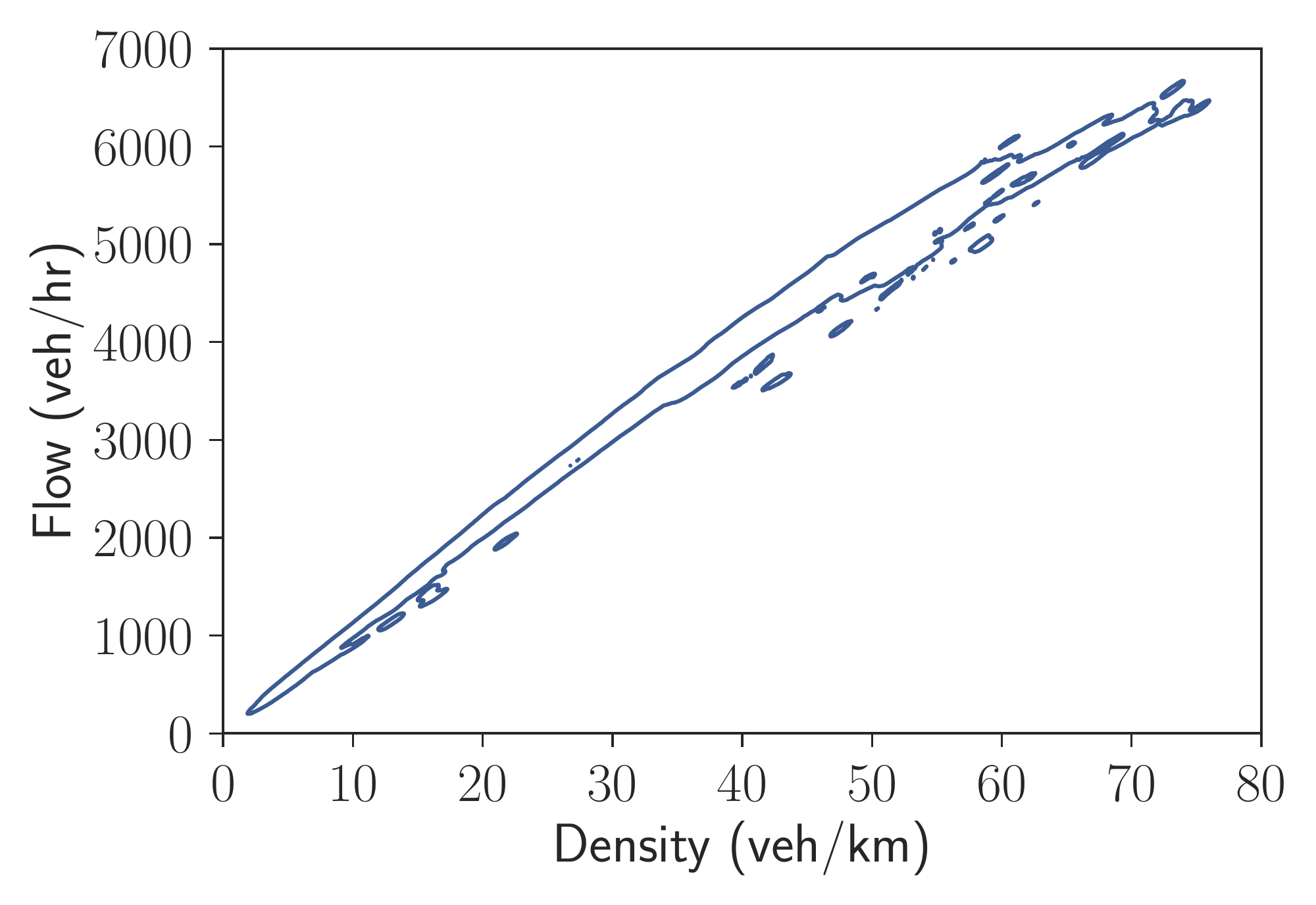}
	\end{subfigure}

	\begin{subfigure}{.32\textwidth}
		\includegraphics[width=\textwidth]{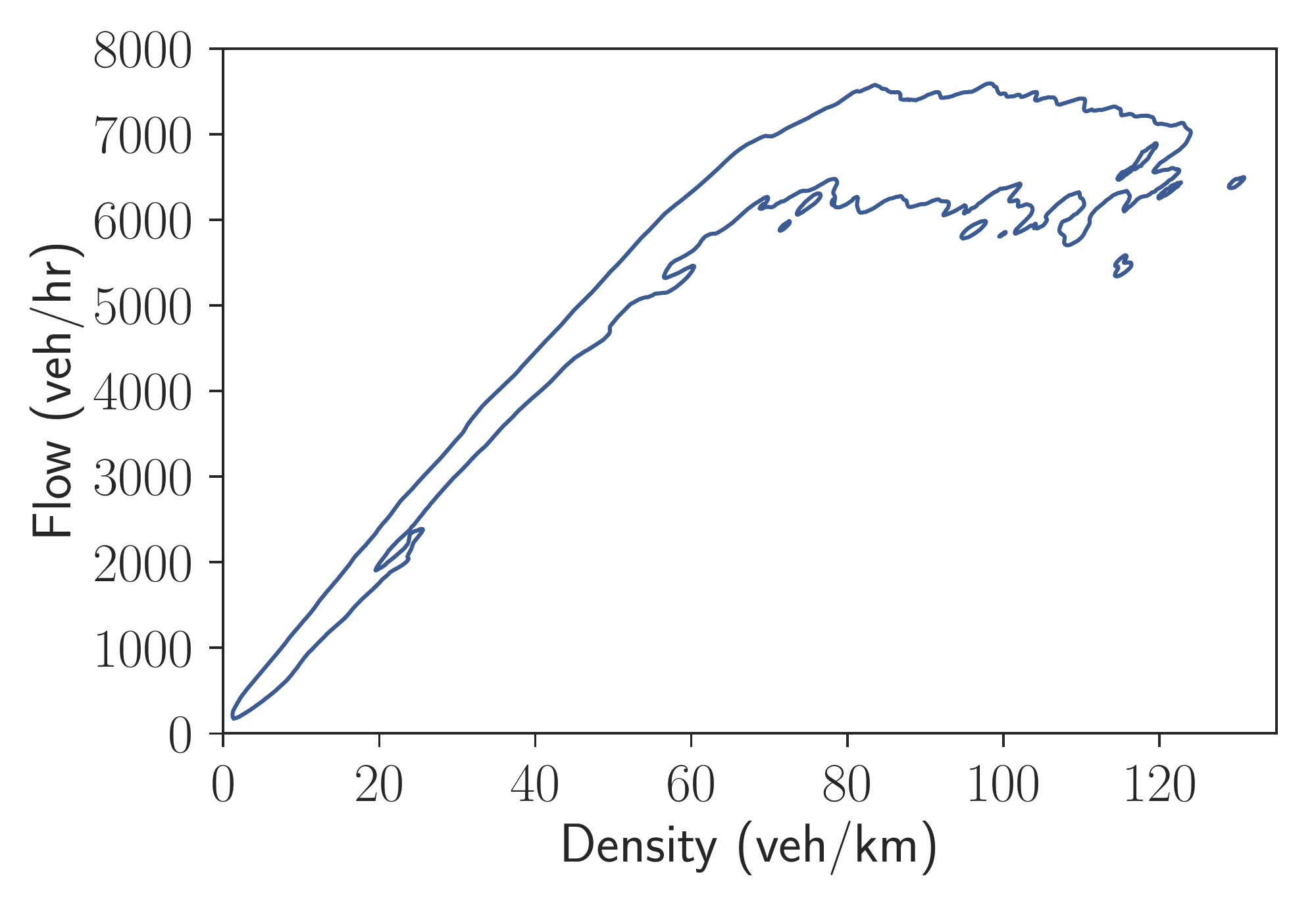}
	\end{subfigure}
	\begin{subfigure}{.32\textwidth}
		\includegraphics[width=\textwidth]{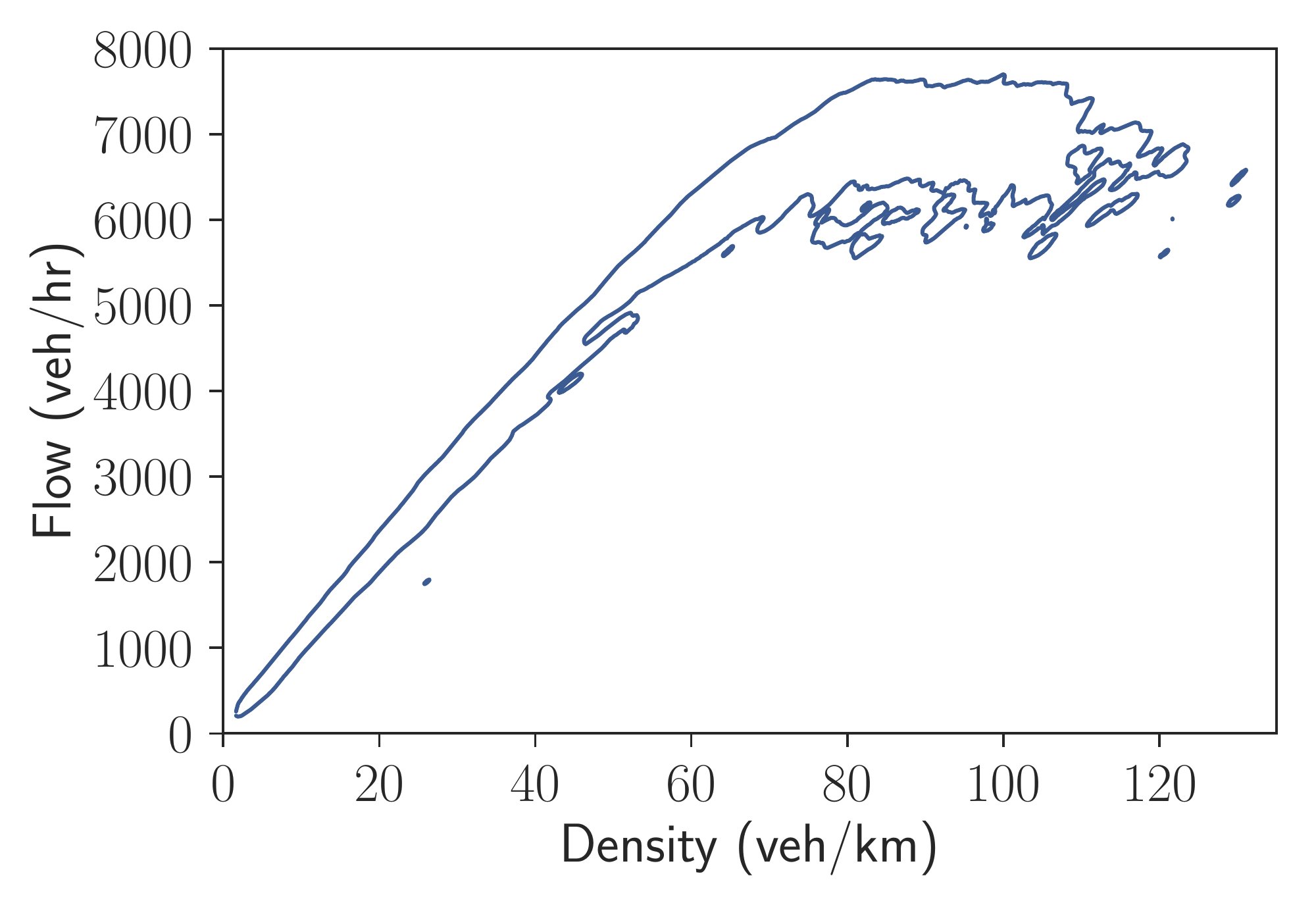}
	\end{subfigure}
	\begin{subfigure}{.32\textwidth}
		\includegraphics[width=\textwidth]{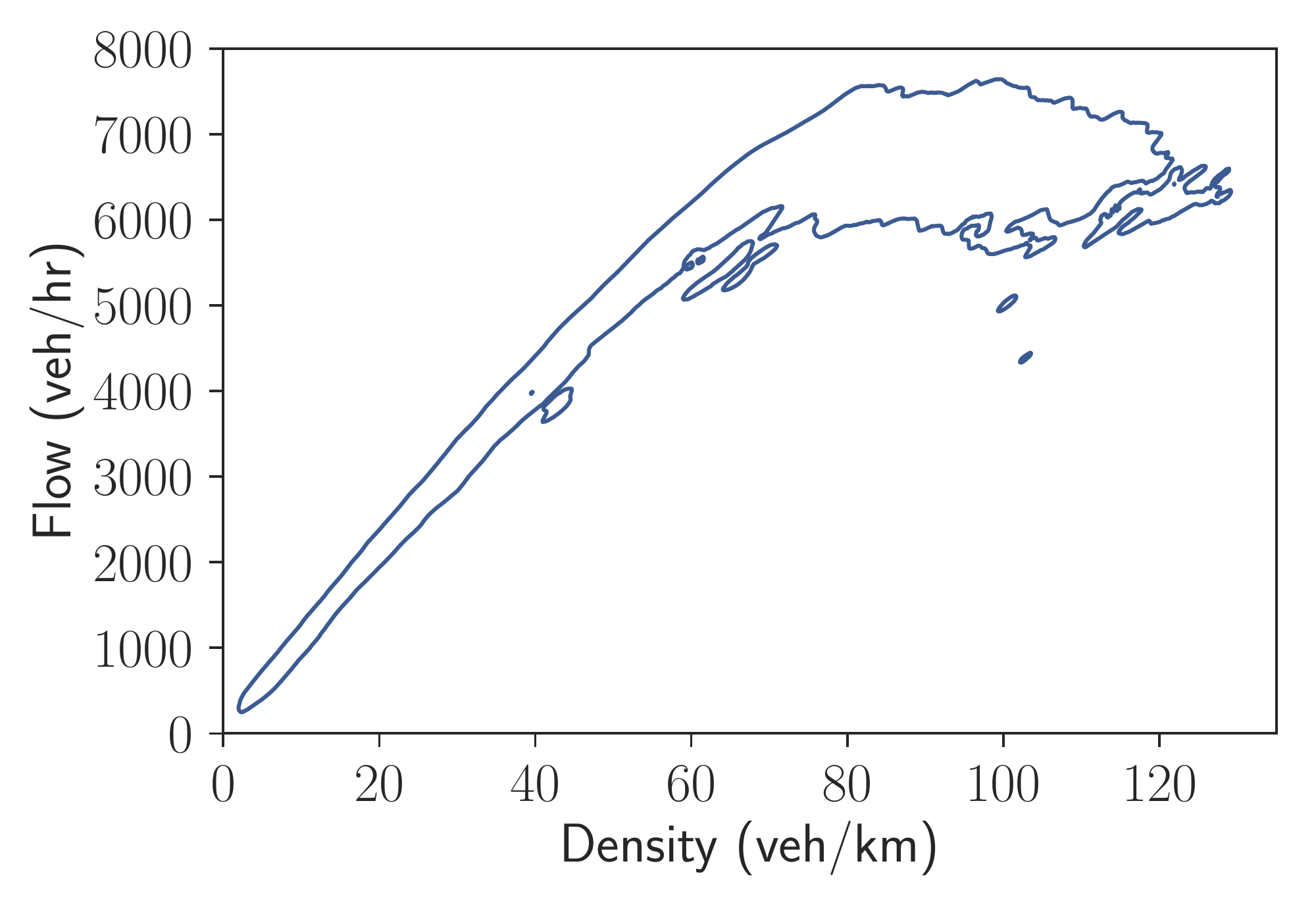}
	\end{subfigure}
	\caption[Stability Contour Plots]{ Contour Plots of 3-week segmented data. Top row: Segmented data taken from a link between junctions 2 and 3, 3635 kilometers in length. Bottom row: data taken from a link between junctions 15 and 16, 7146 kilometers in length. We see general stability using 3-week segmentation of the diagrams, shown by the similar shape of each diagram. Clearly, the top-row diagrams have significantly less flow breakdown than the bottom-row diagrams, however all 6 diagrams shown have accidents, obstructions and abnormal traffic events in, so it appears some links simply experience significant flow breakdown far more often than others. }\label{fig:StabilityPlots}
\end{figure*}

If we are to use this contour approach for any real-time applications, we should first be confident that the system we are observing is stationary over relevant timescales.
For our purposes, stationary means that when we look at subsets of our data, the density-flow relationship and the corresponding contour do not significantly change. 
Whilst we may get different events and severity of events in different periods of time, we would hope that the typical behaviour of the density-flow series remains reasonably constant. 
To test this, we construct the contours of typical behaviour using 3 disjoint 3-week subsets of data.
For each of the 3 week periods, we perform kernel density estimation and determine the 95\% contour.
We plot each for two different links, showcasing how different types of contour, one with a significant amount of data present at high densities can be stationary in Fig. \ref{fig:StabilityPlots}.

Considering Fig. \ref{fig:StabilityPlots}, we see generally similar diagrams across each 3 week-period, both for the contours with only low density (0-60 veh/km) data, and for the one with high-density data (up to 120 veh/km) present.
This showcases that despite clearly different dynamics a link may experience, we can still find a stationary distribution of the density-flow series due to the recurrent nature of traffic. 
All of the windows shown have events in them, however the typical behaviour remains constant on each link.
After verifying stationarity, we consider only contours fit to 3 weeks of data throughout this paper.

\section{Overview of Hypothesis Tests}\label{appendix:Hyptests}

Below we briefly discuss the mathematical concepts for three statistical tests relevant to our work. 
An extensive overview and discussion of statistical tests in general can be found in \cite{testing_statistical_hypotheses}.

\subsection{The Paired $t$-Test}

The paired $t$-test is perhaps the best known method to compare a dataset with paired observations.
We first collect a dataset with observations of some variable $x$.
In a paired setup, we have two groups of observations, each recording $x$ but in such a way that observations are naturally paired.
Examples of this include measuring a patients weight at the start of a study, and then again later after some change has been implemented.
In our case, we have measurements of some performance criteria that are paired due to being measured on the same links, and hence using the same data.
The test then considers each data pair $i \in \{1, \dots, N\}$, with values $x_{1,i}$ and $x_{2,i}$.
If we denote the average differences between pairs as $\bar{d}$, and the standard deviation of differences as $\sigma_{\bar{d}}$, the test statistic is computed as: 
\begin{equation}
t = \frac{\bar{d} - \tilde{\mu}_0}{\frac{\sigma_{\bar{d}}}{N}},
\end{equation}
with $\tilde{\mu}_0$ being the initial hypothesis for the average difference between samples.
This statistic has $N-1$ degrees of freedom, and can be compared to standard statistical tables to determine if the observed differences are statistically significant.

When using this test, if the sample size is small ($<30$ stated in \cite{statistical_comparsions_of_classifiers_over_multiple_data_sets}), then the differences in between the two variables compared must be normally distributed.
Additionally, since it uses the average difference, it is influenced heavily by outliers.

\subsection{The Wilcoxon Signed Rank Test}

The Wilcoxon Signed Rank Test is a non-parametric test used to compare paired observations.
Instead of using mean differences, it considers the ranks of variables.
Firstly, one assumes that the variable investigated has some natural ordering, the data is recorded in matched pairs and that each pair is chosen randomly and independently.
The null hypothesis of this test is then that the median difference between pairs is 0, following a symmetric distribution.
To test this, we first compute the absolute difference between paired data-points, and discard any that are equal in value, potentially reducing the sample size.
We then order the remaining data-pairs by the absolute magnitude of their difference.
The test statistic is computed as: 
\begin{equation}
W = \sum_{i=1}^N \left[ sign(x_{2,i} - x_{1,i}) \cdot R_i \right],
\end{equation}
with $R_i$ being the rank of the data-pair, and $sign(x)$ being -1 if $x<0$, 0 if $x=0$ and 1 otherwise.
The value of $W$ can then be compared to standard statistical tables, and a p-value attained from this.
Since this test uses ranks, outliers have less of an impact on the result.
However, the distributions of the differences between data-pairs may not actually be symmetric in practice.

\subsection{The Paired Sign Test}

Another non-parametric test using to compare paired observations is the sign test.
To conduct this, we first assume that the differences between data-points are independent, each difference comes from the same continuous population and data has a natural ordering.
We collect independent pairs of sample data, and then initially assume that given any pair $x_{1,i}, x_{2,i}$, $\mathbb{P}\left(x_{1,i} < x_{2,i}\right) = 0.5$.
This means for any measurement pair, it is equally likely that one groups observation is larger than the other groups.
If we denote the number of pairs where $x_{2,i}-x_{1,i} > 0$ as $S$, then under the null hypothesis $S$ follows a binomial distribution with $N$ trials and probability of success $0.5$.
Again, standard statistical tables can be used to determine a p-value given the measured test statistic.

\bibliographystyle{elsarticle-harv}
\bibliography{refrences}

\end{document}